\documentclass[prd,aps]{revtex4}
\usepackage{graphicx}
\usepackage{latexsym}
\usepackage{amsmath}
\begin{document}
\renewcommand{\theequation}{\arabic{section}.\arabic{equation}}

\title{Crystal, Superfluids, Supersolid and Hetero-Structure in
System of Two-Component Strongly-Correlated Bosons in a Cubic Optical Lattice}

\author{Ikuo Ichinose, Takumi Ishima, Naohiro Kobayashi, and Yoshihito Kuno}
\affiliation{Department of Applied Physics, Nagoya Institute of Technology,
Nagoya, 466-8555, Japan}

\date{\today\\[24pt]}

\begin{abstract}
In the present paper, we study finite-temperature phase structure of 
two-component hard-core bosons in a cubic optical lattice.
The system that we study in the present paper
is an effective model for the Bose-Hubbard model 
with strong on-site repulsions and is called bosonic t-J model. 
This model is a bosonic counterpart of the t-J model 
for the strongly-correlated electron systems like the high-temperature 
superconducting materials.
We study the model by means of path-integral methods and Monte-Carlo simulations.
We found that this system has a very rich phase structure including 
checkerboard-type ``insulating" state, superfluid, phase-separated state, 
inhomogeneous cloudlet state, etc.
We are also interested in the possible supersolid phase with both the
checkerboard order and superfluidity and found that additional nearest-neighbor
inter-species attractive force induces the supersolid state.
In the supersolid state, paired superfluid appears in addition to the superfluid
of single atom.
This result gives important insight into mechanism of the high-temperature
superconductivity of the cuprate.
\end{abstract}

%\pacs{67.85.Hj, 75.10.-b, 03.75.Nt}

\maketitle
%%%%%%%%%%%%%%%%%%%%%%%%%%%%%%%%%%%%%%%%%%%%%%%%%%%%%%%%%%%%%%%%%%%%%%
\section{Introduction}
\label{sec:intro}

In the last decade,  system of  ultra-cold atoms has been 
one of the most actively studied area in atomic and condensed matter physics.
Its various physical properties have been
investigated quite intensively by both experimental and theoretical methods\cite{optical}.
As dimensionality and interactions between particles are highly controllable
and also there are no effects of impurities and defects, cold atomic system
in an optical lattice is regarded as a ``final simulator" for quantum many-body
systems and it sometimes elevates a purely academic theoretical model to a
realistic one.

It is also expected that obtained knowledge of cold atomic
systems gives useful insights into physical properties of strongly-correlated 
many-body systems.
In the present paper, we shall investigate a system of two-component
bosons with strong repulsions.
We shall first introduce a bosonic t-J model,
which is an effective model of the Bose-Hubbard model\cite{BHM} with
strong on-site repulsive interactions between atoms\cite{btJ1,btJ2,btJ3}.
The original t-J model of fermions is a canonical model for the 
high-temperature ($T$) superconducting (SC) materials\cite{ftJ}.
The fermionic t-J model has been studied quite intensively since the discovery of the
high-$T_c$ SC, but precise knowledge of its phase structure etc is
still lacking partially because of difficulty of numerical study of the fermionic
system.
We expect that study on more tractable bosonic counterpart of the t-J model
is useful to understand the strongly-correlated quantum systems including
high-$T_c$ materials.
This is one of our motivation to study the bosonic t-J model.

Another motivation of the present study is to see if some exotic state
like the supersolid (SS), which was theoretically predicted in single-component
boson systems, exists in the multi-component system.
In 2004, it was reported that the SS state of $^4$He, i.e., solid with superfluidity (SF),
was observed by experiment\cite{SSHe}.
Soon after this report, possibility of SS in cold atom systems was
theoretically investigated.
Study of the hard-core bosons on a square lattice 
concluded that the realization of the SS was difficult in that system
unless a long-range interaction exists\cite{square}.
On the other hand for the hard-core bosons on a triangular lattice,
parameter region for possible SS state has been clarified\cite{triangle}.
In the SS of single-component boson, a density wave and SF coexist.
In the present paper, we shall study the possibility of SS in the two-component 
hard-core boson systems.
It should be remarked that the SS state 
{\em with paired superfluid and checkerboard order}
in the bosonic system corresponds to the coexisting phase of the
antiferromagnetism (AF) and SC in the electron system.
Recently this AF and SC coexisting phase was really observed for highly homogeneous
samples of high-$T_c$ materials\cite{exp}.
Then it is quite interesting to see how the SS emerges in the
present system.

This paper is organized as follows.
In Sec.II, we shall briefly review the derivation of the t-J model from
the Bose-Hubbard model.
Slave-particle operators are introduced to express faithfully the hard-core
nature of two-component bosons.
Then the model is studied by the mean-field theory.
Obtained phase structure at the total filling factor less than unity is
shown.
In particular, the phase diagram contains the parameter region 
corresponding to the SS.
In Sec.III, we study the system by means of Monte-Carlo (MC) simulations.
The system in grand-canonical ensemble is studied rather in detail.
At vanishing hopping of bosons, the system reduces to a spin model.
As it was already shown in the previous paper\cite{btJ3},
this system has three phases at finite $T$, i.e., antiferromagnetic (AF), 
ferromagnetic (FM) and paramagnetic (PM) phases
in terminology of spin model.
Then we turn on the hopping amplitude to investigate how the SF state
appears.
All the above three phases result in the SF phase as the hopping amplitude
is increased.
We are particularly interested in phase transition from the checkerboard
insulating phase (AF phase) to the SF.
It is verified that the phase transition is of first order with a large
hysteresis loop.
This result indicates that a coexisting phase of the AF solid and the SF
may appear.
We show that a phase separated (PS) state of the above two phases really exists 
at certain parameters.
We investigate whether the Josephson-like tunneling of SF through 
AF solid takes place. 

In Sec.IV, we study the system in canonical ensemble with fixed total
particle number.
Starting from the AF solid, 
we increase the hopping amplitude and study how 
phase of the system evolves.
We found that the system behaves rather differently depending on
the density of atoms.
At low-density region, there exists a single phase transition to the SF.
On the other hand for high-density region, 
we found another phase besides those observed in grand-canonical ensemble.
This new phase is composed of SF cloudlets (finite magnitude of droplets) 
in the AF solid background.
The system evolves as AF crystal $\rightarrow$ cloudlet state 
$\rightarrow$ PS state $\rightarrow$ SF.
Physical properties of each phase are investigated by calculating correlation
functions and snapshots.
In Sec.V, we study effects of nearest-neighbor (NN) interaction between atoms.
In particular we are interested in relation between a paired SF (PSF)
\cite{PSF} and the AF order.
In the absence of the NN interspecies attractive force, the PSF does not appear and 
the SF of one-body atom has no AF long-range order.
However we found that by adding the NN interactions between atoms, the state with
both the PSF and AF order emerges.
This result gives an important implication to mechanism of 
the high-$T_c$ SC of the cuprates.
Section VI is devoted for conclusion.

%%%%%%%%%%%%%%%%%%%%%%%%%%%%%%%%%%%%%%%%%%%%%%%%%%%%%%%%%%%%%%%%%%%%%%
\section{Model Hamiltonian and slave-particle representation}
\label{sec:model}
\setcounter{equation}{0}

As explained in the introduction, we shall study the t-J model 
of hard-core bosons in the cubic lattice at finite 
temperature ($T$)\cite{btJ3,btJ2d}.
Hamiltonian of the t-J model is derived from the 
Bose-Hubbard model whose Hamiltonian is given as,
\begin{eqnarray}
H_{\rm Hub}&=&-\sum_{r,i=1}^3 t_a(a^\dagger_{r+i}a_i+\mbox{h.c.})
-\sum_{r,i=1}^3 t_b(b^\dagger_{r+\mu}b_r+\mbox{h.c.})
+U\sum_r(n_{ar}-{1\over 2})(n_{br}-{1\over 2})  \nonumber  \\
&&+{1 \over 2}\sum_{r,\alpha=a,b}V_\alpha n_{\alpha r}(n_{\alpha r}-1)
-\sum_{r,\alpha=a,b}\mu_{c\alpha}n_{\alpha r},
\label{Hub}
\end{eqnarray}
where $r$ denotes site of the cubic lattice, $i(=1,2,3)$ is the 
unit vector in the $i$-th direction (it also sometimes denotes the
direction index), and $a_r$ and $b_r$ are boson annihilation operators.
$n_\alpha$ is the number operator of the boson $\alpha$, and therefore
$U$ and $V_\alpha$ are inter-species and intra-species interactions, respectively.
In the present paper, we shall consider the case $t_a, t_b \ll U,V_a,V_b$,
and {\em the total filling factor of bosons at each site less than unity}\cite{altman}.
Furthermore we set $t_a=t_b=t$ and $V_a=V_b$ and also the chemical
potential $\mu_{ca}=\mu_{cb}$.
Recently studied $^{85}$Rb-$^{87}$Rb atomic system
is a typical example corresponding to the present model\cite{Rb}.
More general cases like $t_a\neq t_b$ will be studied in a future publication.

Effective Hamiltonian in the large on-site repulsion limit, which is called
bosonic t-J model, can be 
obtained by the standard methods of expansion in powers of $t/U$ and
$t/V$,
\begin{eqnarray}
H_{\rm tJ}&=&-\sum_{r,i=1}^3 t(a^\dagger_{r+i}a_r
+b^\dagger_{r+i}b_r+\mbox{h.c.})
+J_z\sum_{r,i}S^z_{r+i}S^z_r  \nonumber  \\
&& -J_{\bot}\sum_{r,i}(S^x_{r+i}S^x_r+S^y_{r+i}S^y_r)
-\sum_{r}\mu_{cr}(1-n_{ar}-n_{br}),
\label{HtJ}
\end{eqnarray}
where pseudo-spin operator $\vec{S}_r={1 \over 2}B^\dagger_r\vec{\sigma}B_r$ with
$B_r=(a_r,b_r)^t$, $\vec{\sigma}$ is the Pauli spin matrices,
and up to the second order of the expansion
\begin{equation}
J_z={4t^2 \over U}-{4t^2 \over V}, \;\;\;
J_{\bot}={2t^2 \over U}.
\label{Js}
\end{equation}
In Eq.(\ref{HtJ}) $\mu_{cr}$ is the chemical potential of hole {\em at site} $r$.
In the following discussion, we shall treat $t,\; J_z$ and $J_{\bot}$ as free
parameters. 
After obtaining the critical couplings etc, we shall return to the
expression (\ref{Js}) and discuss the relation to the Hubbard model.
In the system $H_{\rm tJ}$ in Eq.(\ref{HtJ}), the physical state at each site $r$
is expanded by three orthogonal basis state vectors 
$\{|0\rangle, |a\rangle=a^\dagger_r|0\rangle, |b\rangle=b^\dagger_r|0\rangle\}$,
where $|0\rangle$ is the empty state of the bosons. 

%%%%%%%%%%%%%%%%%%%%%%%%%%%%%%%%%%%%%%%%%%%%%%%%%%%%%%%%%%%%%%%
%FIG.1
\begin{figure}[h]
\begin{center}
%\vspace{0.7cm}
\includegraphics[width=6cm]{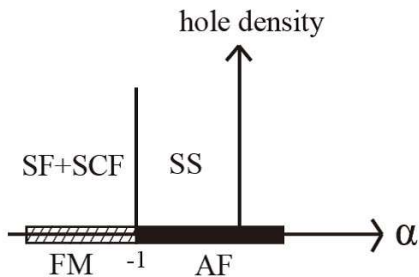}
%\vspace{-0.5cm}
\caption{%(Color online)
Phase structure of model in $\alpha-$hole density plane obtained by MFT,
where $\alpha=-J_{\bot}/J_z$. 
SCF stands for supercounter flow and SS for supersoild.
}
\label{fig:PD0}
\end{center}
\end{figure}
%%%%%%%%%%%%%%%%%%%%%%%%%%%%%%%%%%%%%%%%%%%%%%%%%%%%%%%%%%%

Before going into detailed numerical study on the system,
it is useful to investigate the groundstate properties of the model
by the mean-field theory (MFT).
We use a variational wave function of bosons that has 
a site-factorized form,
\begin{equation}
|\Psi\rangle=\prod_r\Big[\sin {\theta_r \over 2}
\Big(\sin {\chi_r \over 2}a^\dagger_r+\cos {\chi_r \over 2}b^\dagger_r
\Big)+\cos {\theta_r \over 2}\Big] |0\rangle.
\label{vwf}
\end{equation}
It is straightforward to calculate the expectation value of 
the Hamiltonian $H_{\rm tJ}$ (\ref{HtJ}) in the state $|\Psi\rangle$.
From $E_{\rm tJ}=\langle H_{\rm tJ}\rangle$, we can obtained a phase
diagram of the bosonic t-J model at vanishing $T$. 
Details have been already published in the previous paper\cite{btJ3}, 
and therefore here
we only show the phase diagram in Fig.\ref{fig:PD0}.
At the total filling factor unity, the system $H_{\rm tJ}$ reduces to 
the spin model, the anisotropic Heisenberg model.
We introduce parameter $\alpha=-J_\bot/J_z$ for the anisotropy.
At $\alpha=-1$, phase transition takes place, and for $\alpha>-1$ 
antiferromagnetic (AF) phase with 
$\chi_{r=\mbox{\small even}}\neq \chi_{r=\mbox{\small odd}}$
appears whereas for $\alpha<-1$ the pure
ferromagnetic (FM) phase with $\chi_r={\pi \over 2}$ is realized.
For the original atomic model, the AF phase corresponds to the state with
the checkerboard symmetry of $a$ and $b$ atoms.
On the other hand, the FM long-range order (LRO) means nonvanishing
condensation $\langle a^\dagger_r b_r\rangle \neq 0$, and sometimes
it is called supercounter flow (SCF) in the original atomic system.
As holes are doped, SF state with single-atom condensation 
$\langle a_r \rangle=\langle b_r \rangle\neq 0$ takes place.
In particular, the state with AF$+$SF is called supersolid (SS) 
as it has not only the checkerboard symmetry but also off-diagonal LRO
of the SF.

For numerical calculation, it is necessary
to express the physical state $|\mbox{phys}\rangle$ in the constrained Hilbert 
space faithfully.
To this end,
we introduce the following slave-particle representation\cite{btJ3},
\begin{eqnarray}
&& a_r=\phi^\dagger_rc_{r1}, \;\;\; 
b_r=\phi^\dagger_rc_{r2},  \label{slave}  \\
&& \Big(\phi^\dagger_r\phi_r+c^\dagger_{r1}c_{r1}+c^\dagger_{r2}c_{r2}-1\Big)
|\mbox{phys}\rangle =0,
\label{const}
\end{eqnarray}
where $\phi_r$ is a hard-core boson and $c_{r\ell=1,2}$ is an ordinary boson.
Equation (\ref{const}) is the local constraint on the physical state.
From Eq.(\ref{slave}), the hard-core boson $\phi_r$ creates the empty state of the
original bosons $a_r$ and $b_r$, whereas $c_{r1}$ ($c_{r2}$) creates 
the state of $a^\dagger_r|0\rangle$ ($b^\dagger_i|0\rangle$), i.e.,
\begin{equation}
|0\rangle \leftrightarrow \phi^\dagger_r|\Omega\rangle, \;\;
a^\dagger_r|0\rangle \leftrightarrow c^\dagger_{r1}|\Omega\rangle, \;\;
b^\dagger_r|0\rangle \leftrightarrow c^\dagger_{r2}|\Omega\rangle,
\label{slave2}
\end{equation}
where $|\Omega\rangle$ is the empty state of the slave particles.

In order to express the local constraint (\ref{const}) in more 
convenient way, we introduce a CP$^1$ boson (the Schwinger boson) $z_{r\ell}$,
\begin{eqnarray}
&& c_{r\ell}=(1-\phi^\dagger_r\phi_r)z_{r\ell}, \;\; (\ell=1,2)  \nonumber \\
&& \Big(\sum_{\ell=1,2}z^\dagger_{r\ell}z_{r\ell}-1\Big)
|\mbox{phys}\rangle_{z} =0.
\label{CP1}
\end{eqnarray}
It is easily verified that Eq.(\ref{const}) is satisfied by Eq.(\ref{CP1}).
The hard-core boson $\phi_r$ itself can be expressed in terms of 
another CP$^1$ boson $w_{r\ell}$ as follows,
\begin{equation}
\phi_r=w^\dagger_{r2}w_{r1}, \;\;
\Big(\sum_{\ell=1,2}w^\dagger_{r\ell}w_{r\ell}-1\Big)
|\mbox{phys}\rangle_{w} =0.
\label{w}
\end{equation}
It is easily verified $[\phi_r,\phi^\dagger_r]_+=1$, and operators 
$\phi^\dagger_r,\; \phi_{p\neq r}$ etc commute with each other.
From Eq.(\ref{w}), it is obvious that 
$|0\rangle_\phi=w^\dagger_{r2}|0\rangle_w$ and 
$\phi^\dagger_r|0\rangle_\phi=w^\dagger_{r1}|0\rangle_w$.
The effective Hamiltonian $H_{\rm tJ}$ can be easily expressed in terms
of the CP$^1$ bosons $z_{r\ell}$ and $w_{r\ell}$.

%%%%%%%%%%%%%%%%%%%%%%%%%%%%%%%%%%%%%%%%%%%%%%%%%%%%%%%%%%%%%%%%%%%%%%
\section{System in grand-canonical ensemble Revisited: Monte-Carlo
simulations}
\label{sec:gce}
\setcounter{equation}{0}

\begin{figure}[h]
\begin{center}
%\vspace{0.7cm}
\includegraphics[width=6cm]{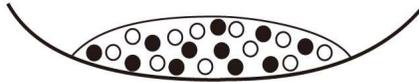}
%\vspace{-0.5cm}
\caption{%(Color online)
Physical picture of GCE, in which density of atoms is controlled by
chemical potential.
}
\label{fig:pic_GCE}
\end{center}
\end{figure}

In this section, we study the system in the grand-canonical ensemble(GCE).
Some results were already reported in the previous papers\cite{btJ3}.
In the present section, we shall study the system in an {\em inhomogeneous 
chemical potential} $\mu_{cr}$ and physical properties of a hetero-structure
of the AF state and superfluid.
In the cold atom system in an optical lattice, the system
in the GCE corresponds to the case of nearly flat confining potential. 
See Fig.\ref{fig:pic_GCE}.
Generally speaking, around a first-order phase transition point at which particle density 
changes drastically, very subtle adjustment of the chemical potential
is required to study the critical behavior of the system.
Study of the system in the canonical ensemble (CE) with fixed particle 
density is complementary and helpful for investigation of first-order phase transition.
In the present paper, the study of the system in the CE will be reported 
in the following section.

%%%%%%%%%%%%%%%%%%%%%%%%%%%%%%%%%%%%%%%%%%%%%%%%%%%%%%%%%%%%%%%
%FIG.2
%%%%%%%%%%%%%%%%%%%%%%%%%%%%%%%%%%%%%%%%%%%%%%%%%%%%%%%%%%%

Partition function of the system in the GCE at temperature $T$,
$Z_{\rm GCE}$, is given by the path-integral methods as follows\cite{btJ3},
\begin{equation}
Z_{\rm GCE}=\int [D\bar{w}DwD\bar{z}Dz]
e^{-\beta H_{\rm tJ}(\bar{w},w,\bar{z},z)},
\label{ZGCE}
\end{equation}
where $\beta=1/(k_{\rm B}T)$ and the Hamiltonian
$H_{\rm tJ}(\bar{w},w,\bar{z},z)$
is obtained from $H_{\rm tJ}$ in  Eq.(\ref{HtJ}) by substituting 
the CP$^1$ variables $z_{r}$ and $w_{r}$ for the CP$^1$ operators. 
We have  numerically studied the system (\ref{ZGCE}) and obtained 
phase diagram\cite{btJ3} by calculating the internal energy $E$ and 
the specific heat $C$, which are defined by 
\begin{equation}
E={1 \over L^3} \langle \beta H_{\rm tJ}(\mu_{cr}=0) \rangle, \;\;\;
C={1 \over L^3}\langle (\beta H_{\rm tJ}(\mu_{cr}=0)-E)^2 \rangle,
\label{UC}
\end{equation}
where $L$ is the linear system size.
We impose the periodic boundary condition for practical calculation,
and use the following {\em dimensionless parameters} for describing
phase diagram etc,
\begin{equation}
c_1=\beta J_z, \;\; c_3=\beta t, \;\; \alpha=-J_\bot/J_z.
\label{parameter}
\end{equation}

%%%%%%%%%%%%%%%%%%%%%%%%%%%%%%%%
\subsection{Phase diagram}

%%%%%%%%%%%%%%%%%%%%%%%%%%%%%%%%%%%%%%%%%%%%%%%%%%%%%%%%%%%%%%%
%FIG.3
\begin{figure}[h]
\begin{center}
%\vspace{0.7cm}
\includegraphics[width=6cm]{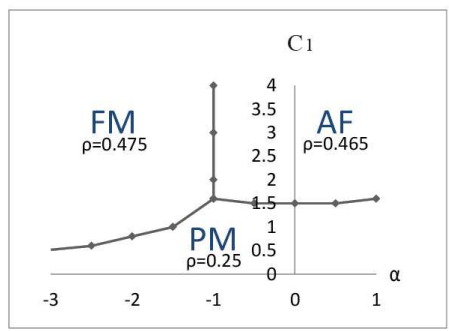}
%\vspace{-0.5cm}
\caption{%(Color online)
Phase structure of model for $t=\mu_c=0$ in $\alpha-c_1$ plane,
where $\alpha=-J_{\bot}/J_z$ and $c_1=\beta J_z$. 
There are three phases, PM, AF and FM phases.
Physical meaning of each phase is explained in the text.
Typical value of particle density $\rho=\langle a^\dagger_r a_r \rangle=
\langle b^\dagger_r b_r \rangle$ in each phase is also shown.
}
\label{fig:PD1}
\end{center}
\end{figure}
%%%%%%%%%%%%%%%%%%%%%%%%%%%%%%%%%%%%%%%%%%%%%%%%%%%%%%%%%%%

We first consider the case of $t=0$, i.e., the system without 
particle hopping.
In Fig.\ref{fig:PD1}, we show the obtained phase diagram in the 
$\alpha-c_1$ plane for ${\mu}_{cr}=0$ and $c_1>0$.
Details of the calculation have been already reported in the previous
paper\cite{btJ3}.
%Similar phase diagrams are obtained for other constant chemical potentials.
There are three phases, paramagnetic (PM), $z$-component-AF and 
$xy$-component-FM phases.
At high-$T$, the system is in the PM phase without any long-range order, 
as it is expected.
As $T$ is lowered, phase transition to the AF or FM phase takes place.
The AF state corresponds to the checkerboard state and the FM state to
the state with the SCF $\langle a^\dagger_rb_r\rangle \neq 0$, 
as we explained in Sec.II.
These phase transitions are of second order.
In the low-$T$ region, 
there exists a line of the phase boundary between 
the AF and FM phases  at $\alpha=-1$ as predicted by the MFT.
On the line  $\alpha=-1$, the pseudo-spin symmetry is enhanced to SU(2)
as the inter-species and intra-species interactions between atoms are
the same, otherwise the symmetry is U(1)$\times Z_2$, i.e., global
rotation of $(S^x_r, S^y_r)$ and reflection $S^z_r \rightarrow -S^z_r$.
The enhancement of the symmetry at $\alpha=-1$ is seen by the 
redefinition of operators
as $a_r\rightarrow -a_r$ for $r\in$ odd site.
By this redefinition, $(S^x_r, S^y_r) \rightarrow -(S^x_r, S^y_r)$ for  $r\in$ odd site,
and then $J_{\bot}\rightarrow -J_{\bot}$.
For  ferroelectric materials, similar
transition line to the present $\alpha=-1$ one is called a 
morphotropic phase boundary and plays a very important role\cite{MPB}.

%%%%%%%%%%%%%%%%%%%%%%%%%%%%%%%%%%%%%%%%%%%%%%%%%%%%%%%%%%%%%%%
%FIG.4
\begin{figure}[h]
\begin{center}
\includegraphics[width=5cm]{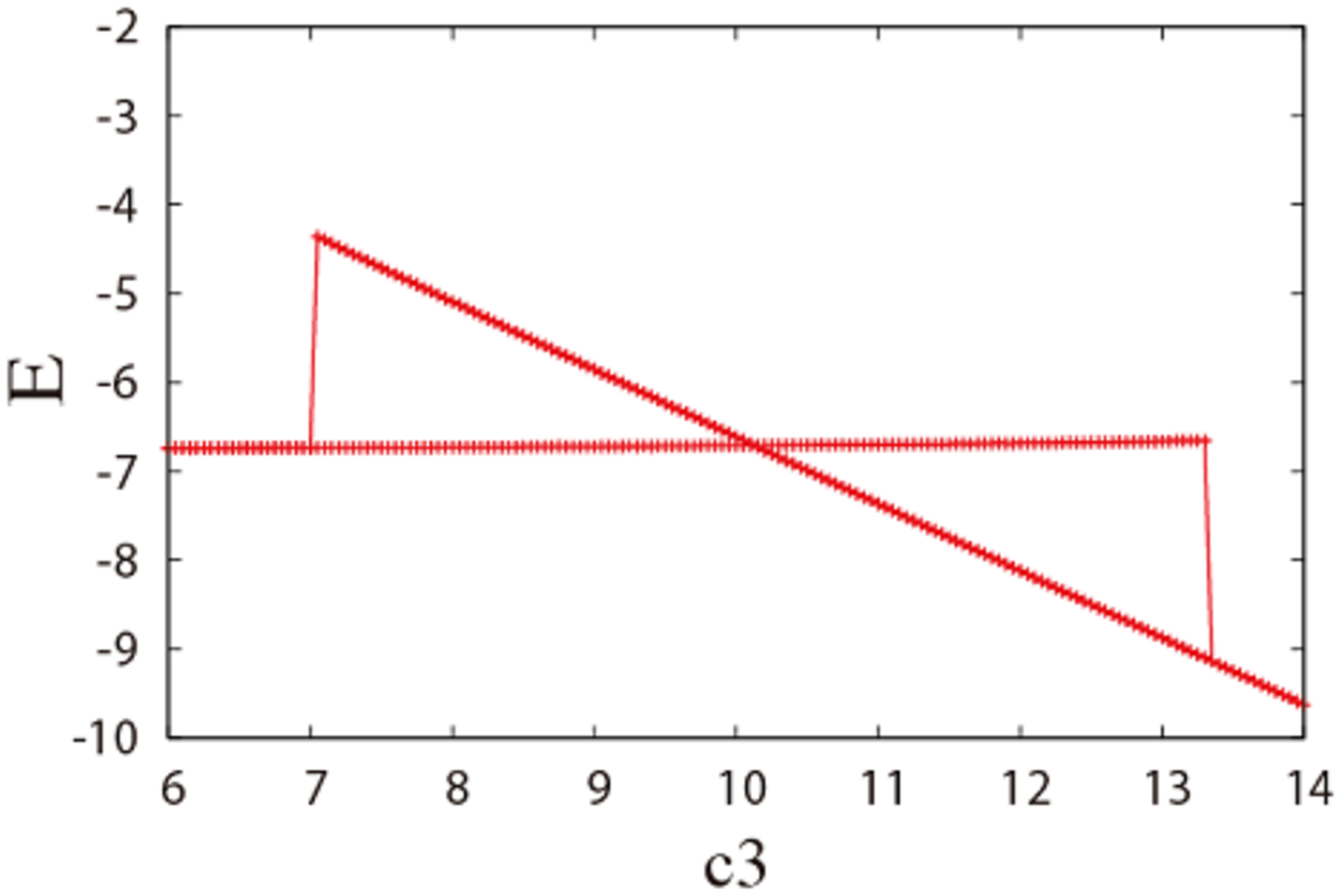}
\hspace{0.5cm}
\includegraphics[width=5cm]{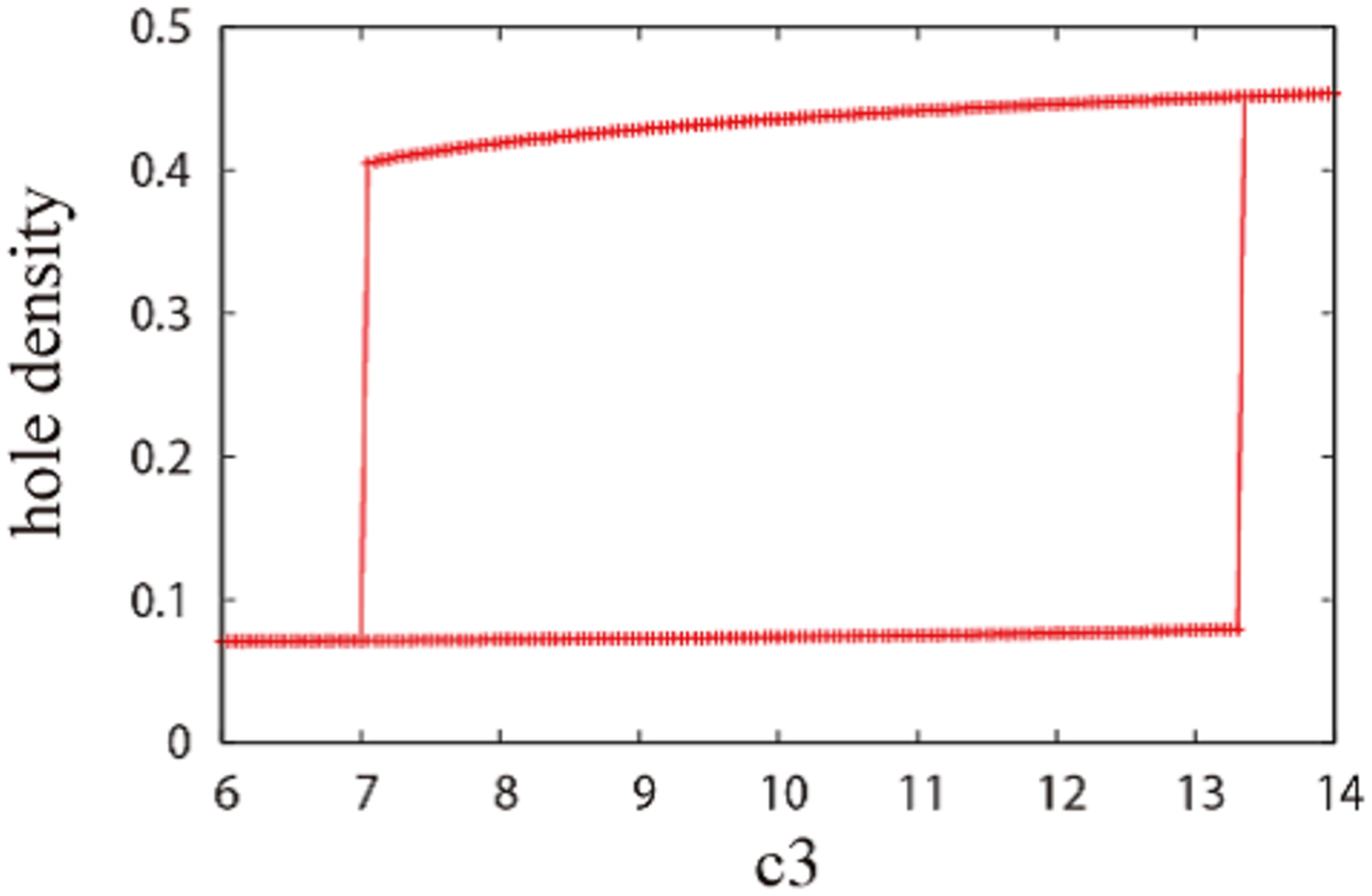}
\vspace{-0.5cm}
\caption{%(Color online)
Internal energy and hole density as a function of $c_3=\beta t$ for 
$c_1=3.0$ and $\alpha=-0.5$.
Phase transition from AF state to SF is of first order.
}
\label{figAF-SF}
\end{center}
\end{figure}
%%%%%%%%%%%%%%%%%%%%%%%%%%%%%%%%%%%%%%%%%%%%%%%%%%%%%%%%%%%

Let us turn on the hopping amplitude $t$.
We choose typical three points in the  $\alpha-c_1$ plane
for each of the three phases existing at $t=0$, and
investigated how the phases
change as the value of $c_3=\beta t$ is increased.
We have found that all of the three phases make phase transition to the
SF phase. 
In Fig.\ref{figAF-SF}, we show $E$ and hole density for the phase transition 
from the AF phase to SF. 
These results were obtained by the MC simulations of
local-update Metropolis algorithm.
More elaborate update methods slightly decreases the region of
the hysteresis loop\cite{btJ3}.
From the above calculations, 
it is obvious that the phase transition is of first order.
To verify the finite superfluidity, we calculated the boson 
correlation function $\langle B^\dagger_0\cdot B_r \rangle$, and found 
it has a nonvanishing correlation as $r \rightarrow$ large
after the phase transition.
We also have measured the spin correlation function and found that the
SF phase has a FM spin correlation. 
Physical meaning of the appearance of the FM correlation has been
explained in the previous paper\cite{btJ3}(see also discussion in Sec.IV).
Emergence of the first-order phase transition stems from the fact
that both of the phases separated by the phase transition
have the LRO's, i.e., the AF and SF+FM orders, respectively.

Existence of the first-order phase transition suggests an interesting
possibility of various phases in the parameter region of hysteresis loop,
e.g.., (i)phase separation (PS) of AF state and SF, 
(ii) supersolid with coexistence of the global AF order and SF, 
(iii) {\em quantum superposed state of the AF and SF cloudlets}.
We have studied the above problem in both the GCE and the
canonical ensembles.
In the GCE of a constant chemical potential, in which the average 
number density is {\em not} conserved as the parameters in the Hamiltonian vary, 
we have found that in most of the parameter region
only either of the AF solid or SF is stable in the {\em flat potential}
with $\mu_{cr}=$ constant,
and the values of parameters $t$ and $\mu_{cr}$ control which phase appears.
In fact,
we choose an inhomogeneous state as an initial state and update 
the configuration by means of the standard Metropolis algorithm.
After $10^5$ updates, the whole system tends to become either
the  AF state with checkerboard order or the SF with intermediate 
homogeneous hole density. 
Only at very specific value of $c_{3c}=9.362$, the stable PS state of the 
AF and SF phases appears.
This means that the genuine first-order phase transition takes place at  $c_{3c}=9.362$,
and the system is composed of immiscible AF and SF domains at the 
phase transition point.

Here one should remark that the phase transition takes place at
$\beta {t^2 \over U}, \; \beta {t^2 \over U} \sim O(1)$ and
$\beta t \sim O(10)$.
Then ${t \over U}, \; {t\over V} \sim O({1 \over 10})$.
This means that the phase transition observed above in the t-J model
also exists in the original Hubbard model.

%%%%%%%%%%%%%%%%%%%%%%%%%%%%%%%%%%%%%%%%%%%%%%%%%%%%%%%%%%%%%%%%%%%%%%
\subsection{Heterojunction of superfluid and AF crystal}
\label{sec:HJ}
%%%%%%%%%%%%%%%%%%%%%%%%%%%%%%%%%%%%%%%%%%%%%%%%%%%%%%%%%%%%%%%
%FIG.5
\begin{figure}[h]
\begin{center}
\includegraphics[width=6cm]{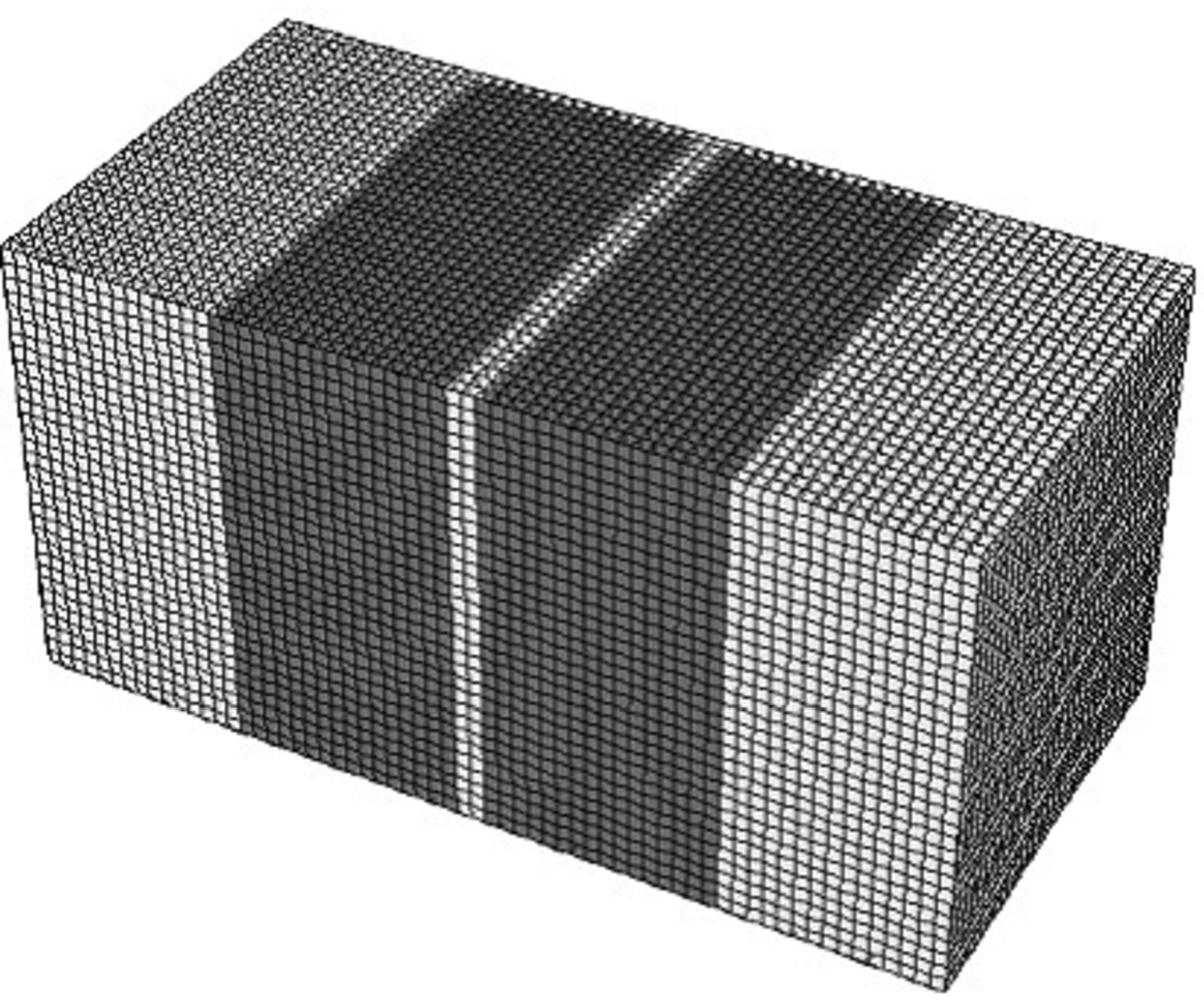}
\hspace{0.5cm}
\includegraphics[width=6cm]{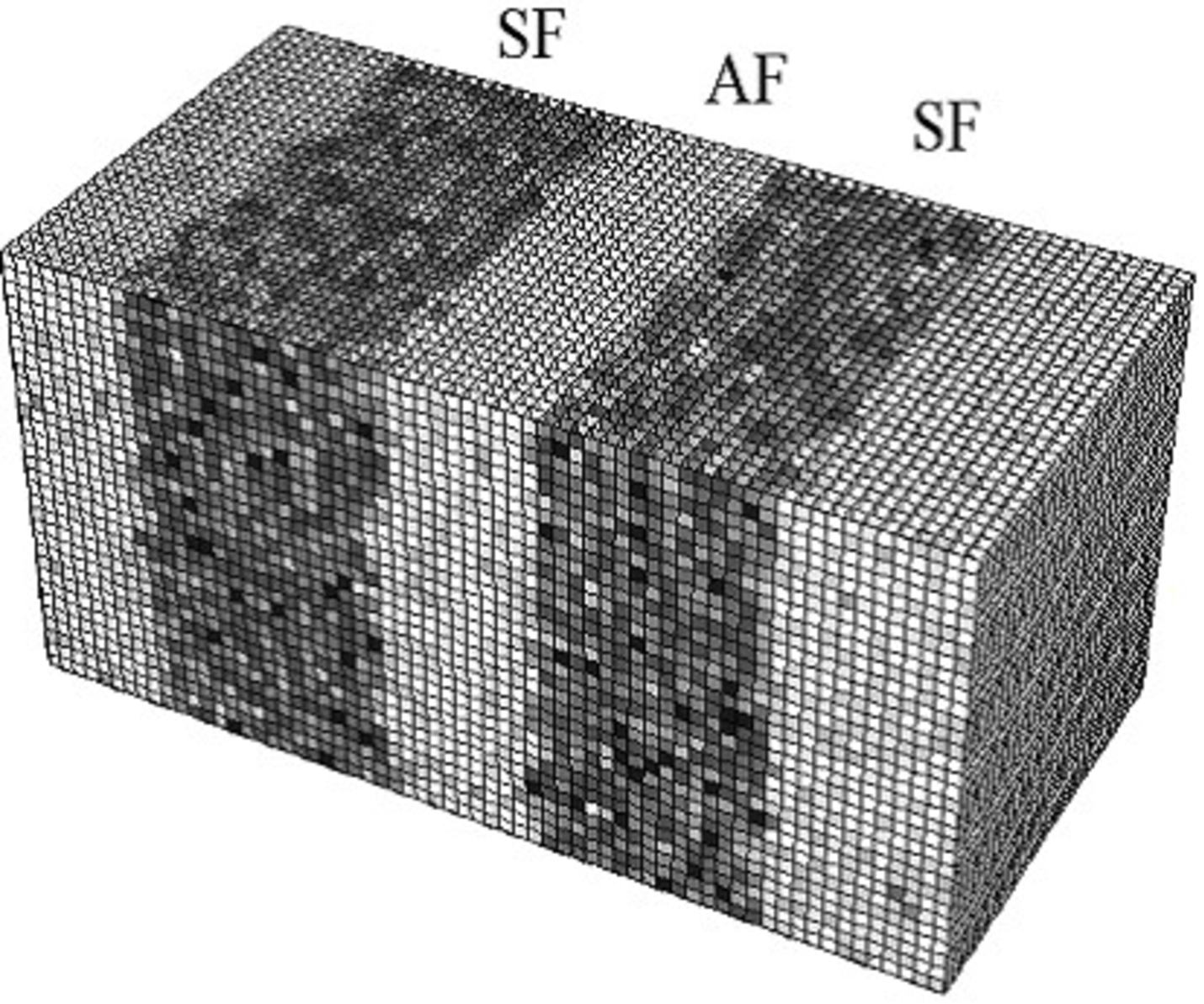}
%\vspace{-0.5cm}
\caption{%(Color online)
Heterojunction of SF and AF state. 
(Left) Initial state before update.
(Right) After 8$\times 10^5$ sweeps.
The central {\em bright} region is the AF solid state with {\em high particle density},
whereas the {\em dark} regions represent state of {\em high hole density}.
Confining potential is given by the chemical potential with 
$\Lambda=40$ and $N_\ell=2$.
}
\label{fig:HJ1}
\end{center}
\end{figure}
%%%%%%%%%%%%%%%%%%%%%%%%%%%%%%%%%%%%%%%%%%%%%%%%%%%%%%%%%%%
%%%%%%%%%%%%%%%%%%%%%%%%%%%%%%%%%%%%%%%%%%%%%%%%%%%%%%%%%%%%%%%
%FIG.6
\begin{figure}[h]
\begin{center}
\includegraphics[width=9cm]{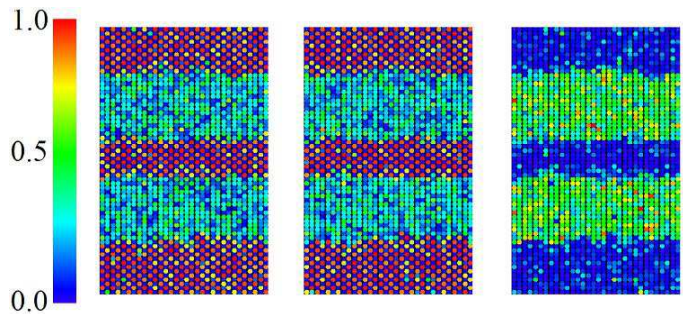}
\caption{%(Color online)
Snapshots in a horizontal plane of heterojunction of SF and AF state in Fig.\ref{fig:HJ1}.
(Left) Density of $a$-atom
(Center)Density of $b$-atom
(Right) Density of hole.
It is obvious that $a$ atom and $b$ atom reside on each sublattice in the AF region.
}
\label{fig:SHHJ}
\end{center}
\end{figure}
%%%%%%%%%%%%%%%%%%%%%%%%%%%%%%%%%%%%%%%%%%%%%%%%%%%%%%%%%%%

In the previous subsection, we showed that both the AF and
SF states can coexist at the first-order phase transition region.
In the GCE with constant chemical potential, 
either the AF or SF state is realized as a ``homogeneous" state
except at the genuine critical coupling $c_{3c}=9.362$.
However in the confinement potential of real experiments, 
the density of the atoms is inhomogeneous.
In the central region in which the atomic density is higher, 
the AF state is expected to appear,
whereas in the outer region the SF state with a lower atomic density
is expected to appear.
If the inhomogeneous PS state is realized around the first-order phase transition point,
it resembles to the classical solid-fluid PS state like coexisting state of ice and water.
However as the present system is a quantum system, 
one may expect that some interesting and characteristic phenomena of
macroscopic quantum system occur in such a coexisting state.
One example is a kind of the Josephson tunneling in a heterojunction of the 
AF ``crystal" state and SF.
In the present case, the both AF state and SF are made of $a$ and $b$ atoms.
Therefore, to study physical properties of heterojunction of these states
is very interesting.

In this subsection, we study the above problem by means of MC simulations.
To specify spatial direction of the AF layer, we introduce a ``confining
potential" in terms of the chemical potential $\mu_{cr}$.
In the practical calculation, we make an AF layer parallel to 
the $y-z$ plane at $x\sim 0$
by adjusting chemical potential as $\mu_{cr}=\Lambda(>0)$ for 
$x=0,\pm 1,\cdots, \pm N_l$, and otherwise zero. 
As expected, a stable AF layer is generated after the MC local 
update of the system.
We show some example in Figs.\ref{fig:HJ1} and \ref{fig:SHHJ}.
Please notice that we are employing the periodic BC.

%%%%%%%%%%%%%%%%%%%%%%%%%%%%%%%%%%%%%%%%%%%%%%%%%%%%%%%%%%%%%%%
%FIG.7
\begin{figure}[h]
\begin{center}
\includegraphics[width=6cm]{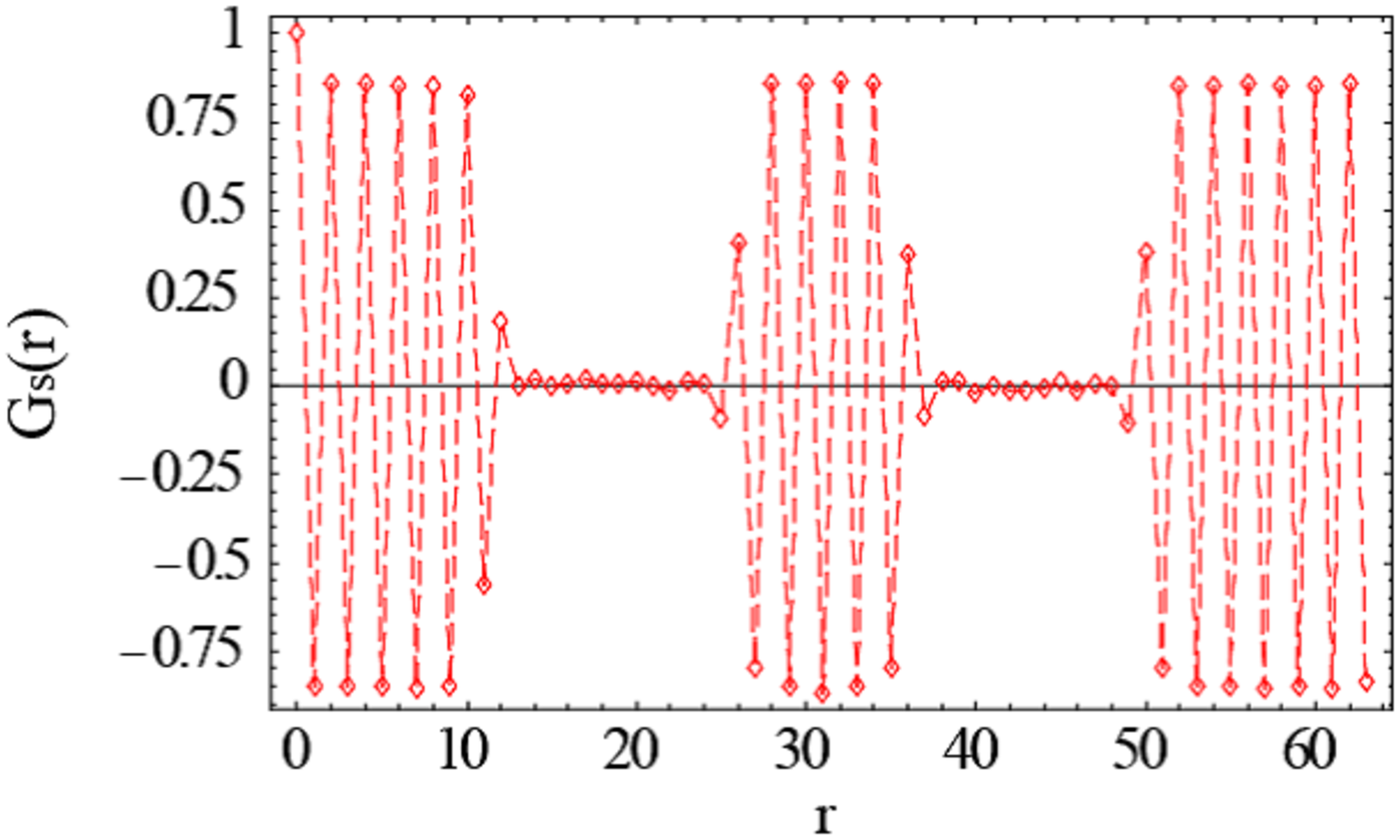}
\hspace{1cm}
\includegraphics[width=6cm]{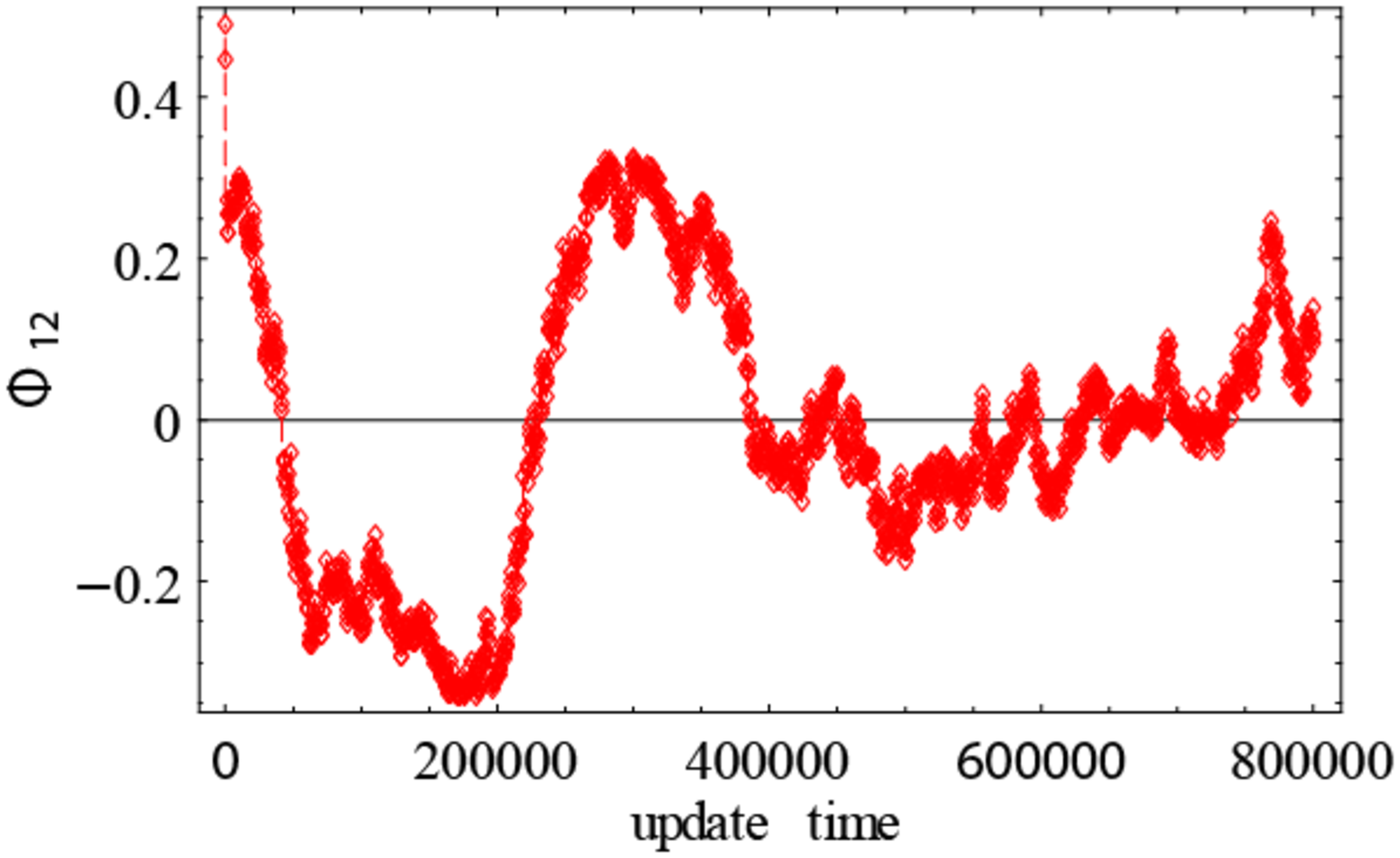}
%\vspace{-0.5cm}
\caption{%(Color online)
Physical properties of 
heterojunction of SF and AF state. 
(Left) Spin correlation $G^\bot_{\rm S}(r)$.
There exists no spin correlations in SF region.
(Right)Phase difference $\Phi_{12}$ of Bose condensates in two SF regions
as a function of update. 
It fluctuates strongly.
}
\label{fig:HJ2}
\end{center}
\end{figure}
%%%%%%%%%%%%%%%%%%%%%%%%%%%%%%%%%%%%%%%%%%%%%%%%%%%%%%%%%%%

We first show the spin correlation function for the above
heterojunction,
\begin{equation}
G^\bot_{\rm S}(r)=\langle \vec{S}_{r_0}\cdot \vec{S}_{r_0+r} \rangle,
\label{GS_x}
\end{equation}
where the site $r_0$ is located in one of the AF layers and 
$r$ is a lattice vector perpendicular to the AF and SF layers.
It is obvious that there exist no spin correlations 
in the SF regions, whereas in the AF regions the checkerboard
configurations of $a$ and $b$ atoms are realized.

In this PS state, it is interesting to investigate if the tunneling of SF current takes
place through the AF layer.
To study this problem, we measured the correlation function of the
bosons through the intermediate AF layer.
If there exists a finite correlation between SF states separated by the
AF layer, ``Josephson tunneling" occurs as in the SC-insulator heterojunction.
In Fig.\ref{fig:HJ2}, we show the result of the SF correlation
$\Phi_{12}$ that is defined as 
\begin{equation}
\Phi_{12}={\rm Im} \Big[\log \Big(\langle B^\dagger_{r_1}\cdot B_{r_2}\rangle\Big)\Big],
\label{Phi12}
\end{equation}
where $r_1$ ($r_2$) is located inside of the left (right) SF layer in 
Fig.\ref{fig:HJ1}.
It is obvious that correlation of two SF states fluctuates randomly 
in the MC updates.
We examined some other cases and obtained similar results.
This result is unexpected because both the AF state and SF are made of
$a$ and $b$ atoms and then it is expected that atoms move almost freely
through the interface.
In order words, there exists a sharp boundary between the two states and 
the ``classical state", i.e., the AF solid, and the quantum state, i.e., SF,
have very different properties.

%%%%%%%%%%%%%%%%%%%%%%%%%%%%%%%%%%%%%%%%%%%%%%%%%%%%%%%%%%%%%%%%%%%%%%
\section{Phase separation and inhomogeneous states in canonical ensemble: 
Monte-Carlo simulations}
\label{sec:ce}
\setcounter{equation}{0}
%%%%%%%%%%%%%%%%%%%%%%%%%%%%%%%%%%%%%%%%%%%%%%%%%%%%%%%%%%%%%%%
%FIG.8
\begin{figure}[h]
\begin{center}
%\vspace{0.7cm}
\includegraphics[width=4cm]{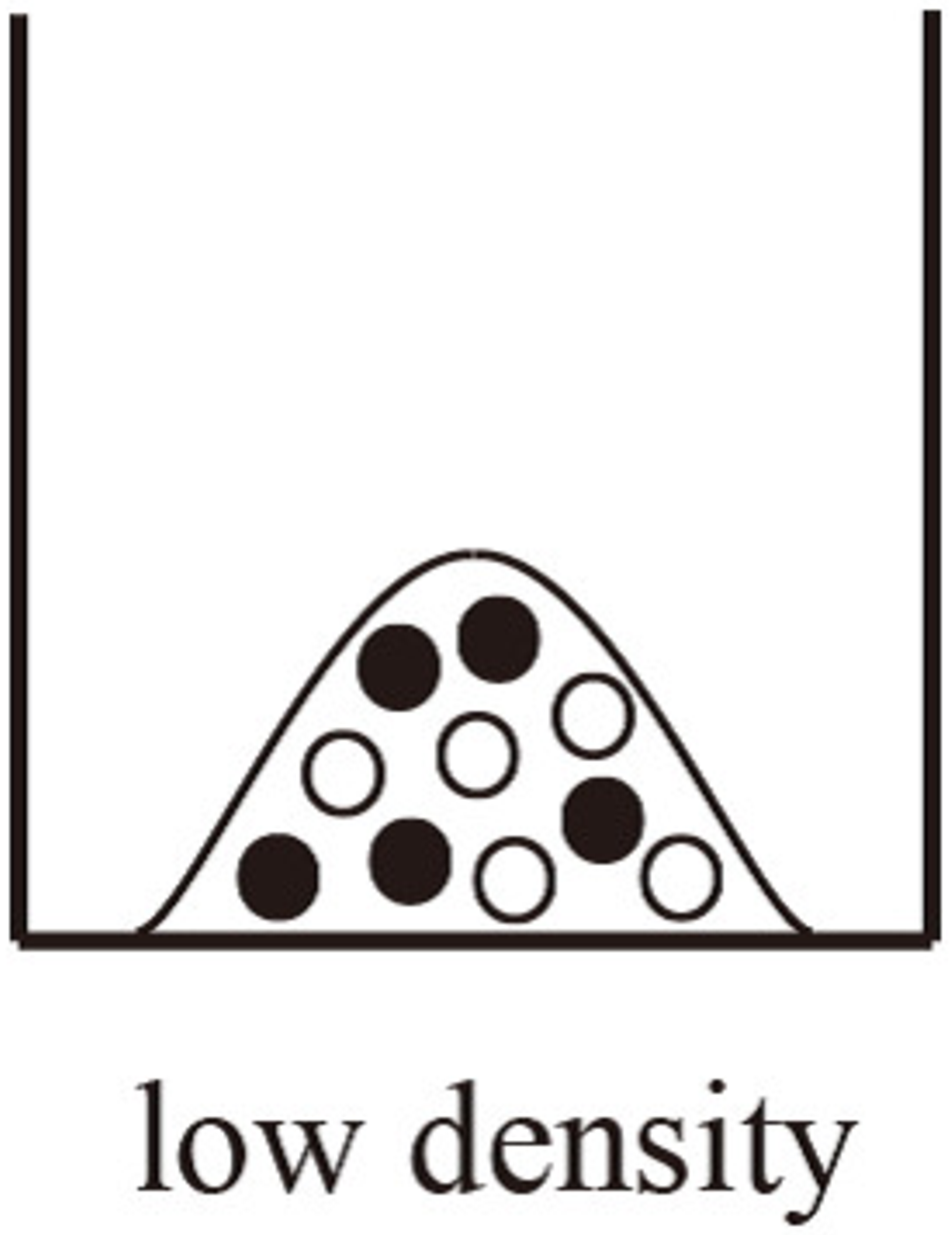}
\hspace{1cm}
\includegraphics[width=4cm]{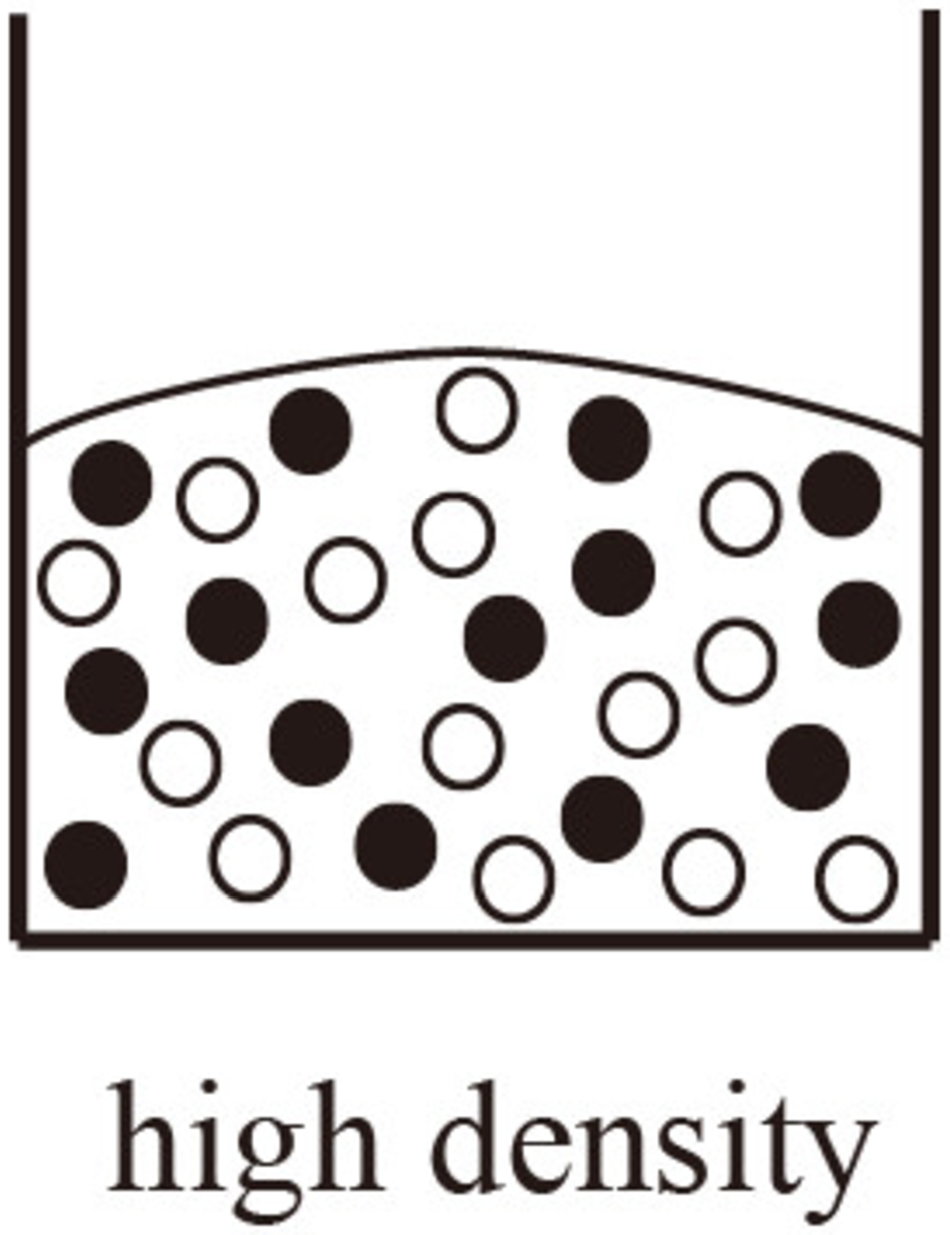}
%\vspace{-0.5cm}
\caption{%(Color online)
Physical picture of CE, in which the average density of atoms is fixed.
}
\label{fig:pic_CE}
\end{center}
\end{figure}
%%%%%%%%%%%%%%%%%%%%%%%%%%%%%%%%%%%%%%%%%%%%%%%%%%%%%%%%%%%

In the previous section, we studied the system in the GCE with fixed
chemical potential.
In particular we are interested in the AF phase and SF that are
separated by the first-order phase transition.
In the GCE, in most of the parameter region it is observed that
either the homogeneous AF state or the homogeneous SF state is stable.
However in the previous studies on some related systems
like hard-core bosons on a triangular lattice, the PS state
is observed in the CE even if it is not found in the GEC\cite{triangle}.
This result stems from the fact that
the CE corresponds to the physical situation of fixed averaged particle density
and therefore atoms are assumed to be confined in a certain flat region
with a sharp boundary like a box,
see Fig.\ref{fig:pic_CE}.
Then it is interesting to study the present system in the CE and
see if a SS state or some related state exists in some parameter region.
To this end, we employ the MC simulations with the standard 
Metropolis algorithm.
The typical sweeps for measurement is $(30000 \sim 50000)\times (10$ samples)
and the acceptance ratio is $40\% \sim 50\%$.
Errors are estimated from 10 samples with the jackknife methods.

%%%%%%%%%%%%%%%%%%%%%%%%%%%%%%%%%%%%%%%%%%%%%%%%
\subsection{Low-density region}

By the practical calculation, we found that physical properties
of the system, including
phase structure itself, changes as the average particle density varies.
In this subsection, we study the system with average atomic density
$\rho_a\equiv {1 \over L^3}\sum_r\langle a^\dagger_r a_r\rangle
=\rho_b\equiv {1 \over L^3}\sum_r\langle b^\dagger_r b_r\rangle=15\%$ 
and the system size $L=24$.
This case corresponds to the low-density region.
For small hopping amplitude $t$ and $\alpha>-1$,
it is expected that the system at low $T$ tends to separate into
the AF solid and ``empty" region,
as the Boltzmann factor dominates the entropy factor at low $T$.
As the hopping amplitude $t$ is increased, a phase transition to the
SF is expected to takes place as observed by the simulations in the GCE in the 
previous section.
This expectation has been actually confirmed by the MC simulations.

\subsubsection{Large $J_z$}

\begin{figure}[h]
\begin{center}
\includegraphics[width=6cm]{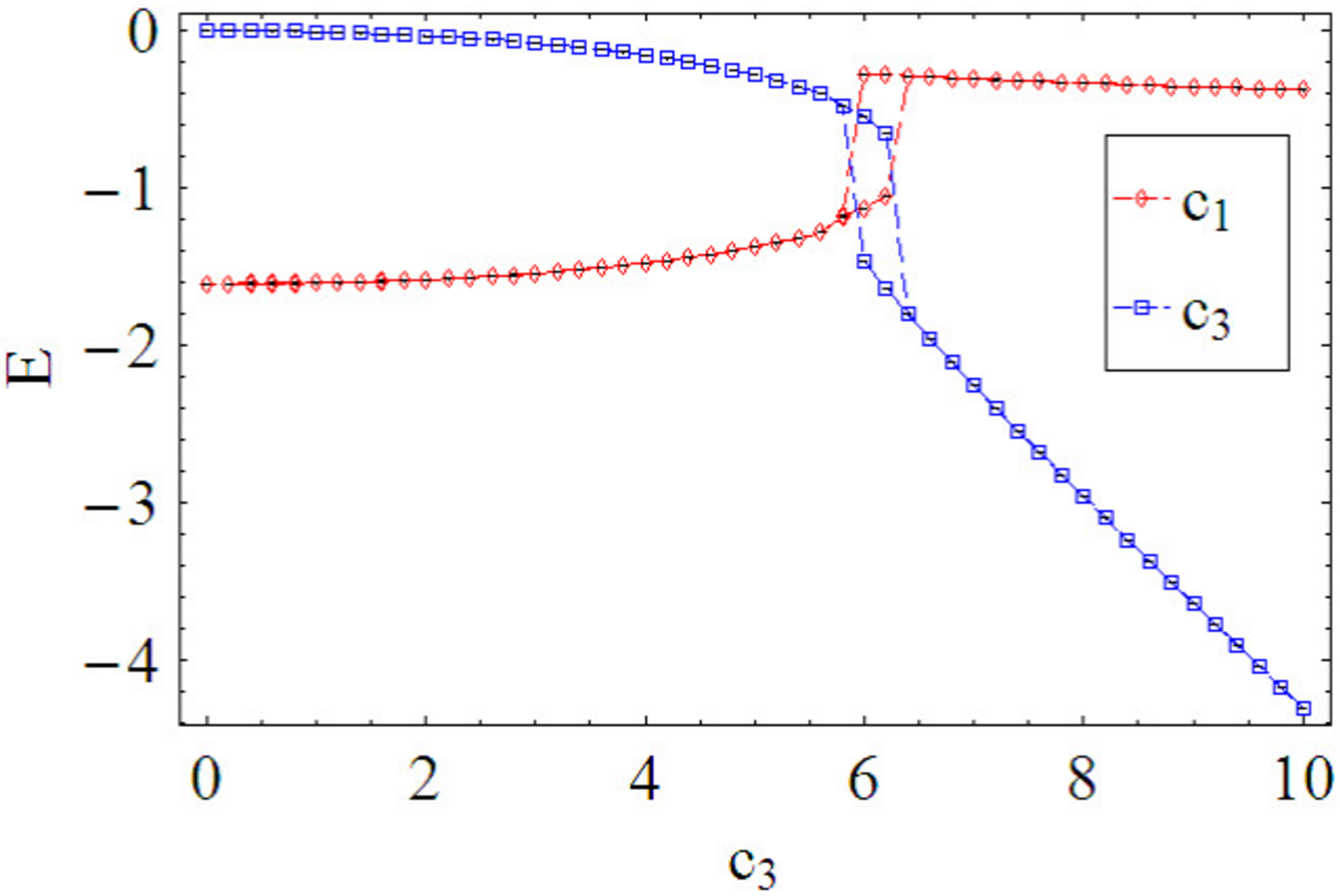}
\hspace{1cm}
\includegraphics[width=6cm]{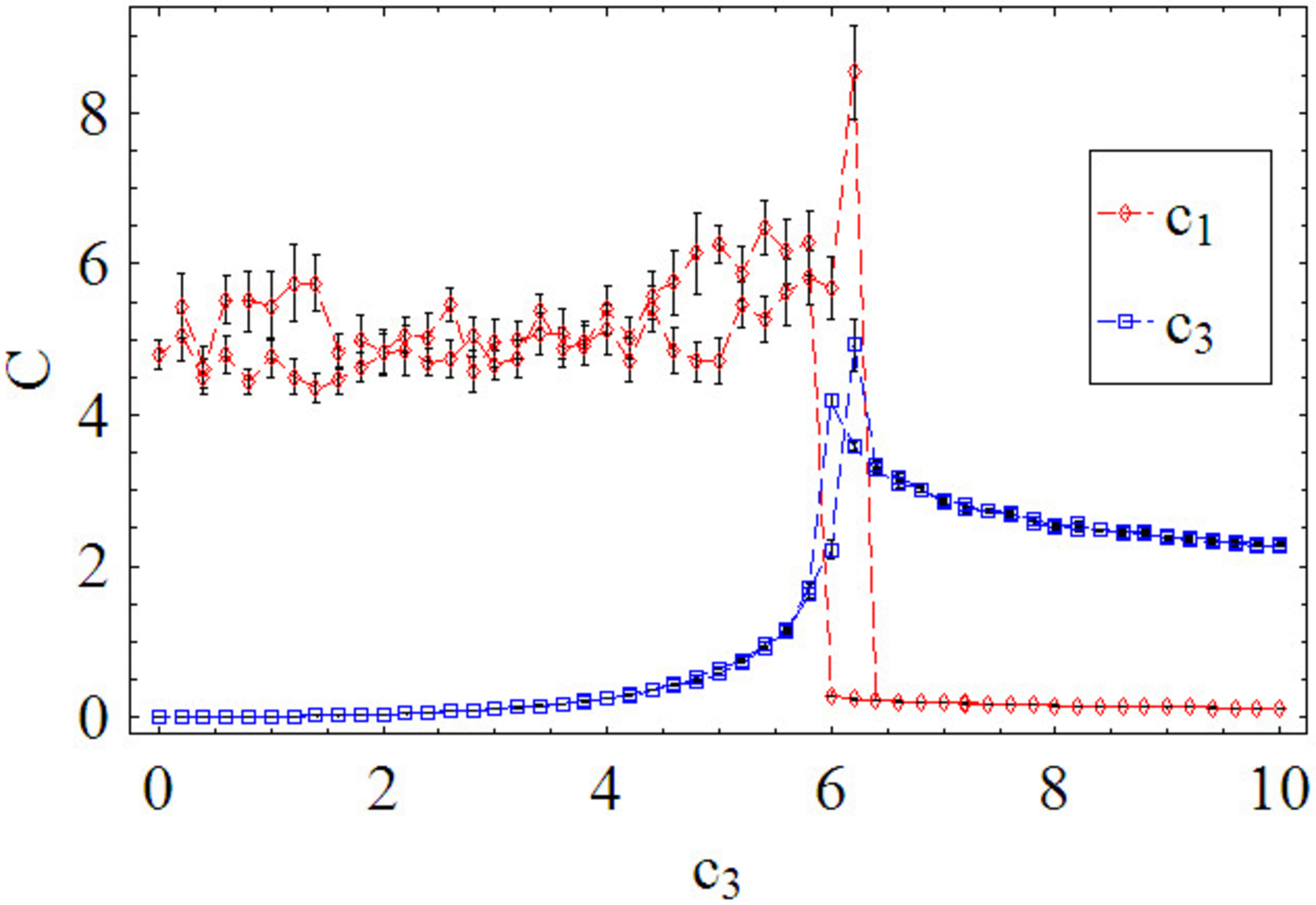}
%\vspace{-0.5cm}
\caption{%(Color online)
Internal energy and specific heat of $c_1$ and $c_3$ terms as a function of $c_3$ 
for system with $\rho_a=\rho_b=15\%$ in CE. 
Results show existence of first-order phase transition at 
$c_3\simeq 6.0$.
}
\label{fig:EC_0.7_1}
\end{center}
\end{figure}

\begin{figure}[h]
\begin{center}
\includegraphics[width=5cm]{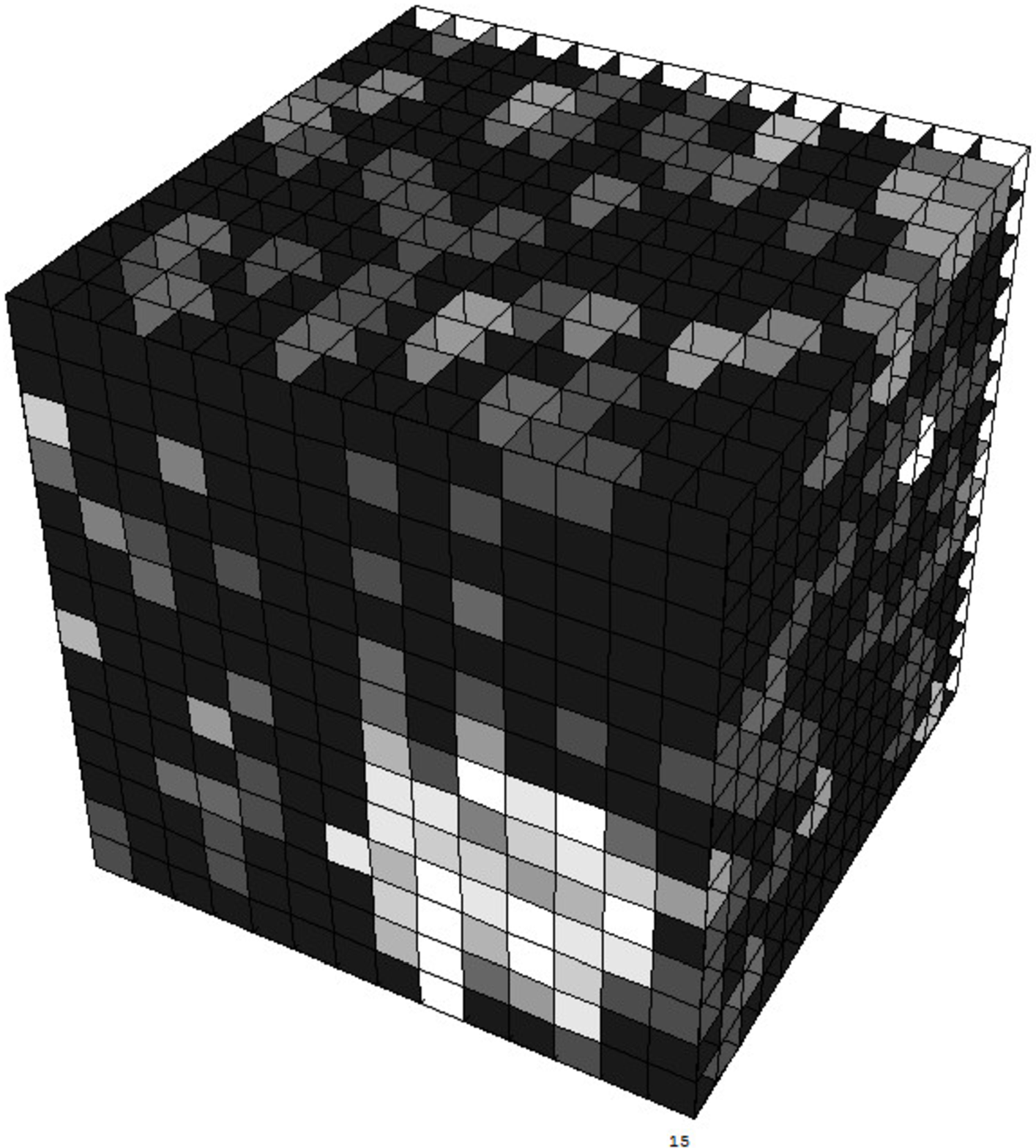}
\hspace{1cm}
\includegraphics[width=5cm]{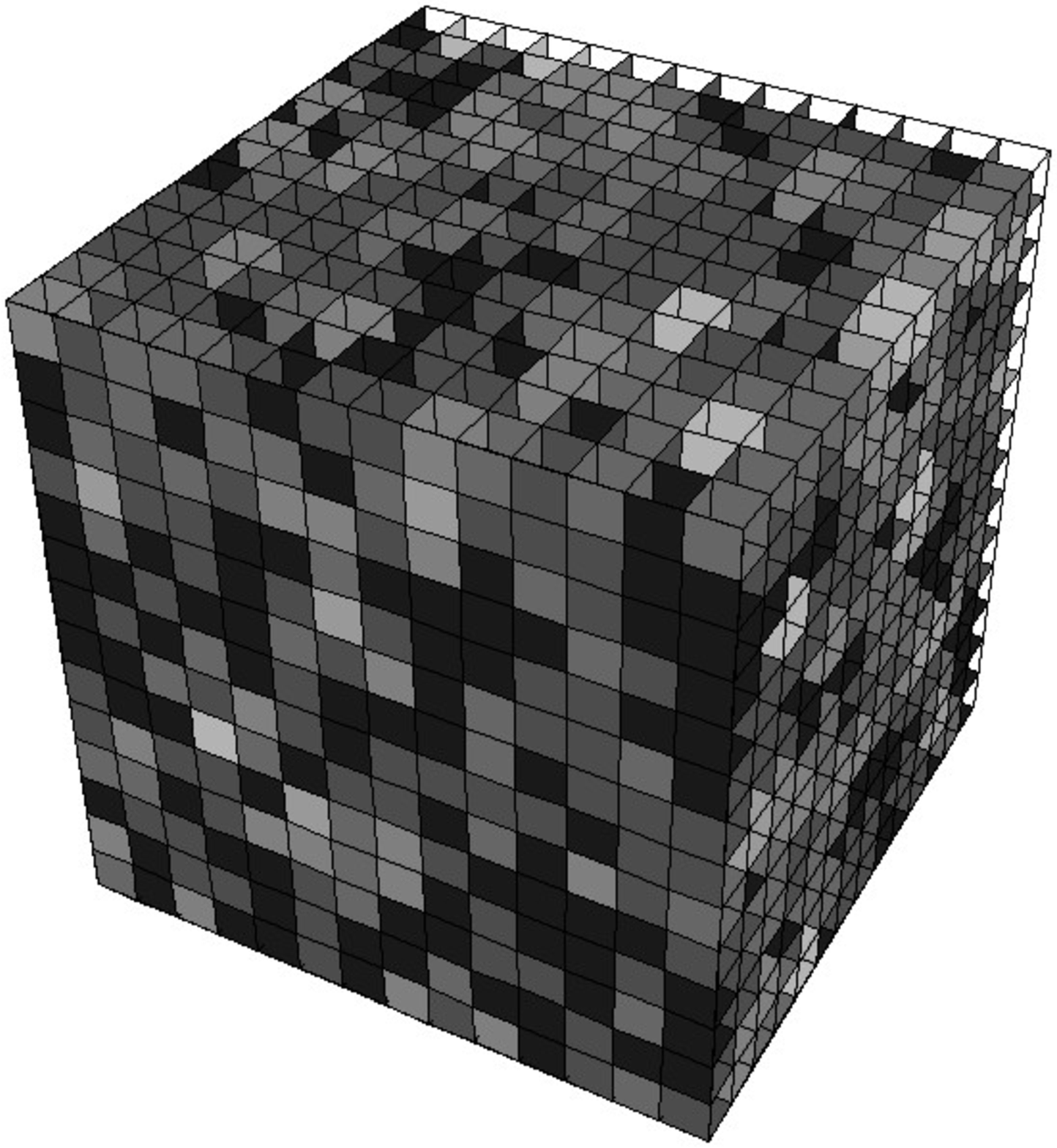} \\
\includegraphics[width=3cm]{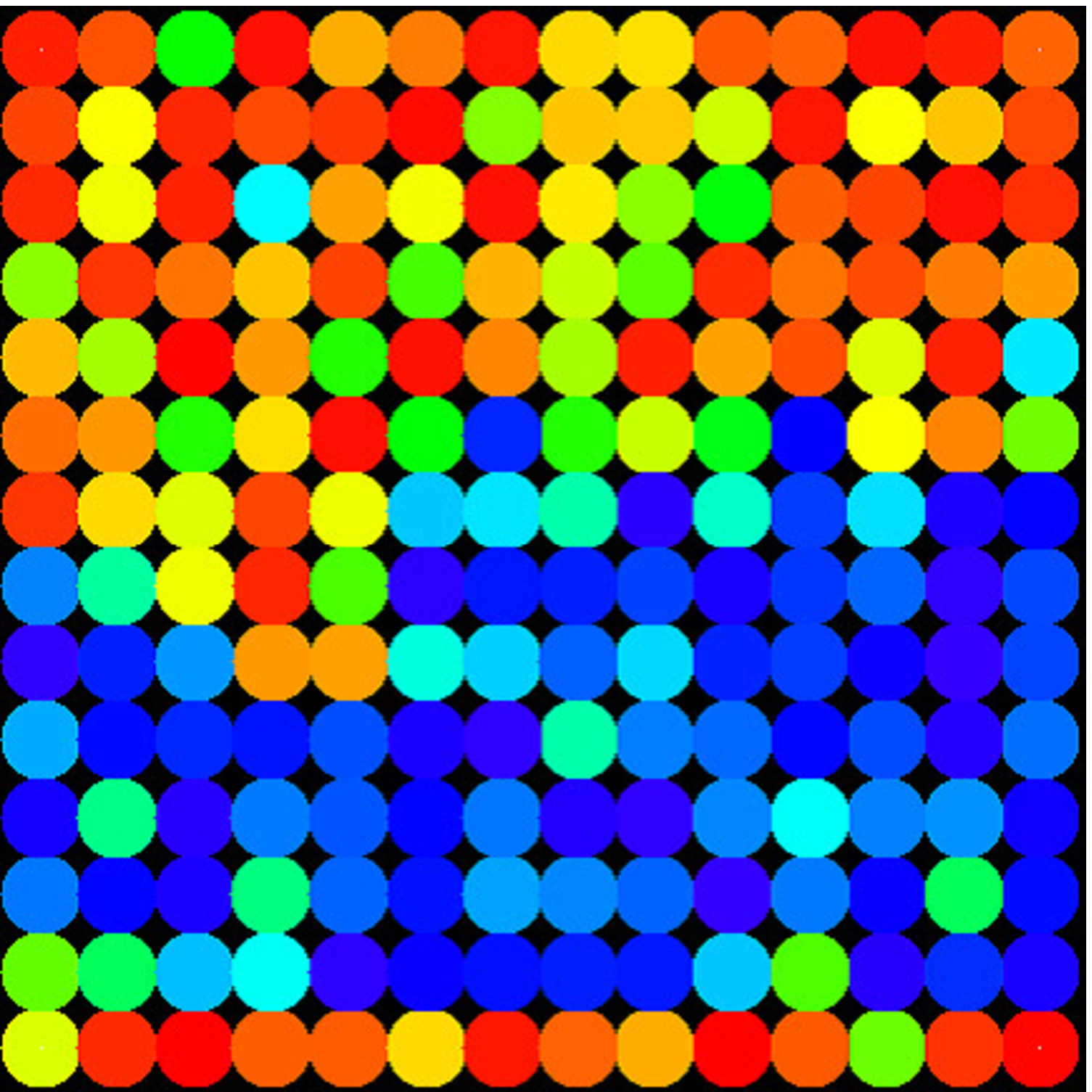} 
\hspace{3cm}
\includegraphics[width=3cm]{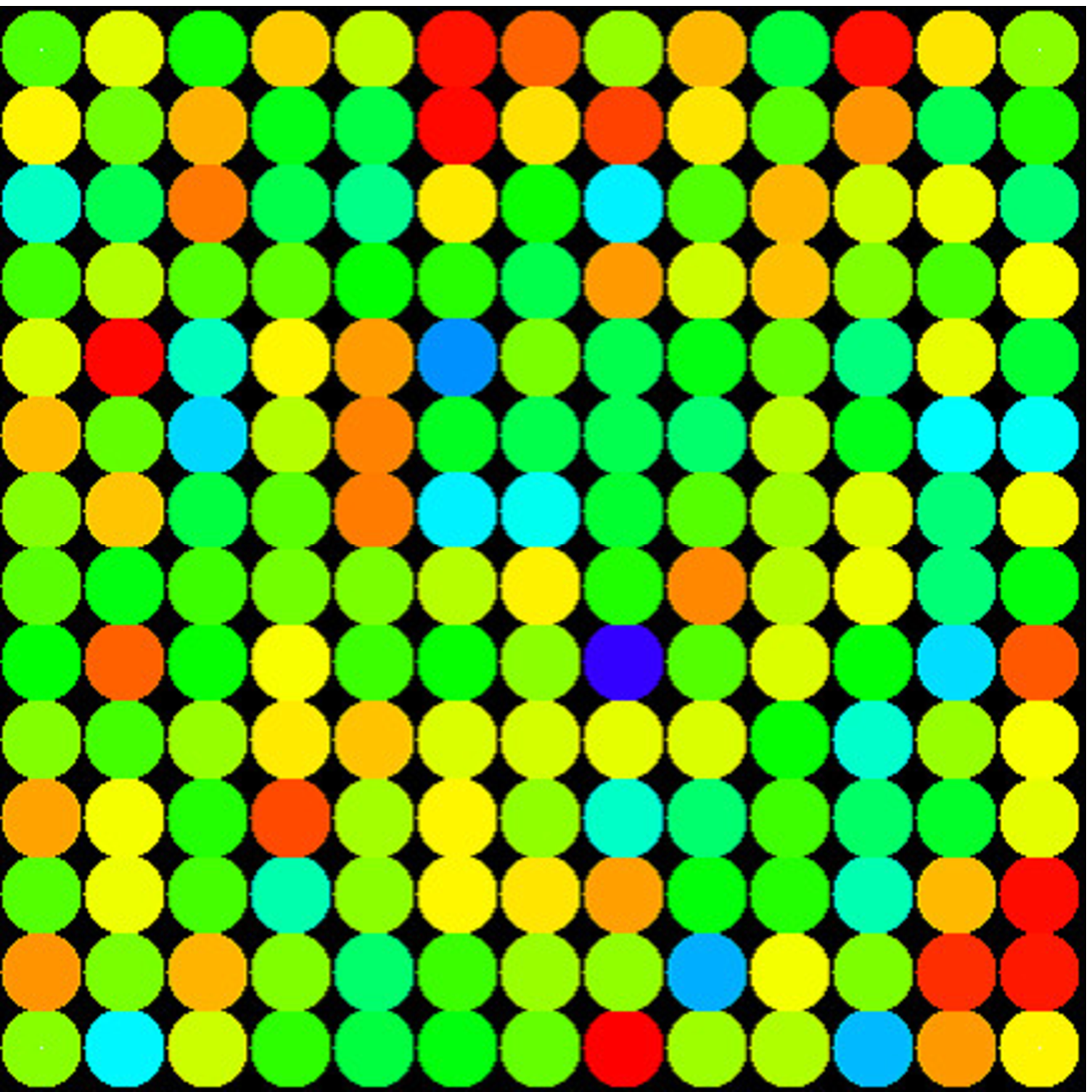} \\
$c_1=5, \; c_3=0$   \hspace{4cm}    $c_1=5, \; c_3=8$
\caption{%(Color online)
Large $J_z$ case.
Snapshot of density of holes in the system with $\rho_a=\rho_b=15\%$
before ($c_3=0$) and after ($c_3=8$) the phase transition at $c_3\simeq 6.0$.
$c_1=5.0$.
Bright (dark) region represents region of high (low) atomic density.
In the low-hopping region ($c_3=0$), system is divided into AF and empty regions.
On the other hand in the high-hopping region ($c_3=8$), homogeneous SF state 
is realized.
The phase transition is of first order.
}
\label{fig:snap_0.7_1}
\end{center}
\end{figure}

We first consider the case $c_1=5$ and $\alpha=-0.5$, relatively large value of $J_z$.
For small hopping amplitude $t$, the system exists in the AF phase.
As $t$ is increased, a phase transition takes place at $c_3\simeq 6.0$.
See Fig.\ref{fig:EC_0.7_1}, in which internal energy and specific heat
of each term in $H_{\rm tJ}$ in (\ref{HtJ}) are shown.
It is useful to see snapshot of the system before and after the phase transition.
See Fig.\ref{fig:snap_0.7_1}.
It is obvious that holes (i.e., atoms) are localized and make some domain 
for $c_3=0$, whereas particles are homogeneously distributed for $c_3=8$.

%%%%%%%%%%%%%%%%%%%%%%%%%%%%%%%%%%%%%%%%%%%%%%%%%%%%%%%%%%%%%%%
%FIG.9
%%%%%%%%%%%%%%%%%%%%%%%%%%%%%%%%%%%%%%%%%%%%%%%
%%%%%%%%%%%%%%%%%%%%%%%%%%%%%%%%%%%%%%%%%%%%%%%%%%%%%%%%%%%%%%%
%FIG.10
%%%%%%%%%%%%%%%%%%%%%%%%%%%%%%%%%%%%%%%%%%%%%%%

\begin{figure}[h]
\begin{center}
\includegraphics[width=5.5cm]{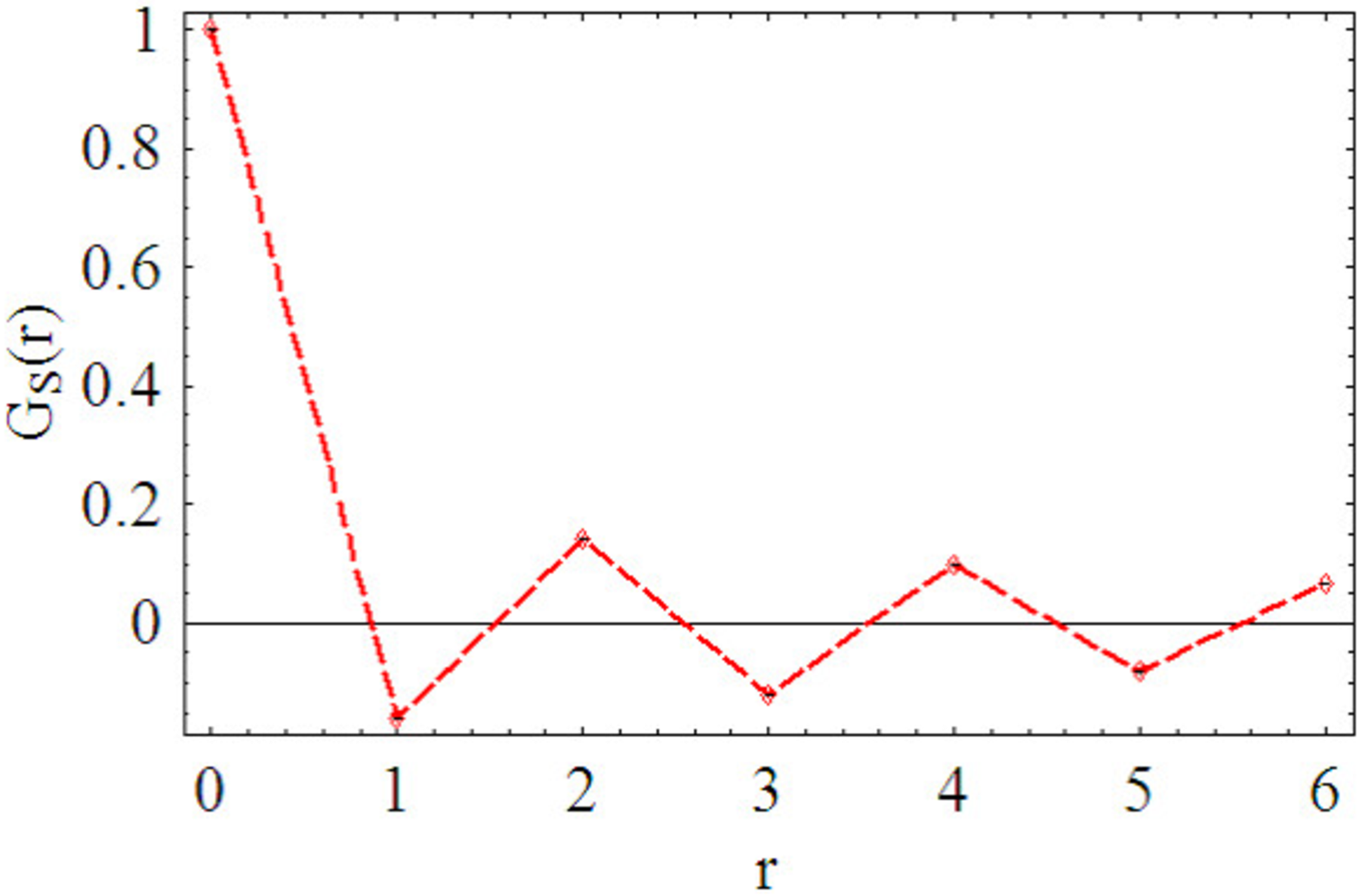}
\hspace{1cm}
\includegraphics[width=5.5cm]{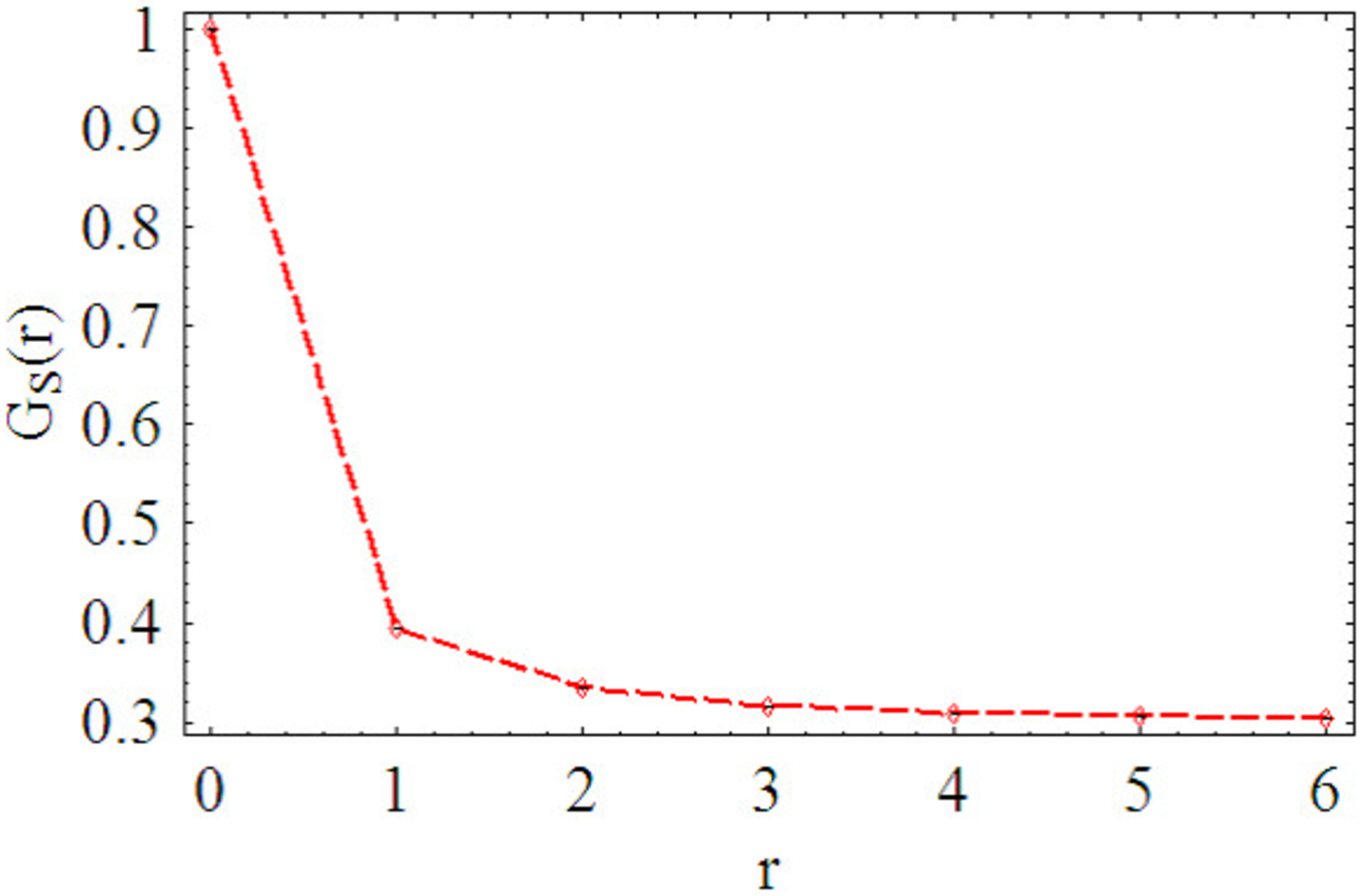} \\
\includegraphics[width=5.5cm]{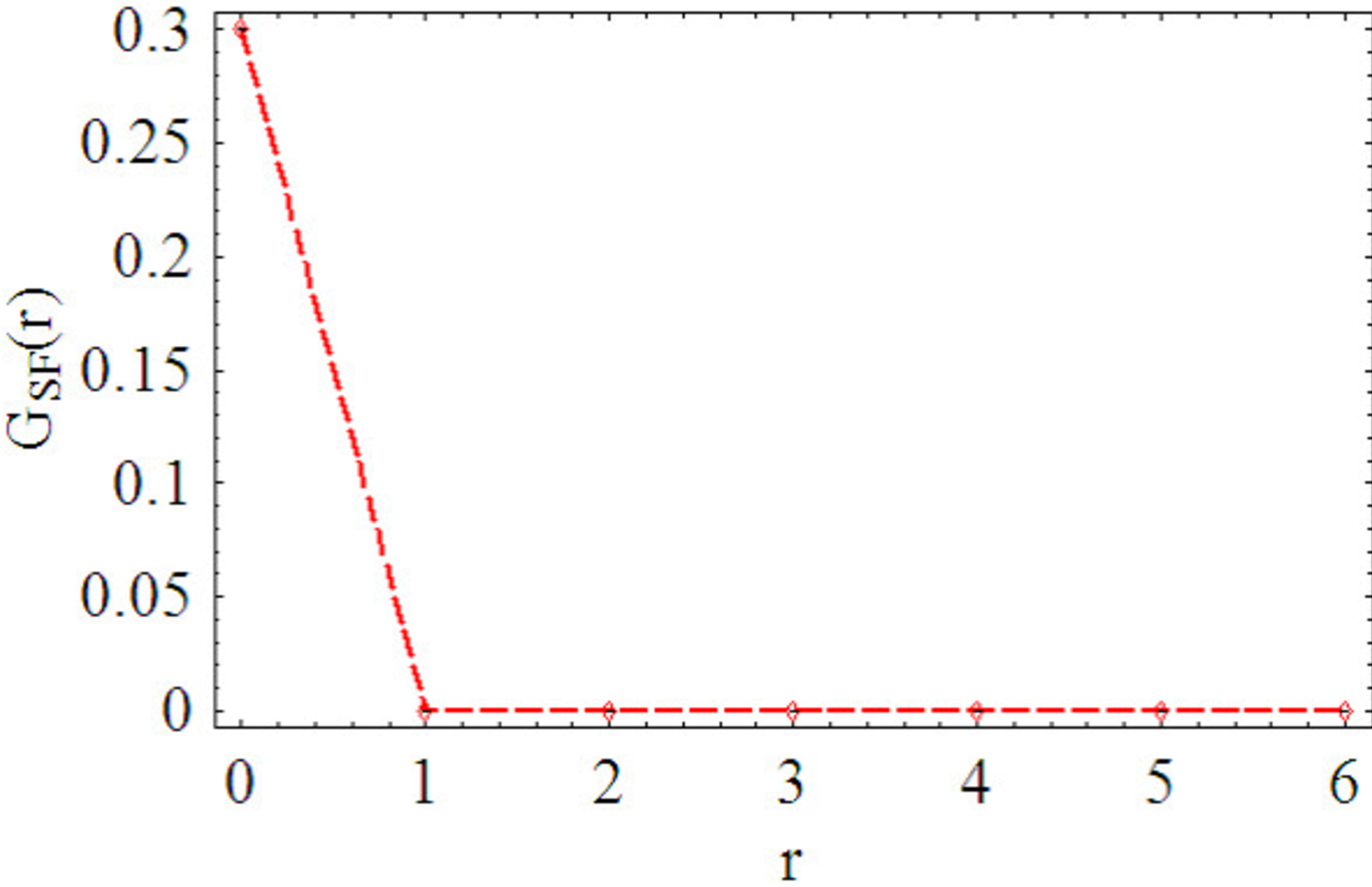} 
\hspace{1cm}
\includegraphics[width=5.5cm]{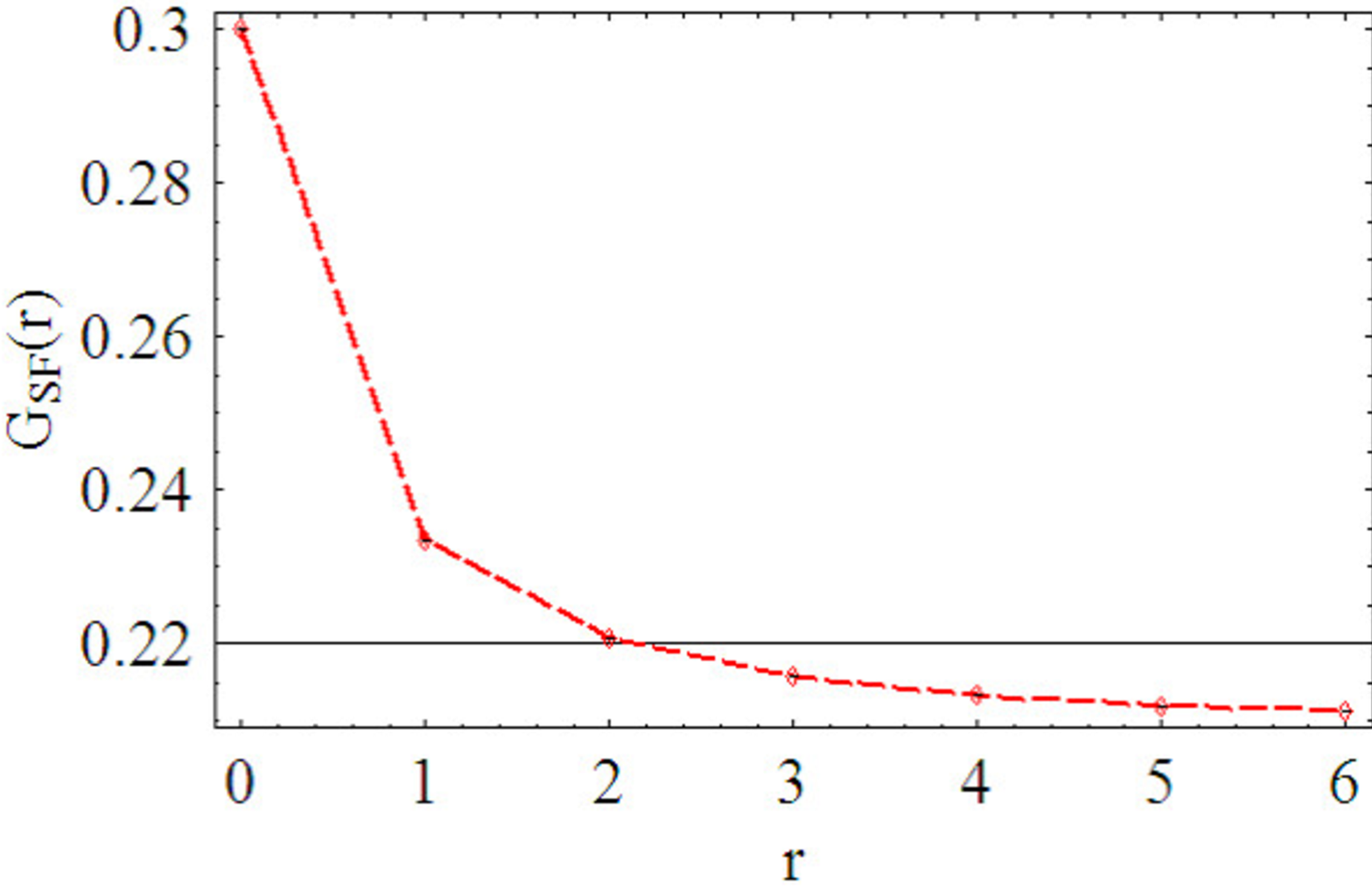} \\
$c_1=5, \; c_3=0$   \hspace{4cm}    $c_1=5, \; c_3=8$
\caption{%(Color online)
Correlation functions $G_{\rm SF}(r)$ and $G_{\rm S}(r)$.
}
\label{fig:corr_0.7_1}
\end{center}
\end{figure}

To verify that the observed phase transition is a transition from the AF solid
to the SF, 
we calculated the correlation function  of the boson operator 
$B_r=(a_r, b_r)^t$, 
\begin{equation}
G_{\rm SF}(r)={1 \over L^3}\sum_{r_0}
\langle B^\dagger_{r_0}\cdot B_{r_0+r} \rangle.
\label{GSF}
\end{equation}
We also calculated the spin correlation function,
\begin{equation}
G_{\rm S}(r)={1 \over L^3}\sum_{r_0}
\langle \vec{S}_{r_0}\cdot \vec{S}_{r_0+r} \rangle.
\label{GS}
\end{equation}
Results are shown in Fig.\ref{fig:corr_0.7_1}.
It is obvious that the SF phase transition at $c_3\simeq 6.0$
accompanies a transition from the AF to FM in the spin degrees of freedom.
As explained in the previous paper\cite{btJ3},
it should be remarked that the FM order in the SF state is
a natural outcome of the two-component SF and it should not be
regarded as the genuine FM order in the magnetism.
In fact, the wave function of the $a$ and $b$-atomic SF is well-described
by the following MFT-type wave function,
\begin{equation}
|{\rm SF}\rangle=\prod_r\Big[\sin {\theta \over 2}
\Big({1 \over \sqrt{2}}a^\dagger_r+{1 \over \sqrt{2}}
e^{i\phi}b^\dagger_r
\Big)+\cos {\theta \over 2}\Big] |0\rangle,
\label{vwf2}
\end{equation}
where we have assumed the translational symmetry and recovered
the relative phase between $a$ and $b$ SF condensates, $e^\phi$,
which was ignored in Eq.(\ref{vwf}). 
By using Eq.(\ref{vwf2}), it is straightforward to calculate
expectation value of the pseudo-spin
operator $\vec{S}_r$ to find that the state $|{\rm SF}\rangle$ exhibits
the FM order.

%%%%%%%%%%%%%%%%%%%%%%%%%%%%%%%%%%%%%%%%%%%%%%%%%%%%%%%
%FIG.11
%%%%%%%%%%%%%%%%%%%%%%%%%%%%%%%%%%%%%%%%%%%%%%%

\subsubsection{Small $J_z$}

\begin{figure}[h]
\begin{center}
\includegraphics[width=6cm]{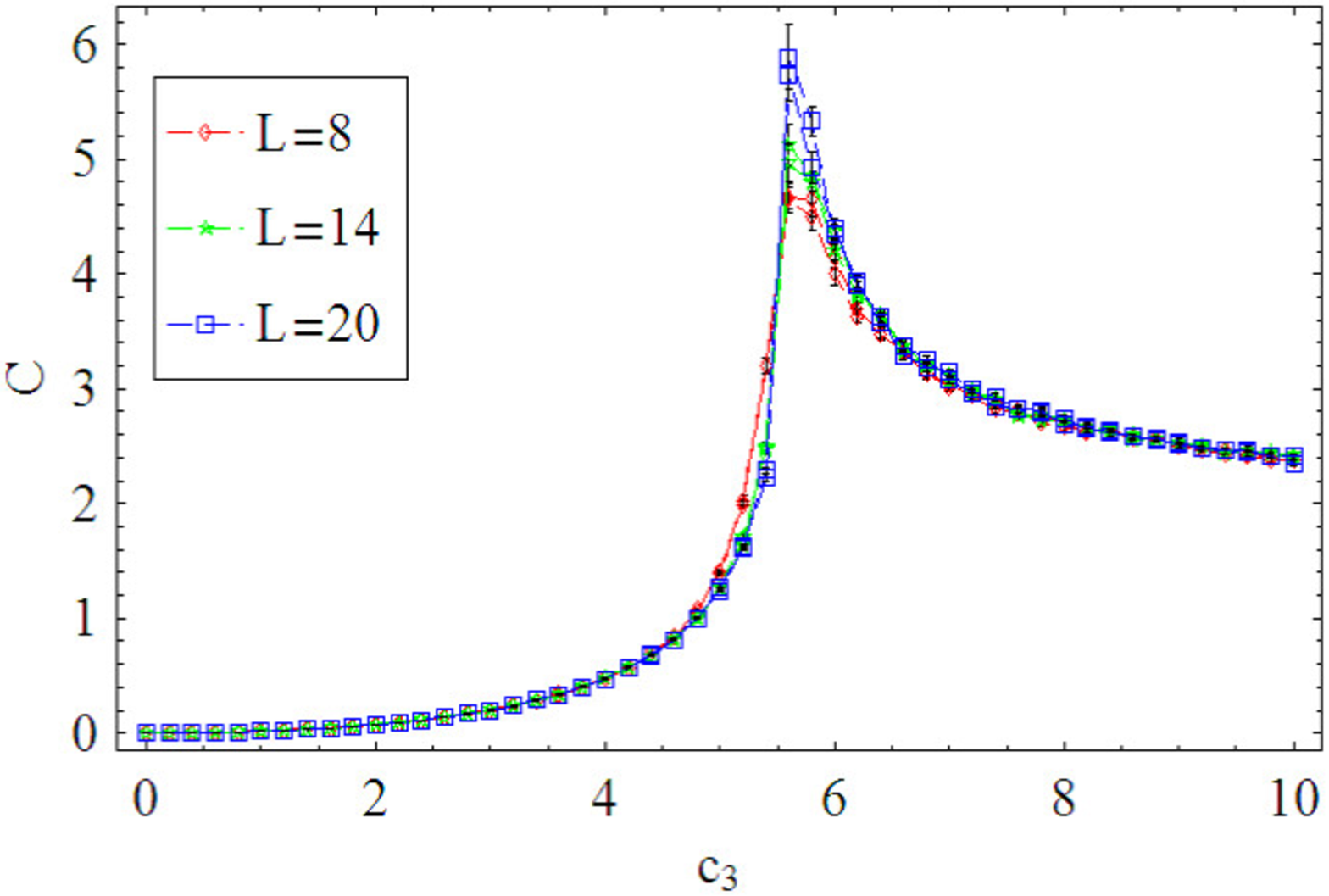}
\includegraphics[width=4cm]{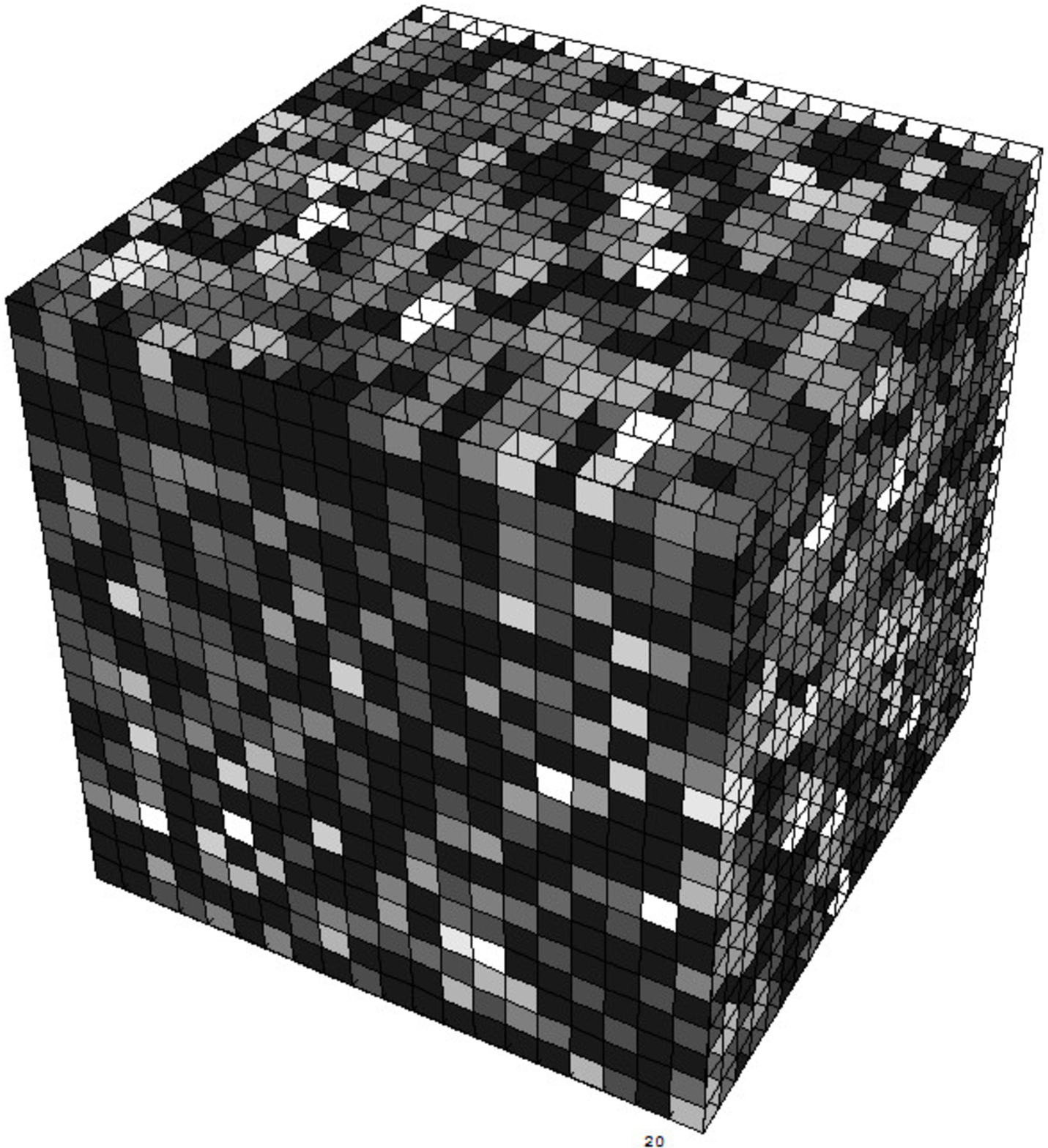}
\hspace{0.5cm}
\includegraphics[width=3cm]{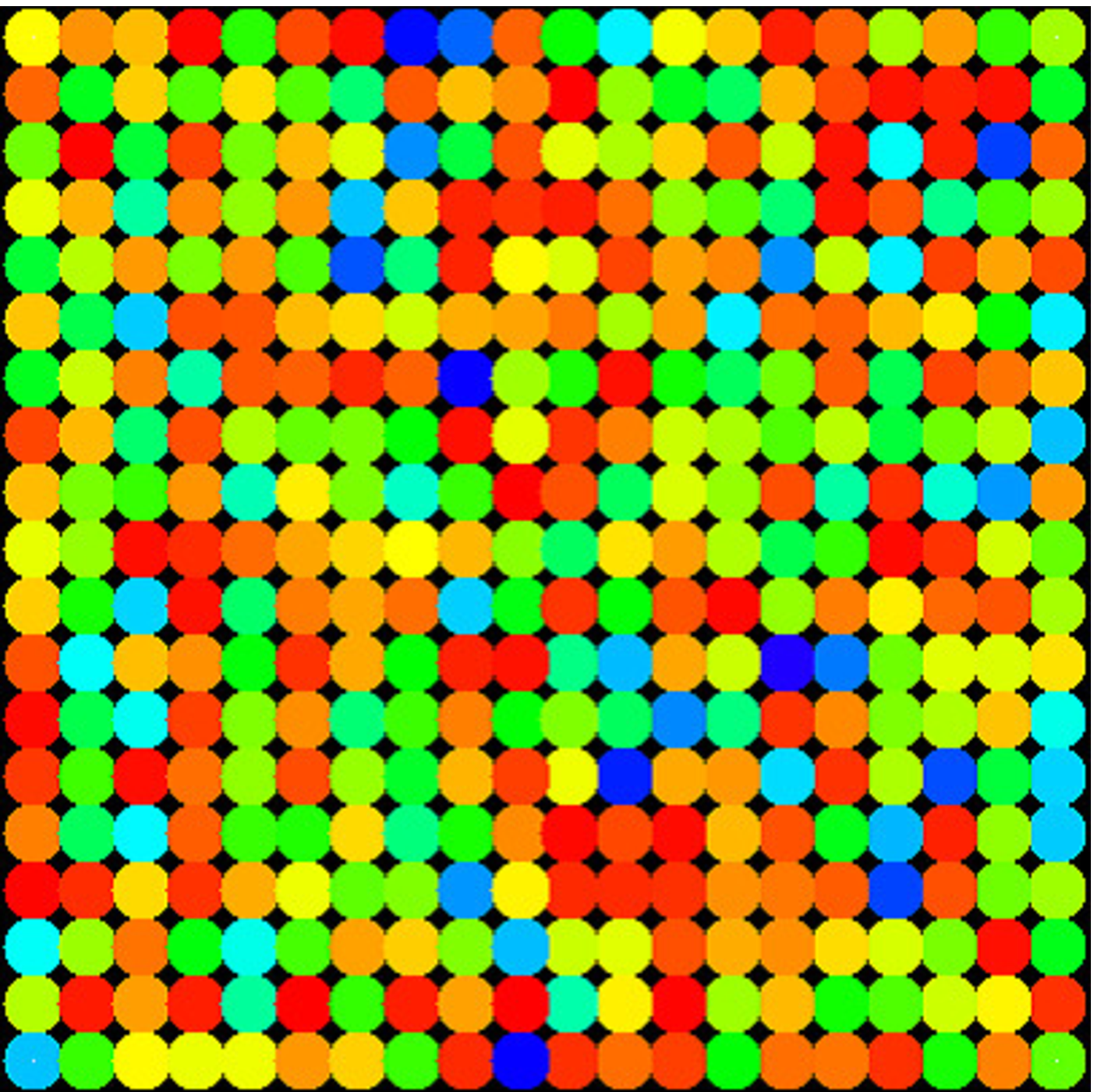}
\caption{%(Color online)
Small $J_z$ case.
Specific heat $C$ for  $\rho_a=\rho_b=15\%$ with $c_1=3$
as a function of $c_3$.
There exists a second-order phase transition at $c_3\simeq5.5$.
We also show snapshots for $c_3=0$.
Compared with the case of $c_1=5$, system in the low-hopping region 
is homogeneous.
}
\label{fig:C_0.7_2}
\end{center}
\end{figure}

We also studied the low-density case with $c_1=3$ and $\alpha=-0.5$, 
relatively small value of $J_z$.
Calculation of the specific heat $C$ is shown in Fig.\ref{fig:C_0.7_2}.
$C$ has a system-size dependence, which indicates that the 
phase transition at $c_3\simeq 5.5$ is of second order.
We also show snapshots of hole density for $c_3=0$.
Compared with the case of $c_1=5$, atoms are distributed rather homogeneously.
We verified that spin correlation function $G_{\rm S}(r)$ in that region 
has no LRO, and therefore the phase transition is from the PM to FM+SF.
This is the reason why the phase transition turns to second order.
The result suggests existence of the tricritical point at which
the phase transition changes from second to first one as $c_1$ is increased.
For large $J_z$, there exists a strong attractive force between $a$ and $b$
atoms sitting nearest-neighbor sites, and therefore the AF solid tends to
be high density.
On the other hand for small $J_z$, the attractive force is not strong
enough to generate the AF solid.

%%%%%%%%%%%%%%%%%%%%%%%%%%%%%%%%%%%%%%%%%%%%%%%%%%%%%%%%%%%%%%%
%FIG.12
%%%%%%%%%%%%%%%%%%%%%%%%%%%%%%%%%%%%%%%%%%%%%%%

%%%%%%%%%%%%%%%%%%%%%%%%%%%%%%%%%%%%%%%%%%%%%%

\subsection{High-density region}

%%%%%%%%%%%%%%%%%%%%%%%%%%%%%%%%%%%%%%%%%%%%%%%%%%%%%%%%%%%%%%%
%FIG.13
\begin{figure}[h]
\begin{center}
\includegraphics[width=6cm]{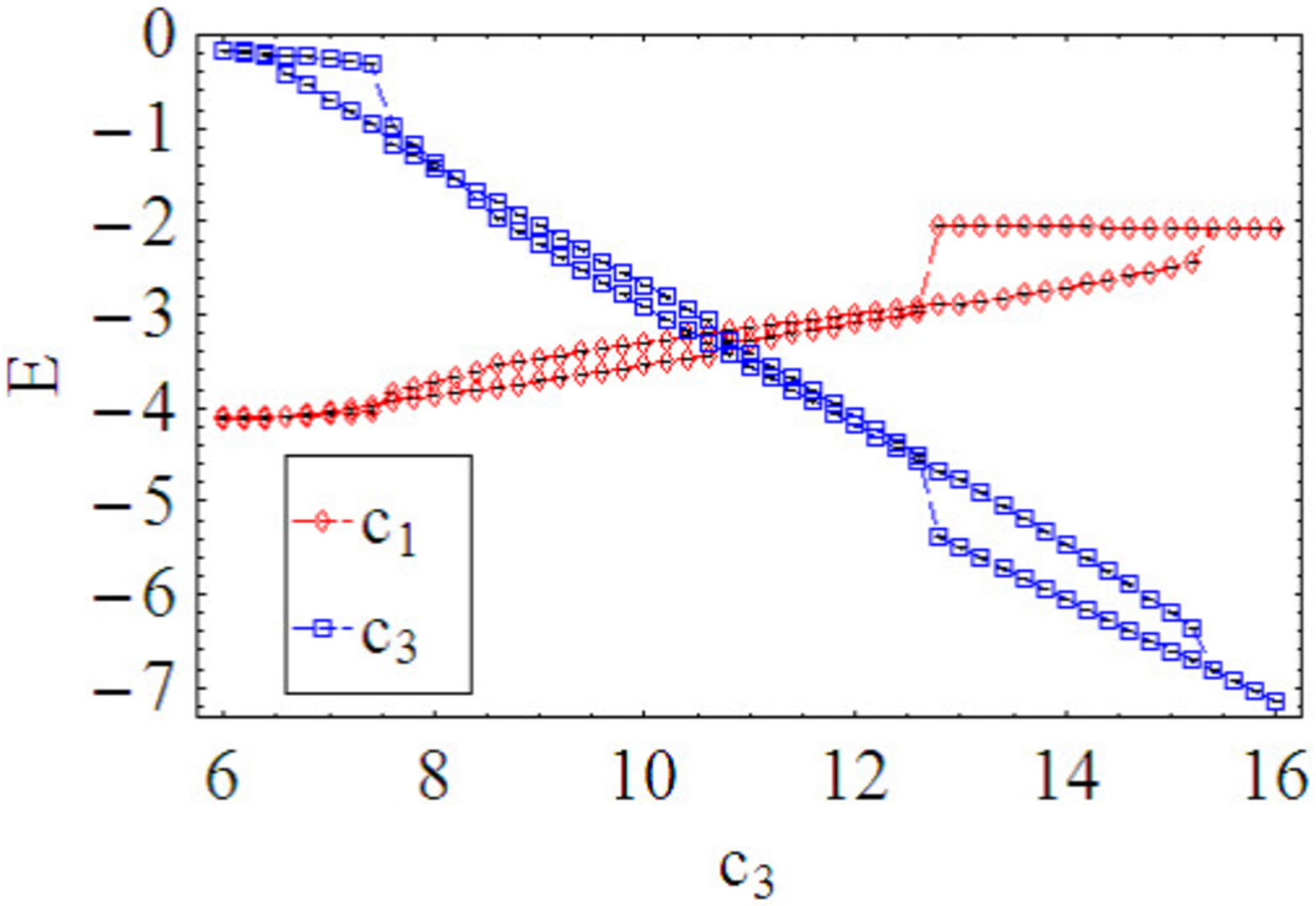}
\includegraphics[width=6cm]{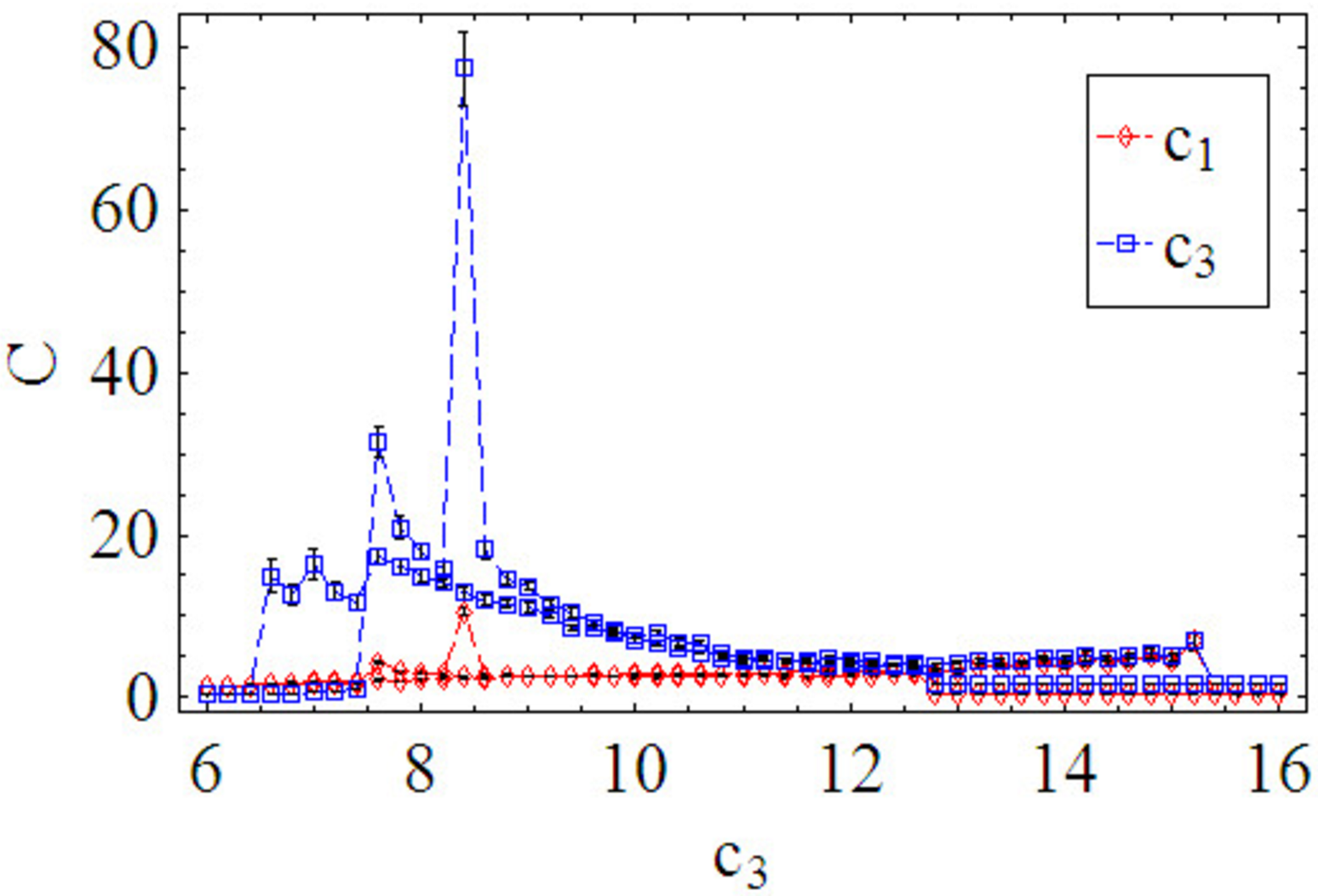}
\caption{%(Color online)
Internal energy and specific heat of each term in $H_{\rm tJ}$ measured in
CE as a function of $c_3$.
$\rho_a=\rho_b=38\%$ with $c_1=3$.
There are three phase transition s at $c_3=7.6, \ 8.4$, and $14.5$.
$L=24$.
}
\label{fig:C_CE1}
\end{center}
\end{figure}
%%%%%%%%%%%%%%%%%%%%%%%%%%%%%%%%%%%%%%%%%%%%%%%%%%%%%%%%%%%

\begin{figure}[h]
\begin{center}
\includegraphics[width=6cm]{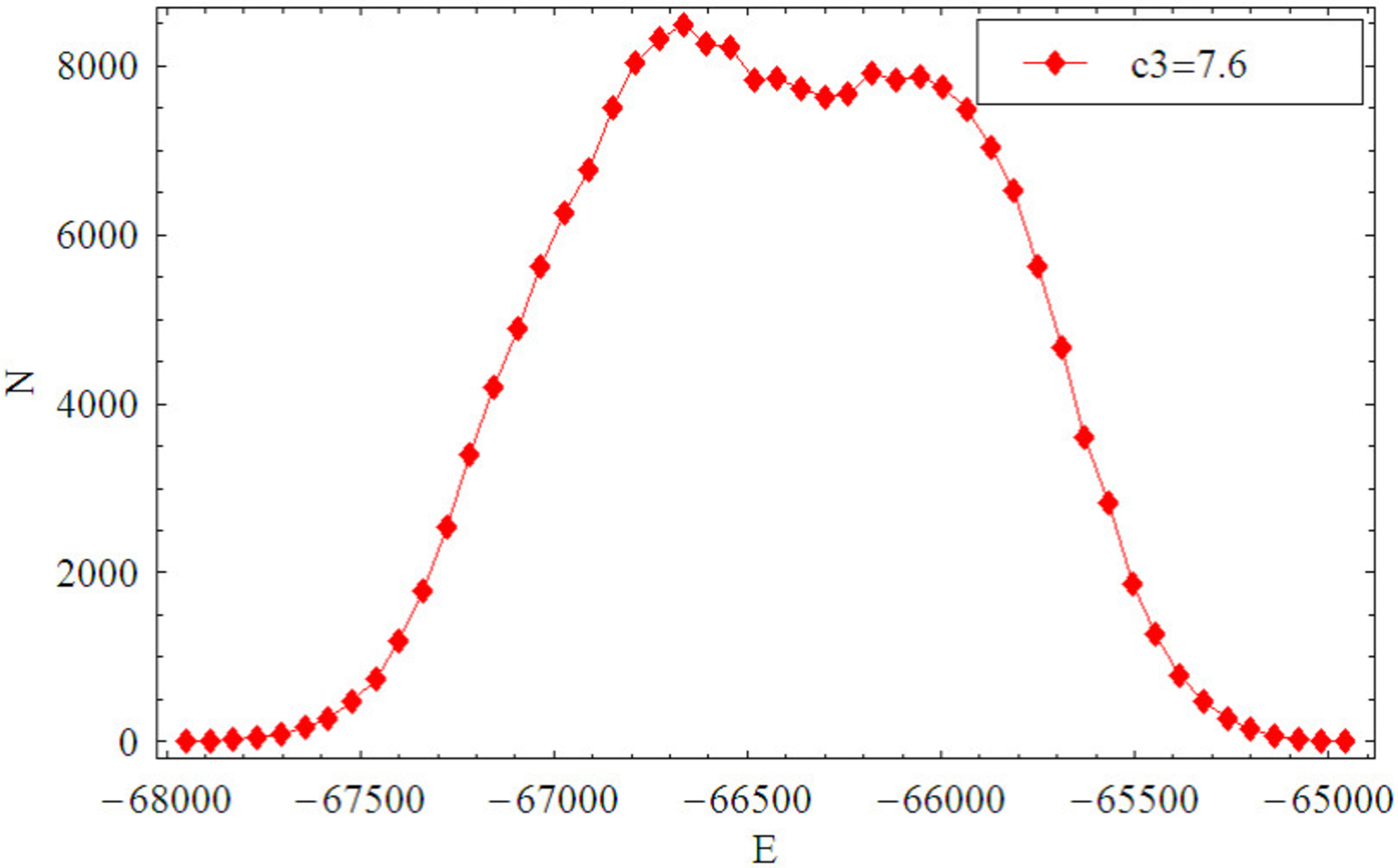}
\includegraphics[width=6cm]{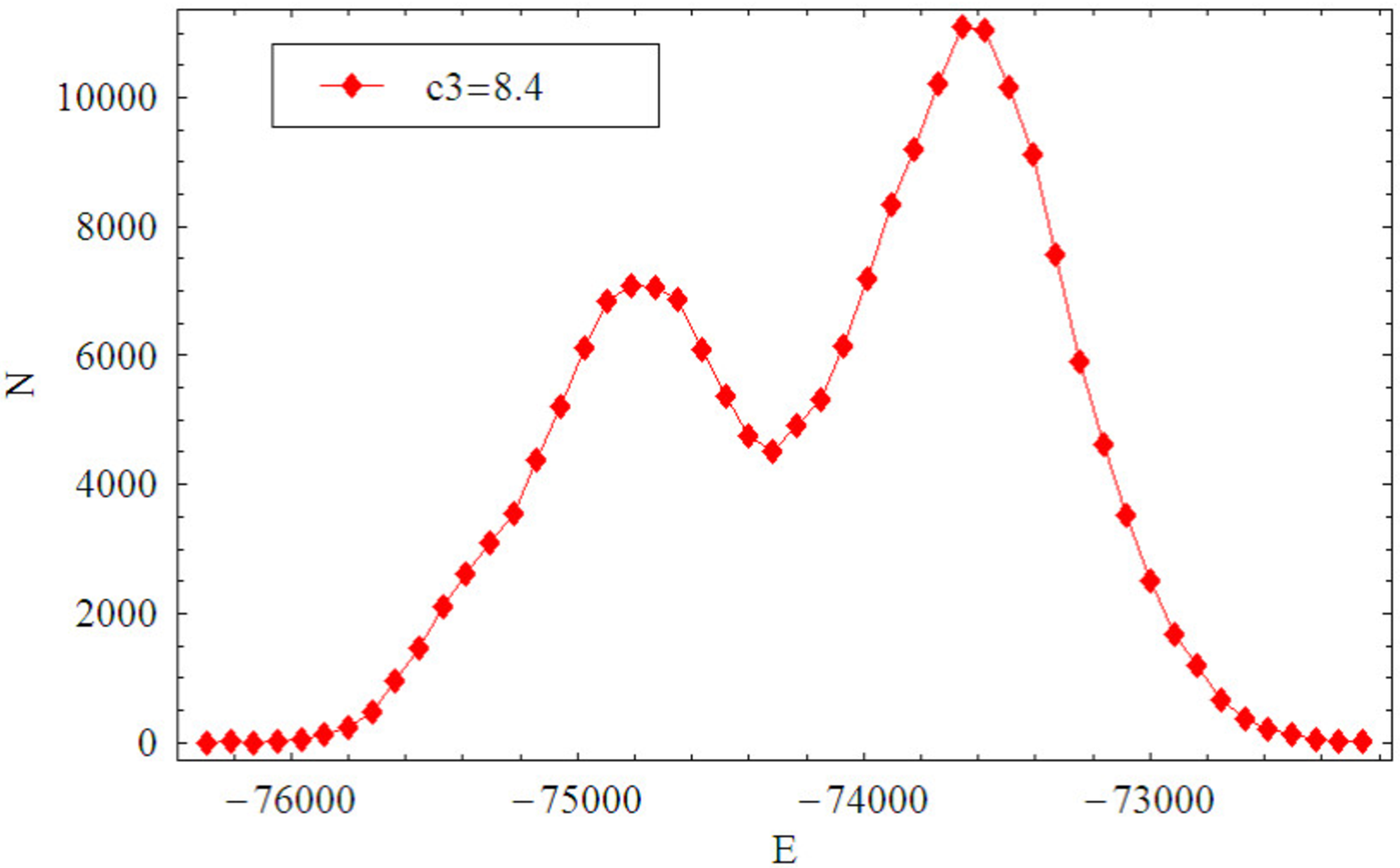}
\caption{%(Color online)
Histograms of $N(E)$.
They have double-peak shape, which indicate the existence of the first-order phase
transition at $c_3=7.6$ and $8.4$.
}
\label{fig:histo_CE1}
\end{center}
\end{figure}

In this subsection, we study the system with average atomic density
$\rho_a=\rho_b=38\%$ with $c_1=3$ and the system size $L=24$.
This case is close to that observed by the GCE in the previous section
and corresponds to the high-density region (see Fig.\ref{fig:pic_CE}).
Numerical study of the internal energy $E$ and the
specific heat $C$ indicates the existence of  three phase transitions
as shown in Fig.\ref{fig:C_CE1}.
All of them are of first order as the density of state $N(E)$ has a 
double-peak shape as shown in Fig.\ref{fig:histo_CE1},
where $N(E)$ is defined as 
\begin{eqnarray}
Z_{\rm CE} &=& \int_{\rm CE} [D\bar{w}DwD\bar{z}Dz]
e^{-\beta H_{\rm tJ}(\bar{w},w,\bar{z},z)},  \nonumber \\
&=& \int dE \; e^{-\beta E} N(E).
\label{ZCE}
\end{eqnarray}
In Eq.(\ref{ZCE}), $\int_{\rm CE}$ denotes the integration over 
configurations with a fixed number of atoms (or fixed average density
of atoms), and $N(E)$ is a function of $\beta$ and the coupling constants.
Furthermore
from $c_3\simeq 13$ to $c_3\simeq 15$, there is a large hysteresis
loop in $E$ indicating existence of a first-order phase transition.
It is obvious that the system with high particle density has a rather
complicated phase structure compared with the low-density case.

\begin{figure}[h]
\begin{center}
\includegraphics[width=3cm]{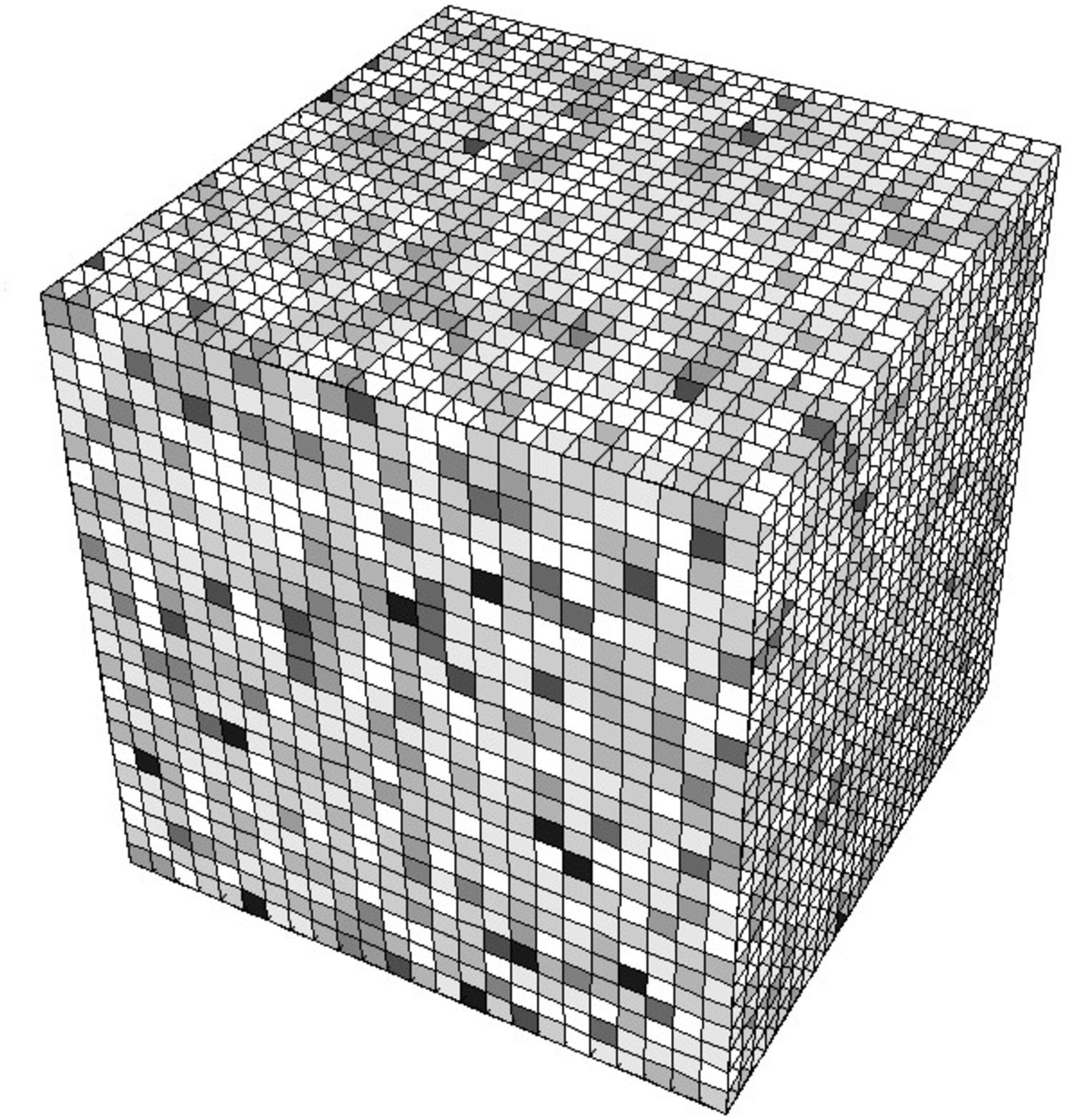}
\includegraphics[width=3cm]{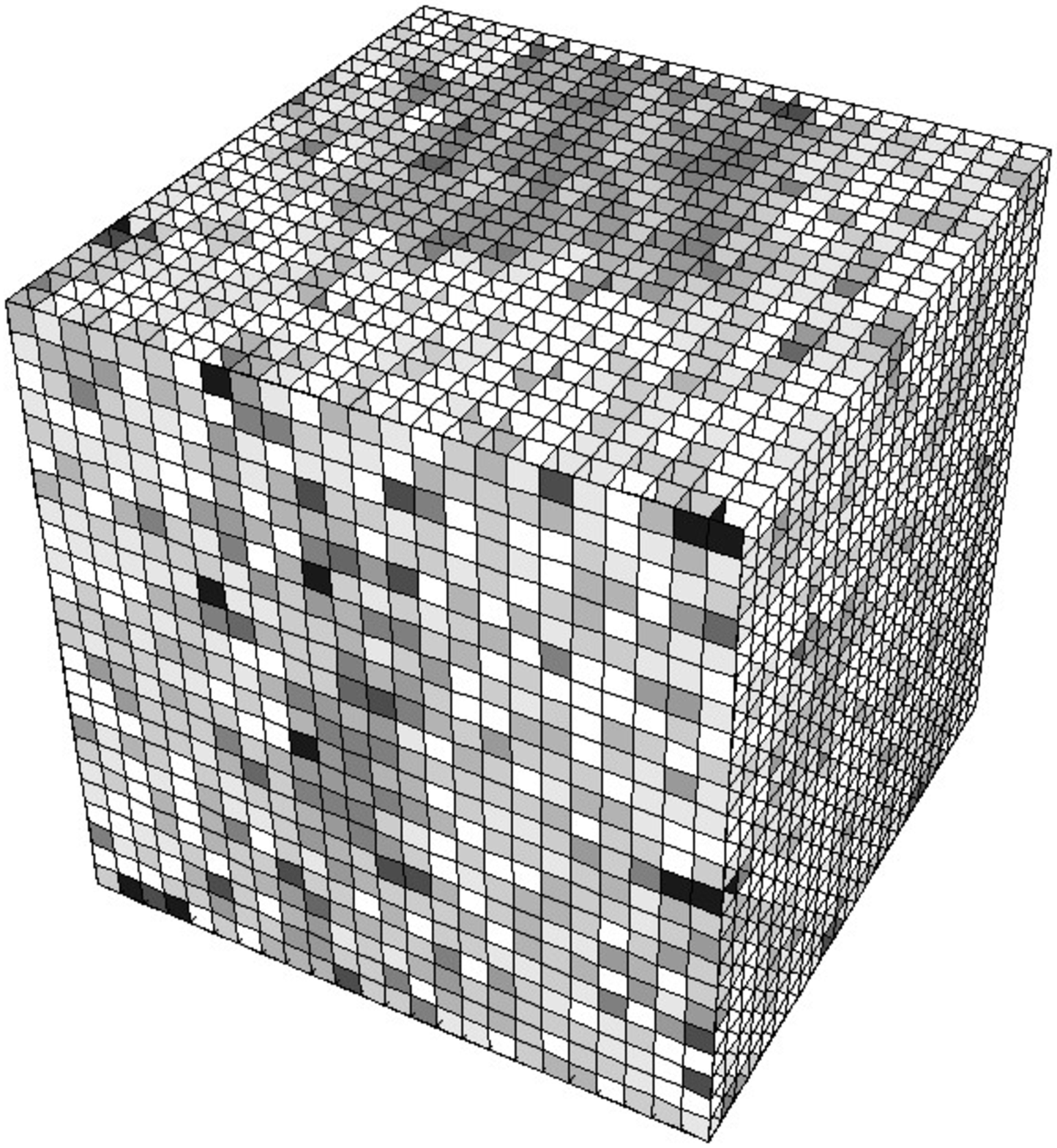}
\includegraphics[width=3cm]{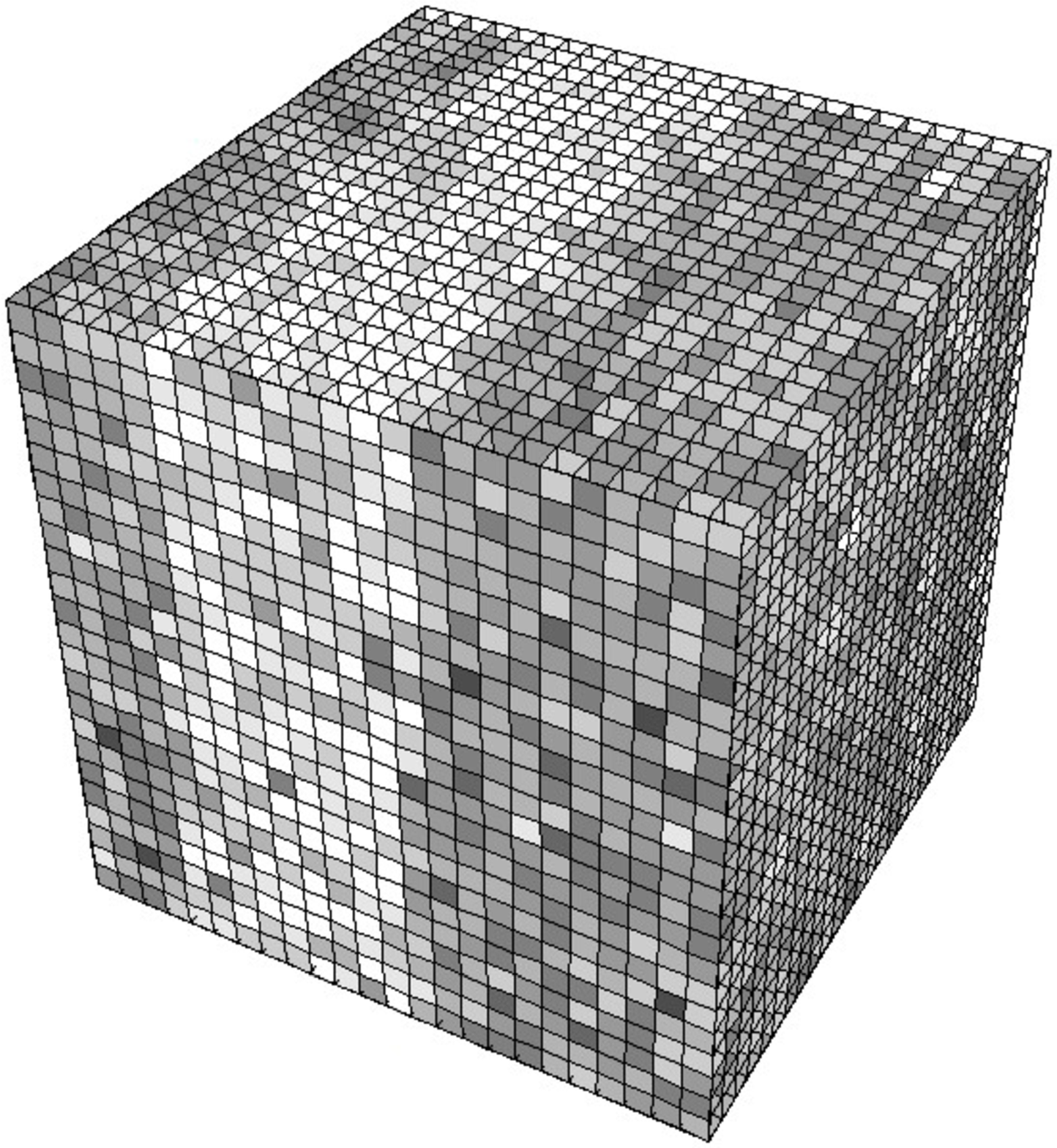}
\includegraphics[width=3cm]{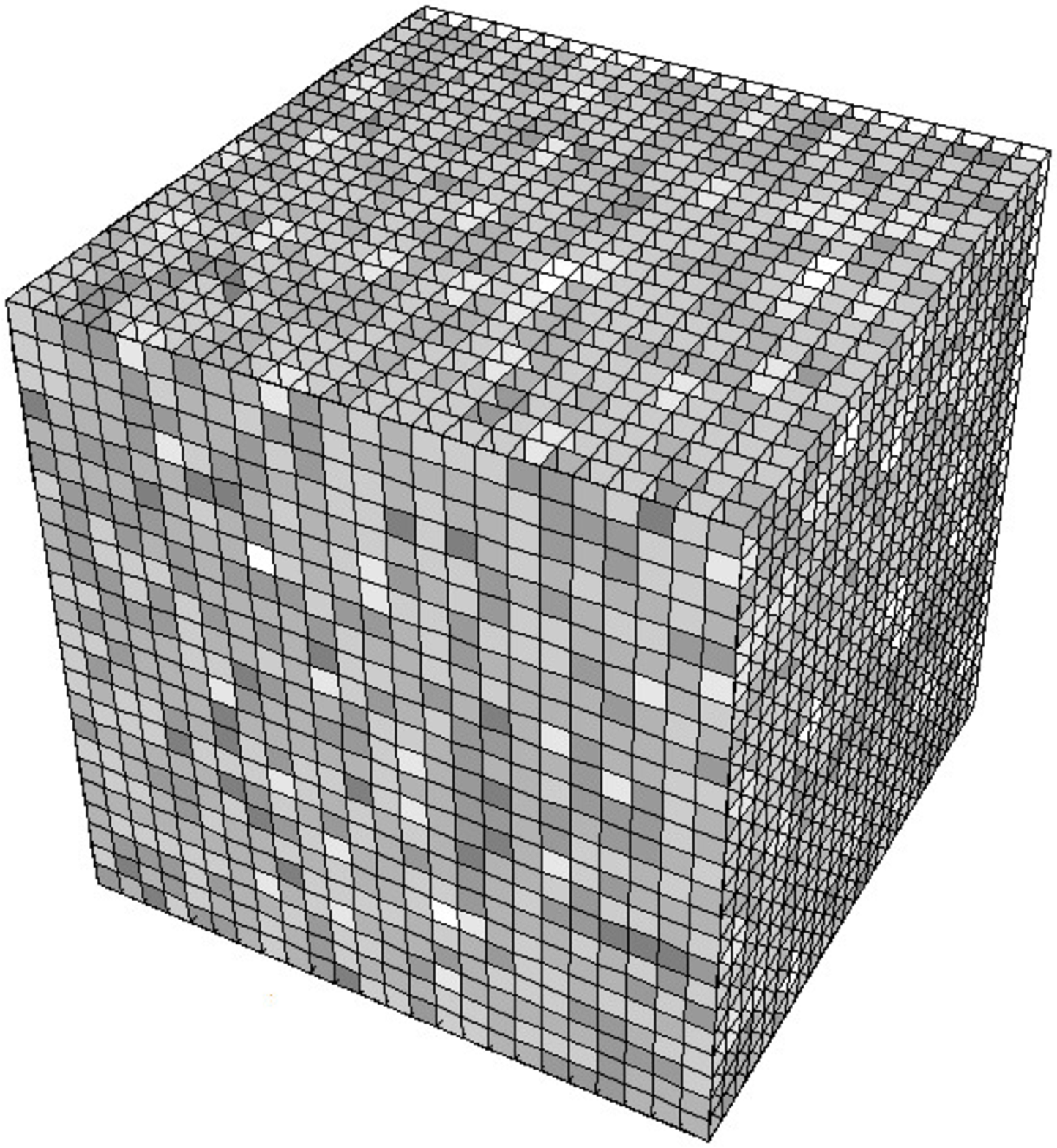}
%\vspace{-0.5cm}
\caption{%(Color online)
Snapshot of system at various values of $c_3$ with $c_1=3$.
(From left to right)$c_3=7, 8, 10.6$ and $16$.
Bright (dark) region corresponds to higher (lower) atomic density.
From left to right, AF, cloudlet state, PS and SF.
}
\label{fig:snap2}
\end{center}
\end{figure}

Snapshots of the system at various $c_3$ shown in Fig.\ref{fig:snap2}
are quite useful to understand physical meaning of the above three phases.
At $c_3$ is increased from zero, a transition from the AF 
checkerboard state to a new phase takes place.
This new phase is inhomogeneous and composed of immiscible 
AF solid region and SF droplet of mesoscopic magnitude, i.e. SF cloudlet.
The appearance of this coexisting phase comes from the fact that
the AF and SF states tend to have a sharp interface as we saw 
in the previous section.
At the second phase transition, a complete PS takes place and the system 
is divided into the AF and SF phases.
Width of each phase is determined by the atomic density and the hopping
amplitude.
%As the value of $c_3$ is increased furthermore, a transition to 
%an inhomogeneous phase composed of immiscible 
%SF region and AF cloudlet takes place.
Finally for large $c_3$, the phase transition into the homogeneous SF takes place. 

%%%%%%%%%%%%%%%%%%%%%%%%%%%%%%%%%%%%%%%%%%%%%%%%%%%%%%%%%%%%%%%
%FIG.14
%%%%%%%%%%%%%%%%%%%%%%%%%%%%%%%%%%%%%%%%%%%%%%%%%%%%%%%%%%%
%%%%%%%%%%%%%%%%%%%%%%%%%%%%%%%%%%%%%%%%%%%%%%%%%%%%%%%%%%%%%%%
%FIG.15
%%%%%%%%%%%%%%%%%%%%%%%%%%%%%%%%%%%%%%%%%%%%%%%%%%%%%%%%%%%

\begin{figure}[h]
\begin{center}
\includegraphics[width=6cm]{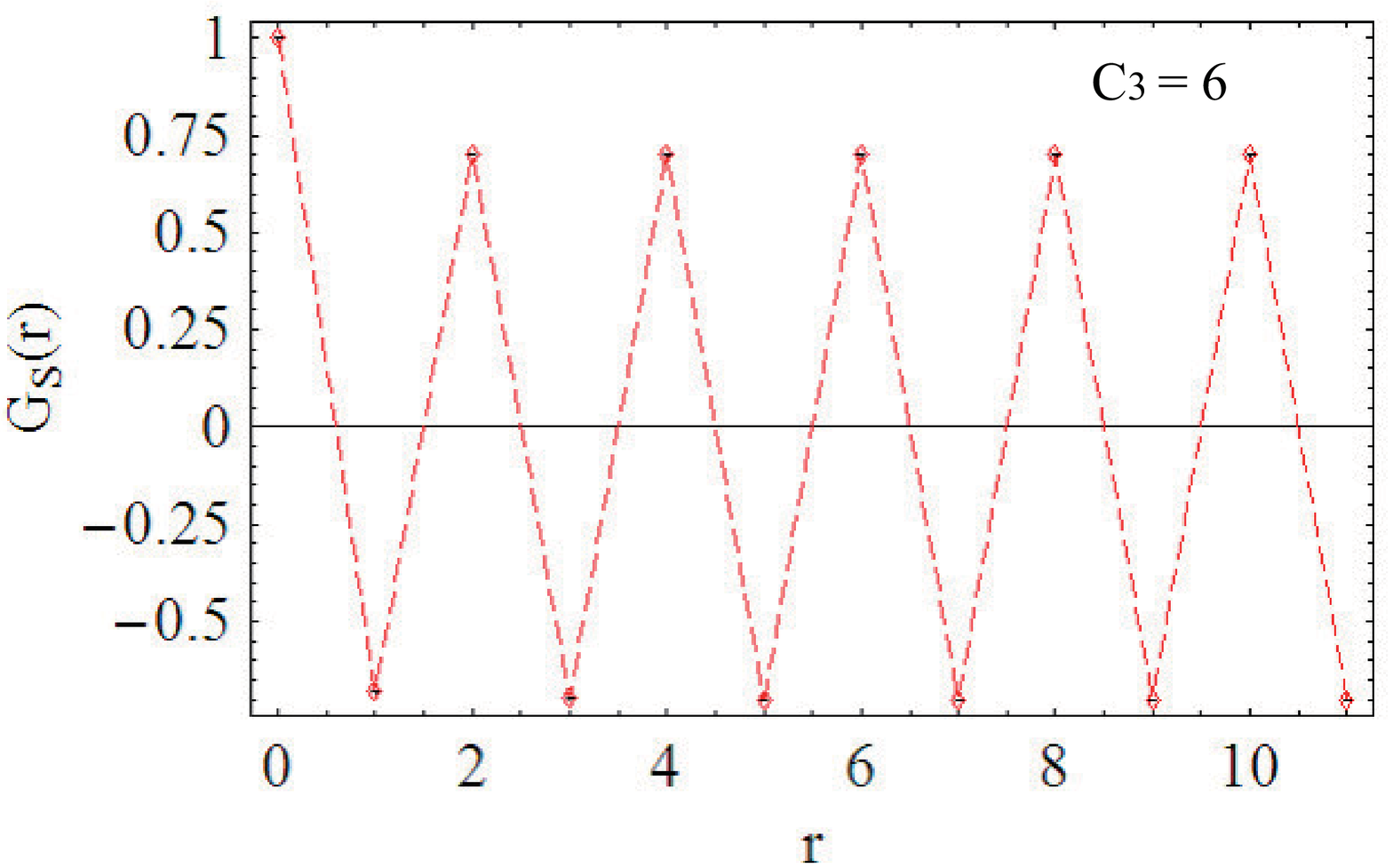}
\includegraphics[width=6cm]{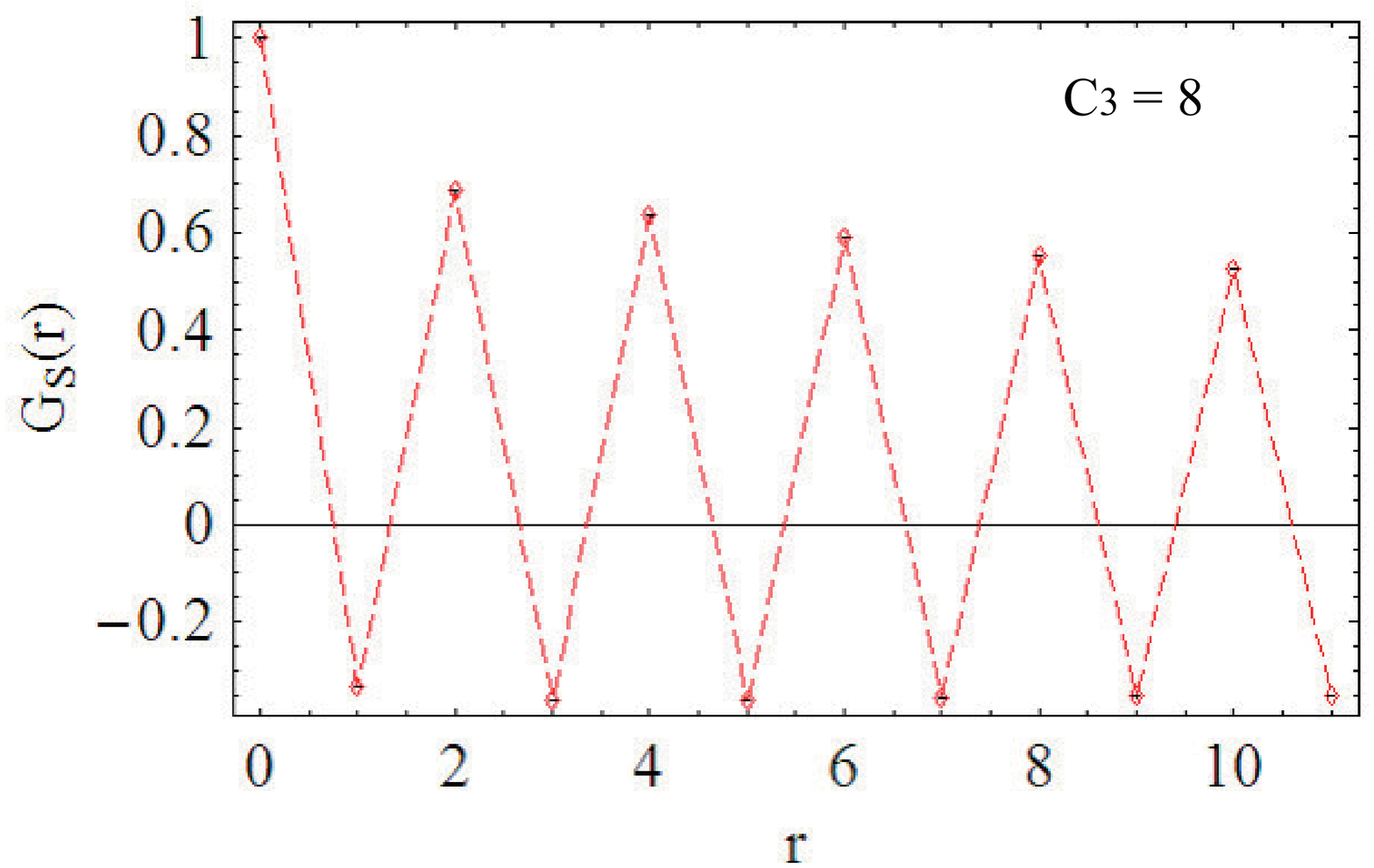} 
\includegraphics[width=6cm]{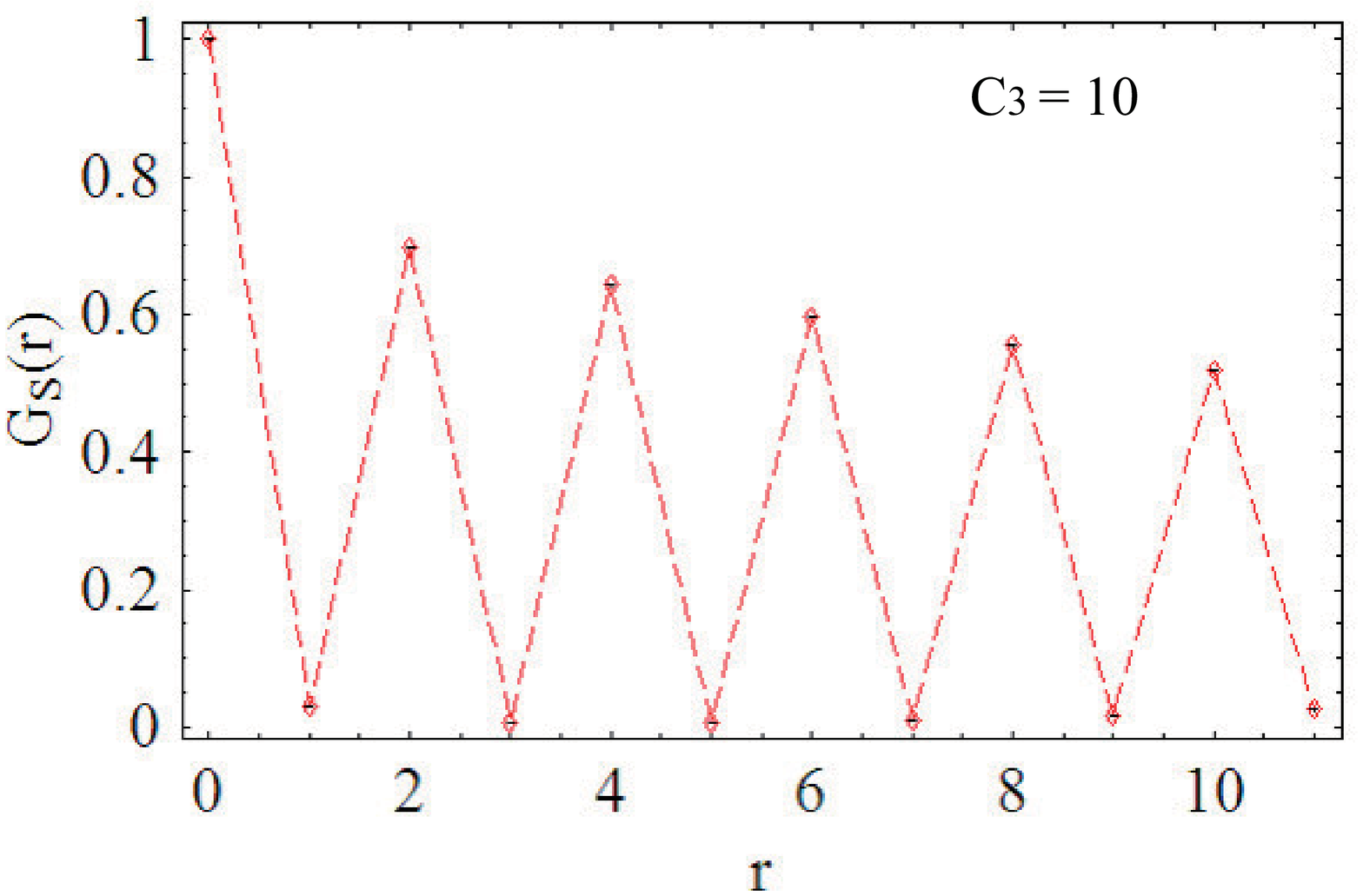}
\includegraphics[width=6cm]{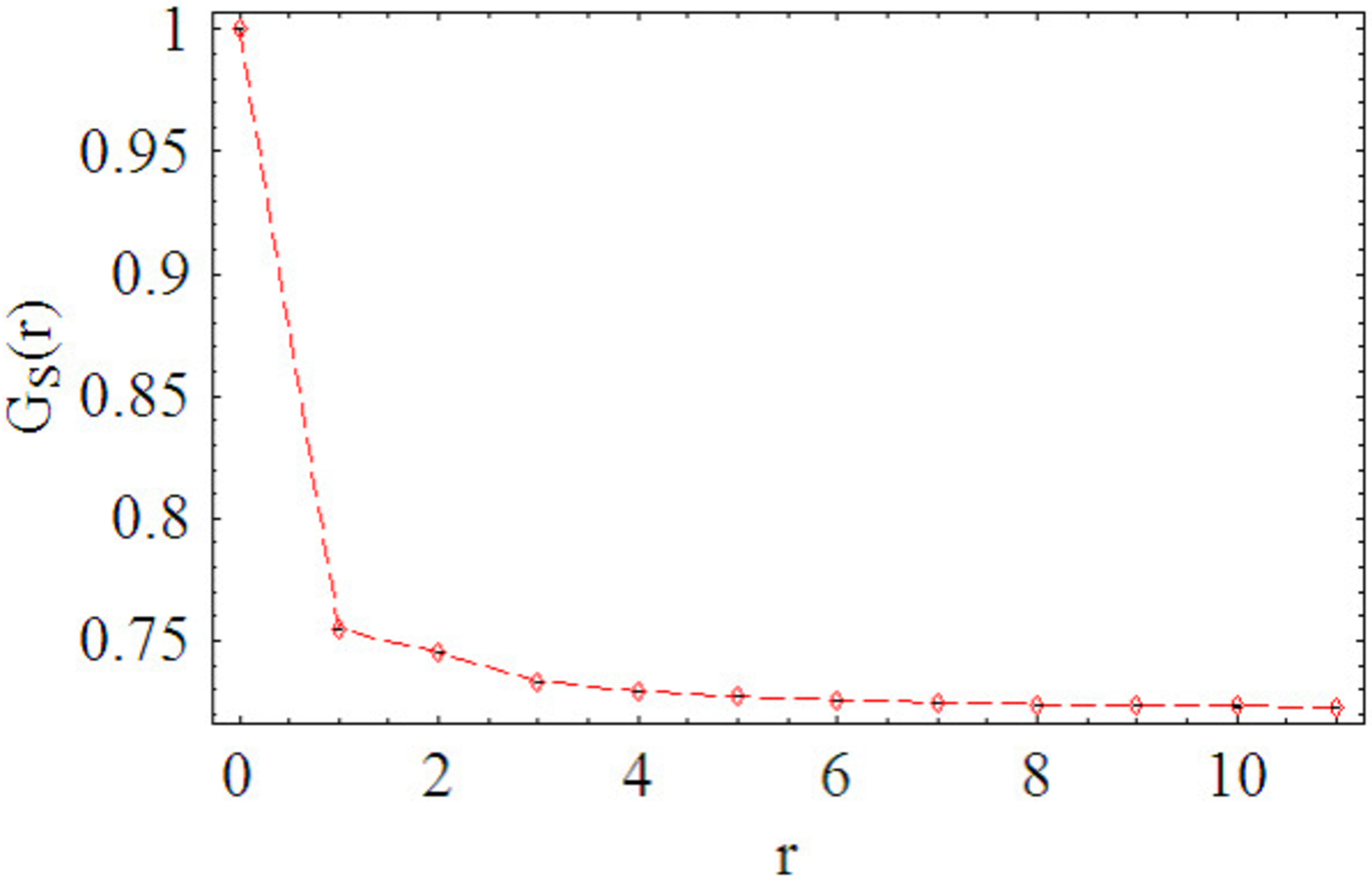}
%\vspace{-0.5cm}
\caption{%(Color online)
Pseudo-spin correlation for $c_3=6, \ 8, \ 10$ and $16$.
In the states of  mixture of AF solid and SF for  $c_3=6, \ 8, \ 10$, 
pseudo-spin correlation is a 
superposition of AF and FM, whereas in the SF, LRO of FM is realized.
}
\label{fig:spincor2}
\end{center}
\end{figure}

It is important to see how the spin correlation function behaves
in the above phases.
We show the results in Fig.\ref{fig:spincor2}.
For $c_3=6$, the system exists in the AF phase, and for $c_3=16$
the system is in the SF phase and the spin correlation
has a FM order.
In the intermediate regions, the spin correlation is a superposition
of the AF and FM LRO's.
Similar behavior has been observed in other systems by the previous 
study\cite{tJHiggs}.
It should be remarked that
even in the SP state, the correlation function averaged over 
positions and directions does not exhibit any singular behavior that
indicates the existence of sharp phase boundary.

%%%%%%%%%%%%%%%%%%%%%%%%%%%%%%%%%%%%%%%%%%%%%%%%%%%%%%%%%%%%%%%
%FIG.16
%%%%%%%%%%%%%%%%%%%%%%%%%%%%%%%%%%%%%%%%%%%%%%%%%%%%%%%%%%%

%%%%%%%%%%%%%%%%%%%%%%%%%%%%%%%%%%%%%%%%%%%%%%%%%%%%%%%%%%%%%%%
%FIG.17
%\begin{figure}[h]
%\begin{center}
%\includegraphics[width=12cm]{fig17_S.eps}
%\includegraphics[width=3cm]{fig17a.eps}
%\hspace{0.4cm}
%\includegraphics[width=3cm]{fig17b.eps}
%\hspace{0.4cm}
%\includegraphics[width=3cm]{fig17c.eps}
%\vspace{-0.5cm}
%\caption{%(Color online)
%Snapshots of density of a-atoms (left), b-atom (center) and holes (right)
%in PS state for $c_3=7.6$.
%}
%\label{fig:snapPS}
%\end{center}
%\end{figure}
%%%%%%%%%%%%%%%%%%%%%%%%%%%%%%%%%%%%%%%%%%%%%%%%%%%%%%%%%%%
%%%%%%%%%%%%%%%%%%%%%%%%%%%%%%%%%%%%%%%%%%%%%%%%%%%%%%%%%%%%%%%
%FIG.18
\begin{figure}[h]
\begin{center}
\includegraphics[width=5.5cm]{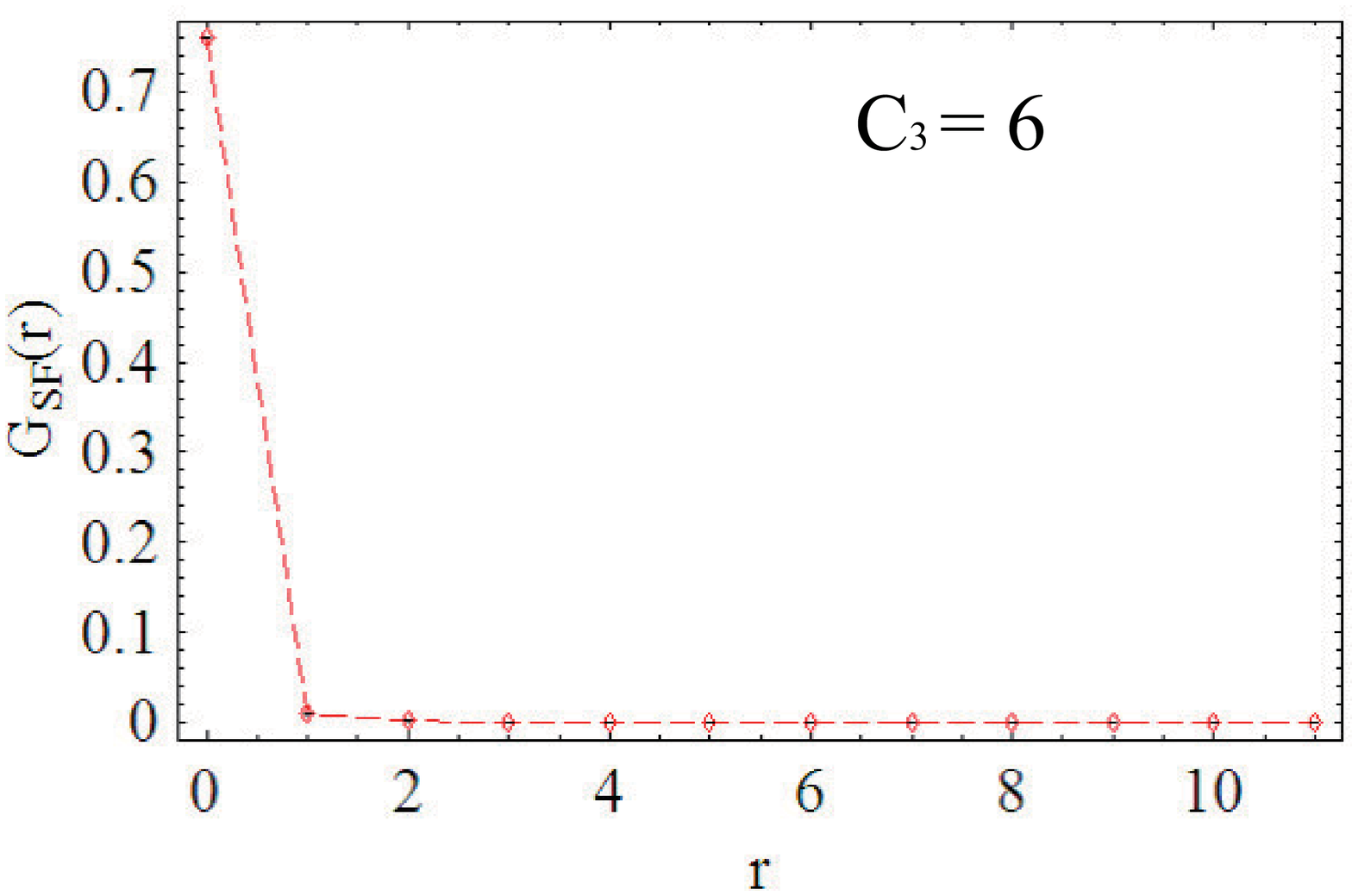}
\includegraphics[width=5.5cm]{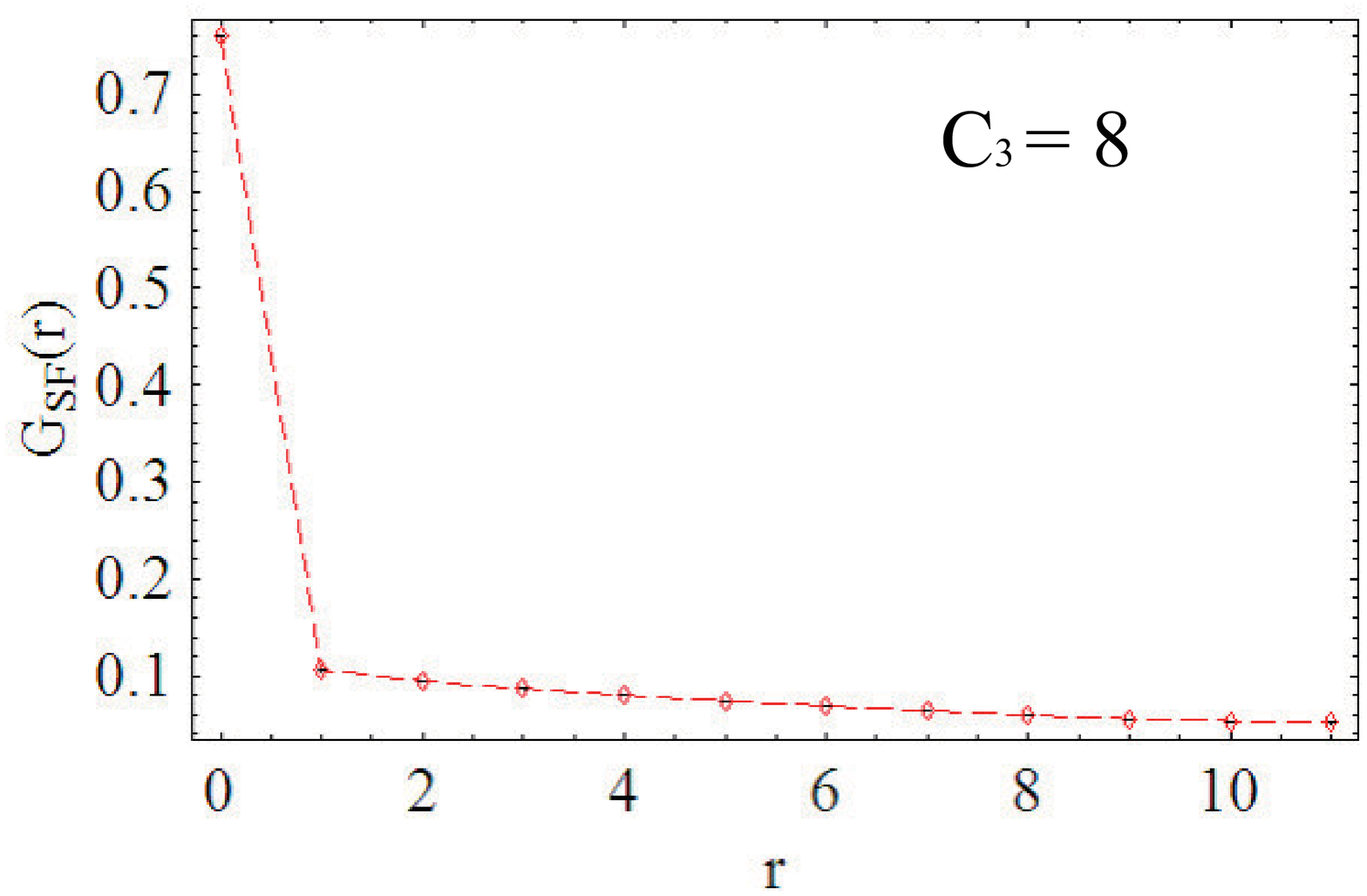} \\
\includegraphics[width=5.5cm]{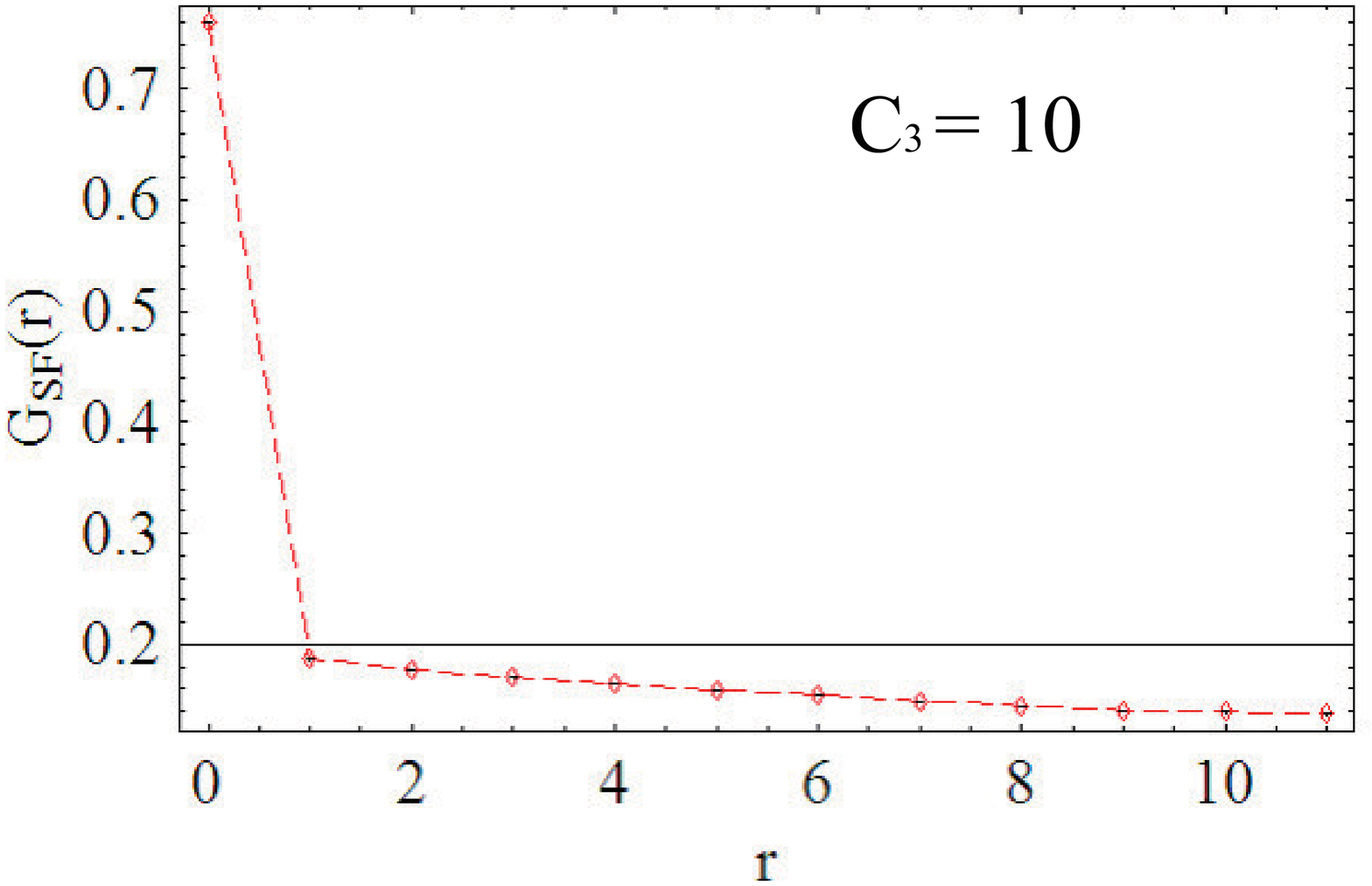}
\includegraphics[width=5.5cm]{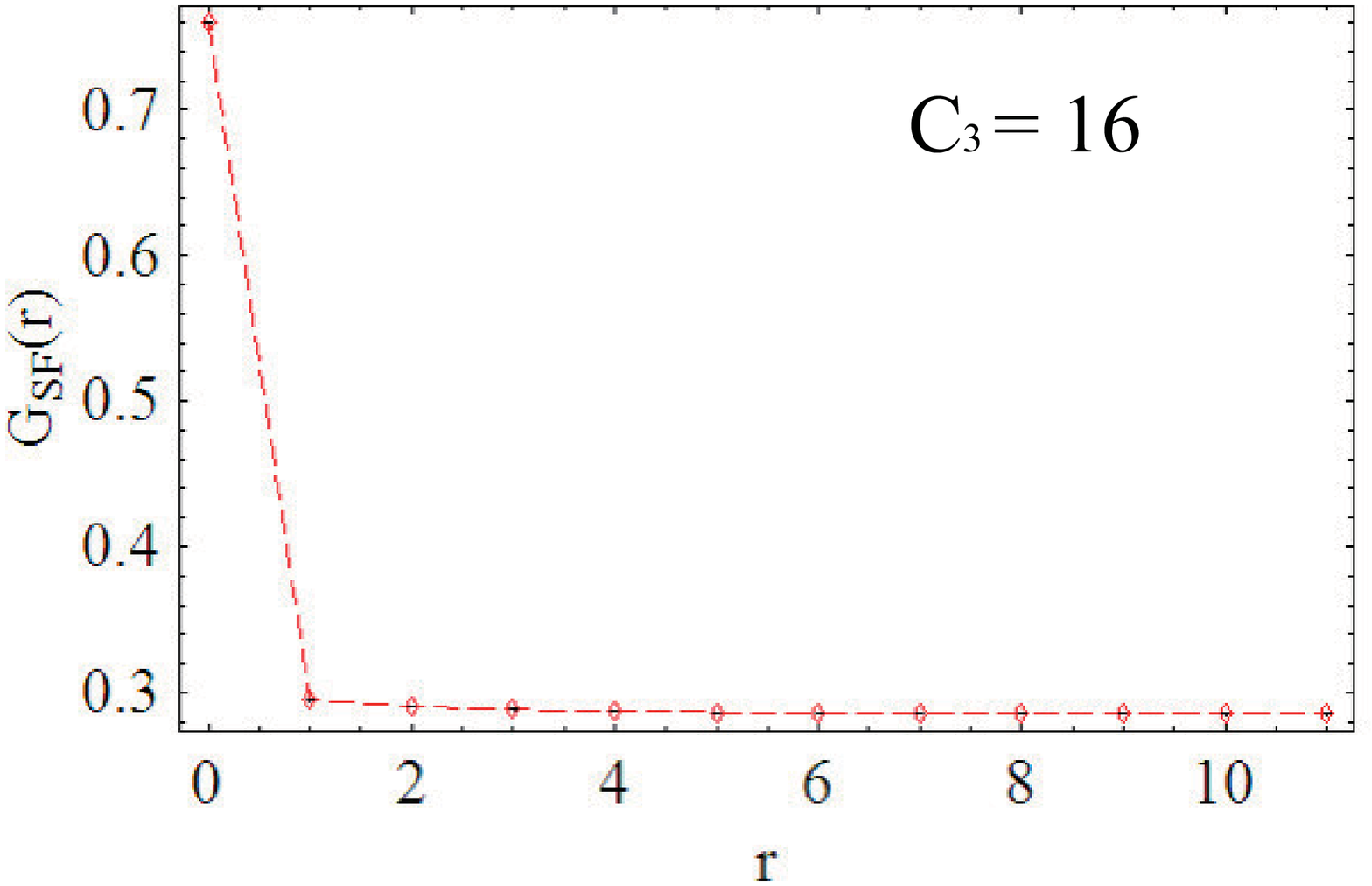}
%\vspace{-0.5cm}
\caption{%(Color online)
Correlation functions of atoms $G_{\rm SF}(r)$ for
$c_3=6, \ 8, \ 10$ and $16$.
}
\label{fig:atomcor2}
\end{center}
\end{figure}
%%%%%%%%%%%%%%%%%%%%%%%%%%%%%%%%%%%%%%%%%%%%%%%%%%%%%%%%%%%

%In Fig.\ref{fig:snapPS}, we show snapshots of density of $a$ atom, $b$ atom 
%and hole for $c_3=7.6$.
% It is clear that in the AF region, checkerboard configuration of
%$a$ and $b$ atoms is realized.
%On the other hand in the SF region, liquid-like distribution of holes 
%is realized.
%In fact,
%in each snapshot, there is a small fluctuation of hole density in the SF phase,
%but after averaging over various snapshots (configurations), 
%hole density has a quite homogeneous distribution. 

It is interesting to see how SF is realized in each phase.
In order to study it, we calculated the correlation function of 
atoms $G_{\rm SF}(r)$.
% (It is obvious that $G_a(r)=G_b(r)$ in the present case.)
The results in Fig.\ref{fig:atomcor2}
show that in the AF phase the correlation tends to vanish very
rapidly (no correlation even in the NN pair), and in the SF phase
it has a genuine LRO.
In the PS state, the correlator seems to have a small but finite LRO, but 
its finiteness comes from the correlation in the SF layers.
Then the above results are consistent with the observation by the GCE
in the previous section, which indicates that the SF correlation terminates
at AF layers.

The second phase existing between $c_3\simeq 7.6$ and $8.4$ is a new phase.
Counterparts of the other three were observed in the previous study of 
the 2D square lattice at $T=0$\cite{btJ2}.
From the observation that the genuine first-order phase transitions
separate the second phase from the AF solid state and the PS state, 
{\em it is not a simple mixture of the AF and SF}.
The most interesting possibility is that the phase is {\em a quantum superposition
of mesoscopic SF cloudlets in the AF background}\cite{QSP}.
We verified that the cloudlets change their shape, location and magnitude
via MC updates.
This fact supports the above expectation.
Anyway further study is required to clarify physical meaning of this interesting phase.

%%%%%%%%%%%%%%%%%%%%%%%%%%%%%%%%%%%%%%%%%%%%%%%%%%%%%%%%%%%%%%%%%%%%
\section{Effects of Nearest-Neighbor Interactions and Paired Superfluid}
\label{sec:PSF}
\setcounter{equation}{0}

In this section, we shall study effects
of NN attractive force between $a$ and $b$ atoms and investigate
the possibility of the PSF.
This problem is closely related to the superconductivity of the cuprates.
The high-$T_c$ materials are hole-doped AF magnets, and 
a pair of holes sitting on NN sites feel attractive force due to (short-range) 
AF background.
It is widely believed that this attractive force is an origin of the superconductivity.
The PSF in the bosonic t-J model corresponds to the SC in the
fermionic model with the condensation of the Cooper pair, 
whereas the single-atom SF corresponds to the metallic state
with movable electrons.
Then it is very interesting to study if the PSF is realized in the 
present system besides the single-atom SF and to see how the 
emergent PSF relates to the AF order.

%%%%%%%%%%%%%%%%%%%%%%%%%%%%%%%%%%%%%%%%%%%%%%%%%%%%%%%%%%%%%%%
%FIG.19
\begin{figure}[h]
\begin{center}
\includegraphics[width=7cm]{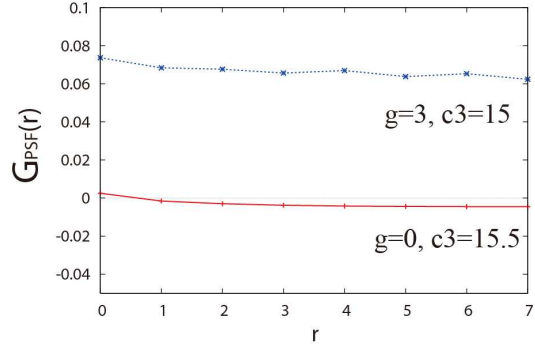}
%\vspace{-0.5cm}
\caption{%(Color online)
Correlation function of PSF in the cases $g=0$ and $g=3$
for $\rho_a=\rho_b=34\%$ and $c_1=3$.
Results indicate existence of small but finite LRO for $g=3$,
whereas no PSF for $g=0$.
}
\label{fig:PFSg=0}
\end{center}
\end{figure}
%%%%%%%%%%%%%%%%%%%%%%%%%%%%%%%%%%%%%%%%%%%%%%%%%%%%%%%%%%

Existence of the PSF is examined by measuring the following correlation function, 
\begin{equation}
G_{\rm PSF}(r)={1 \over 3}
\sum_i\Big(\langle (a^\dagger_r b^\dagger_{r+i})
(a_0b_i)\rangle-\langle a^\dagger_r a_0\rangle
\langle b^\dagger_{r+i} b_i\rangle\Big).
\label{PSF}
\end{equation}
We show the calculation of $G_{\rm PSF}(r)$ in the SF state,
i.e.,  $\rho_a=\rho_b=34\%$ and $c_1=3$, in Fig.\ref{fig:PFSg=0}.
The result obviously indicates that there is no finite density of the 
PSF in the SF of the system $H_{\rm tJ}$.
We also verified that there exists no PSF in other cases.

Then we have studied effects of interaction between a pair of atoms on 
NN sites by adding the following term to the Hamiltonian,
\begin{equation}
\beta H_{\rm NN}=-
g\sum_{r,i}\Big( (a^\dagger_r a_r)(b^\dagger_{r+i}b_{r+i})
+ (b^\dagger_r b_r)(a^\dagger_{r+i}a_{r+i})\Big).
\label{HNN}
\end{equation}
It is obvious that the attractive force $H_{\rm NN}$ enhances the AF and
the PSF.
Experimental realization of ``long-range" interactions like (\ref{HNN}) in cold atom
system was recently discussed\cite{LRI}. 
We have investigated the {\em extended t-J model}
 $H_{\rm tJ}+H_{\rm NN}$ with fixed value of $g$.
First for $c_3=0$ and positive $g$, similar phase diagram to that of $c_3=g=0$ 
in Fig.\ref{fig:PD1} is obtained 
but the parameter region of the AF state is enlarged as a result of $H_{\rm NN}$.

We are again interested in how the AF state evolves as the
hopping parameter $c_3$ is increased.
As in the case of $g=0$ considered in the previous section, we study 
the low and high-density cases separately.

%%%%%%%%%%%%%%%%%%%%%%%%%%%%%%%%%%%%%%%%%%%%%%%%%%%%%%%%%%%%%%%
%FIG.20
\begin{figure}[h]
\begin{center}
\includegraphics[width=6cm]{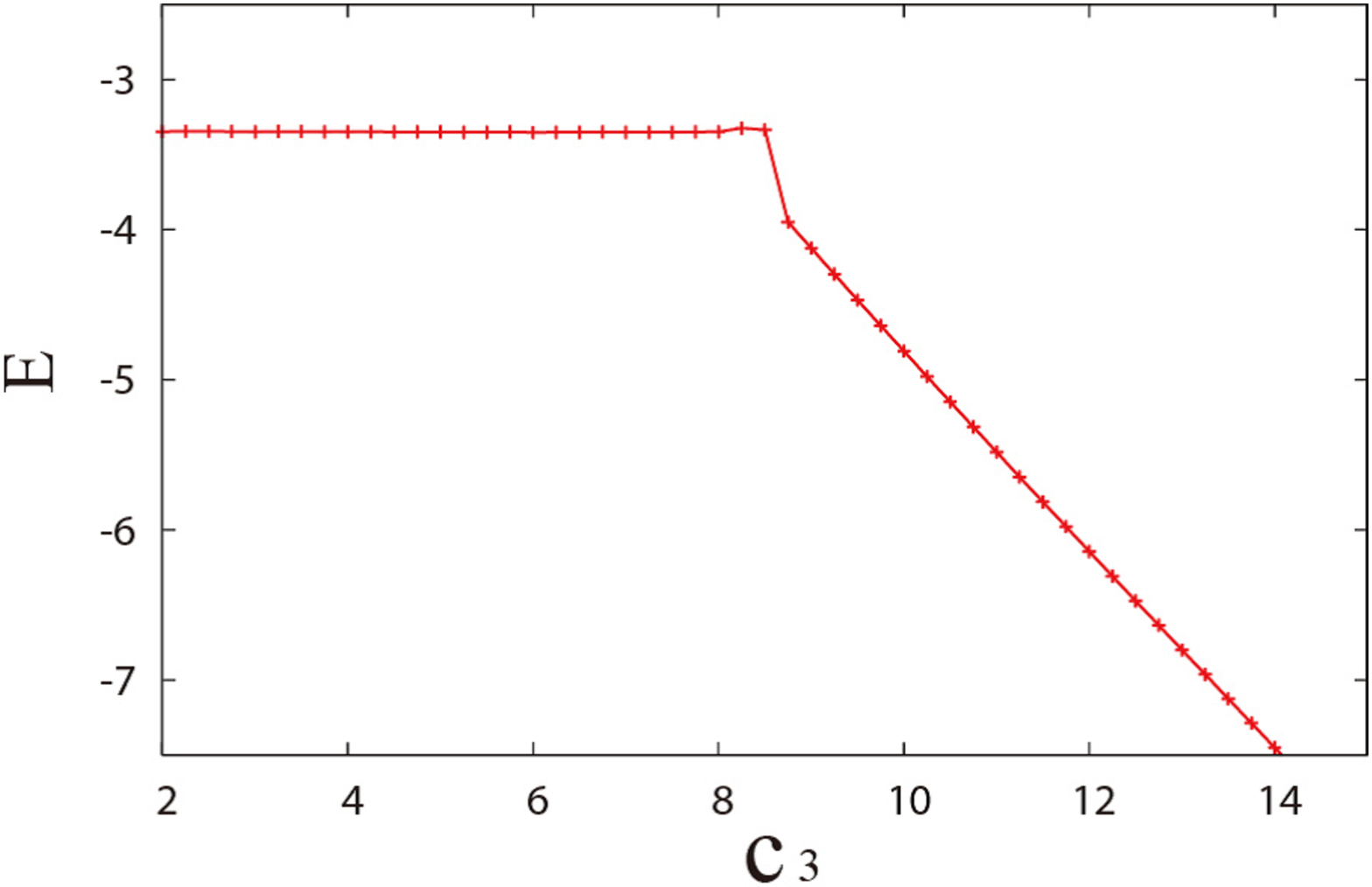}
\hspace{1cm}
\includegraphics[width=6cm]{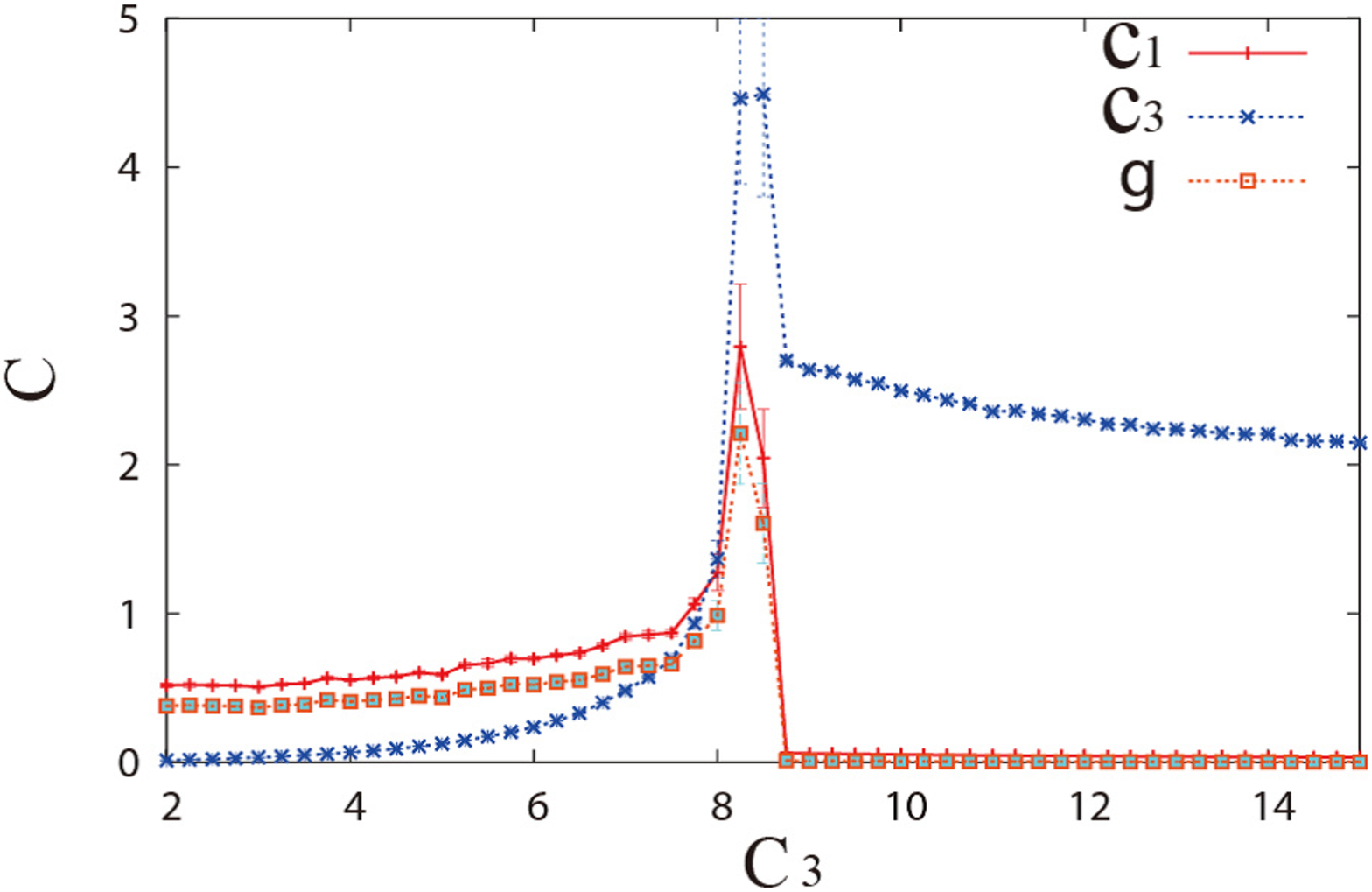}
%\vspace{-0.5cm}
\caption{%(Color online)
System of  $\rho_a=\rho_b=15\%$.
Internal energy $E$ and 
specific heat $C$ as a function of $c_3$ for $g=3,\ c_1=3.0$ and $\alpha=-0.5$.
In $C$, ``specific heat of each term" is shown.
}
\label{fig:ECg=3}
\end{center}
\end{figure}
%%%%%%%%%%%%%%%%%%%%%%%%%%%%%%%%%%%%%%%%%%%%%%%%%%%%%%%%%%

%%%%%%%%%%%%%%%%%%%%%%%%%%%%%%%%%%%%%%%%%%%%
\subsection{Low-density region}

\begin{figure}[h]
\begin{center}
\includegraphics[width=10cm]{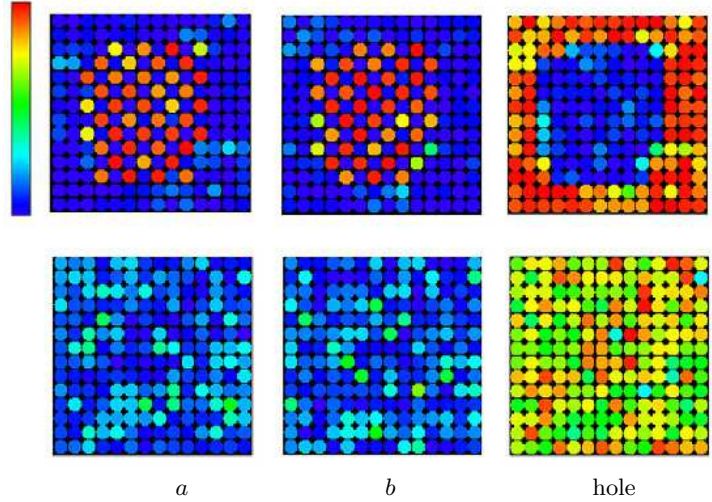} \\
%\hspace{0.5cm}
%\includegraphics[width=4cm]{fig21b.eps}
%\hspace{0.5cm}
%\includegraphics[width=4cm]{fig21c.eps}  \\
%\includegraphics[width=4cm]{fig21d.eps}
%\hspace{-0.5cm}
%\includegraphics[width=4.9cm]{fig21e.eps}
%\hspace{0.5cm}
%\includegraphics[width=4cm]{fig21f.eps} \\
\hspace{0.7cm}$a$ \hspace{2.4cm} $b$ \hspace{2.4cm} hole
\caption{%(Color online)
Snapshots $c_3=0$ (upper panels) and $c_3=8$ (lower panels) in low-density region
with $g=3$.
For $c_3=0$,
particles $a$ and $b$ gather in the central region forming the checkerboard
pattern, whereas the atoms are distributed rather homogeneously for $c_3=8$.
}
\label{fig:snap_low_g=3}
\end{center}
\end{figure}

\begin{figure}[h]
\begin{center}
\includegraphics[width=5cm]{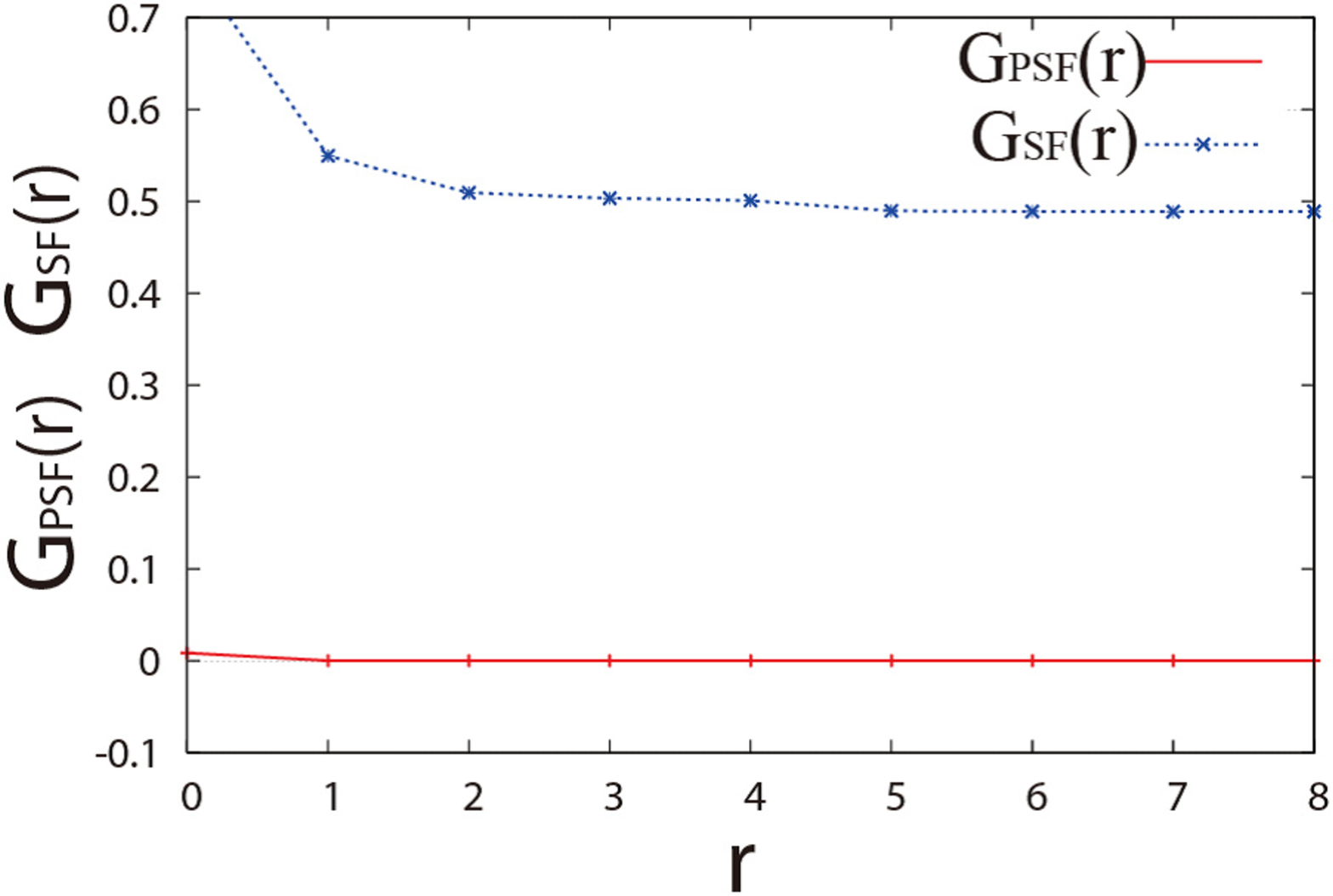}
\hspace{1cm}
\includegraphics[width=5cm]{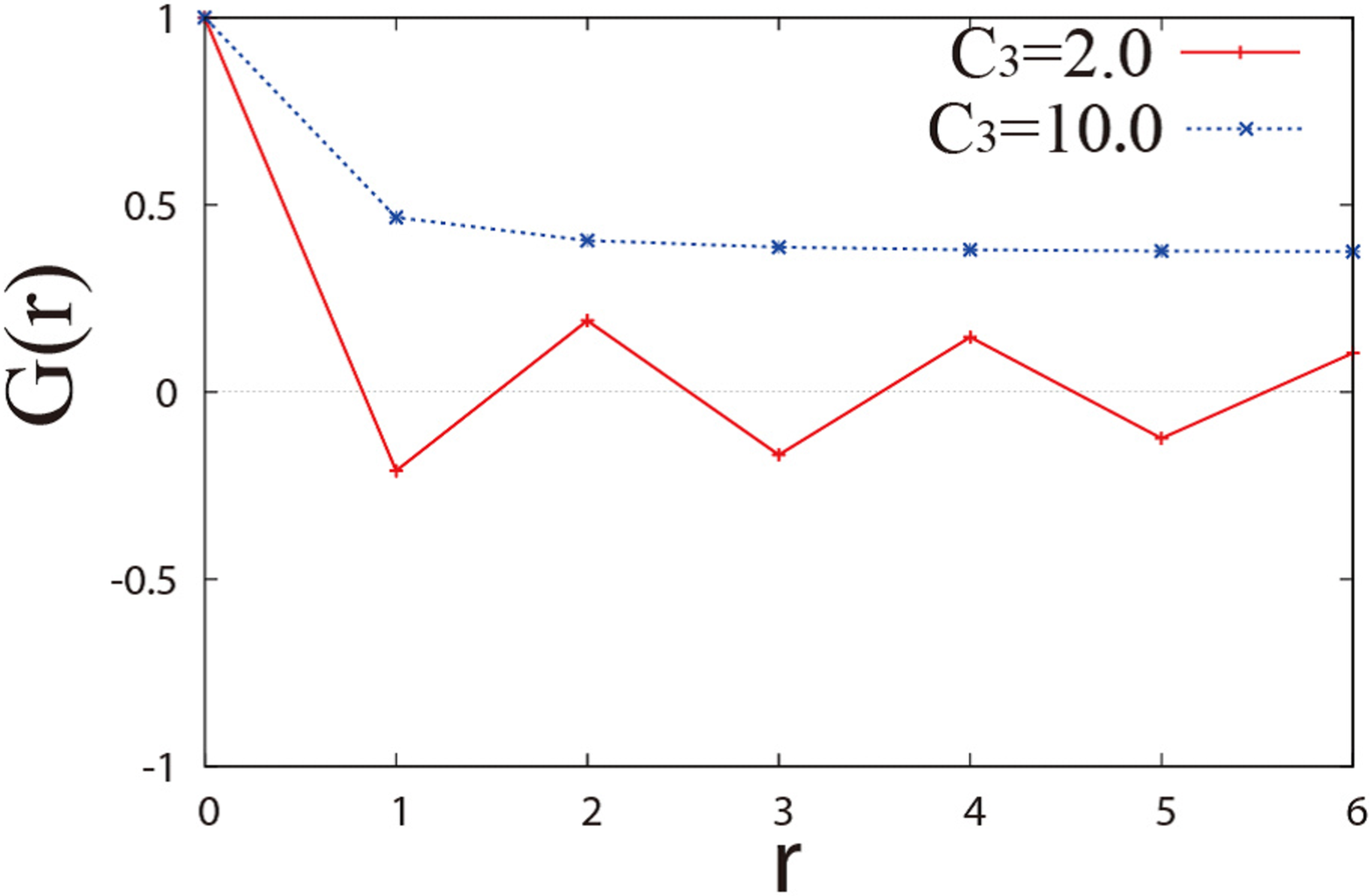}
\caption{%(Color online)
$G_{\rm SF}(r)$ and $G_{\rm PSF}(r)$ (Left panel).
$G_{\rm S}(r)$ (Right panel).
In the SF with low-atomic density, no PSF is generated and 
FM RLO for pseudo-spin correlation is formed. 
}
\label{fig:PSF_low}
\end{center}
\end{figure}

In this subsection, we consider the case with $\rho_a=\rho_b=15\%$
and $g=3$.
We show $E$ and $C$ as a function of $c_3$ in Fig.\ref{fig:ECg=3}.
It is obvious that there exists a first-order phase transition at $c_3\simeq 8.5$.
For small $c_3$, the system is divided into empty state and AF crystal
as in the case $g=0$, and after the phase transition the homogeneous SF state appears.
See Fig.\ref{fig:snap_low_g=3}.
We calculated $G_{\rm SF}(r)$, $G_{\rm PSF}(r)$ 
and $G_{\rm S}(r)$ in the SF phase, and show the results in Fig.\ref{fig:PSF_low}.
It is obvious that no PSF is generated and 
FM LRO for spin correlation is formed in the SF. 
This means that effect of the NN attractive force is negligibly small
in the low-density region.

%%%%%%%%%%%%%%%%%%%%%%%%%%%%%%%%%%%%%%%%%%%%%%%%%%%%%%%%%%%%%%%
%FIG.21
%%%%%%%%%%%%%%%%%%%%%%%%%%%%%%%%%%%%%%%%%%%%%%%%%%%%%%%%%%
%%%%%%%%%%%%%%%%%%%%%%%%%%%%%%%%%%%%%%%%%%%%%%%%%%%%%%%%%%%%%%%
%FIG.22
%%%%%%%%%%%%%%%%%%%%%%%%%%%%%%%%%%%%%%%%%%%%%%%%%%%%%%%%%%

%%%%%%%%%%%%%%%%%%%%%%%%%%%%%%%%%%%%%%%%%%%%
\subsection{High-density region}

%%%%%%%%%%%%%%%%%%%%%%%%%%%%%%%%%%%%%%%%%%%%%%%%%%%%%%%%%%%%%%%
%FIG.23
\begin{figure}[h]
\begin{center}
\includegraphics[width=7cm]{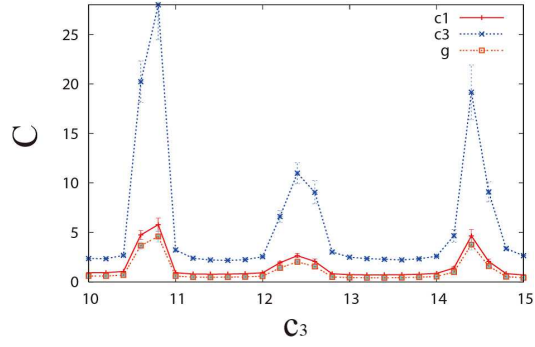}
%\vspace{-0.5cm}
\caption{%(Color online)
Specific heat of each term as a function of $c_3$ for $g=3,\ c_1=3.0$ and $\alpha=-0.5$.
There are three phase transitions at $c_3\simeq 10.8, \ 12.5$ and $14.5$.
}
\label{fig:SHg=1}
\end{center}
\end{figure}
%%%%%%%%%%%%%%%%%%%%%%%%%%%%%%%%%%%%%%%%%%%%%%%%%%%%%%%%%%

%%%%%%%%%%%%%%%%%%%%%%%%%%%%%%%%%%%%%%%%%%%%%%%%%%%%%%%%%%%%%%%
%FIG.24
\begin{figure}[h]
\begin{center}
\includegraphics[width=5cm]{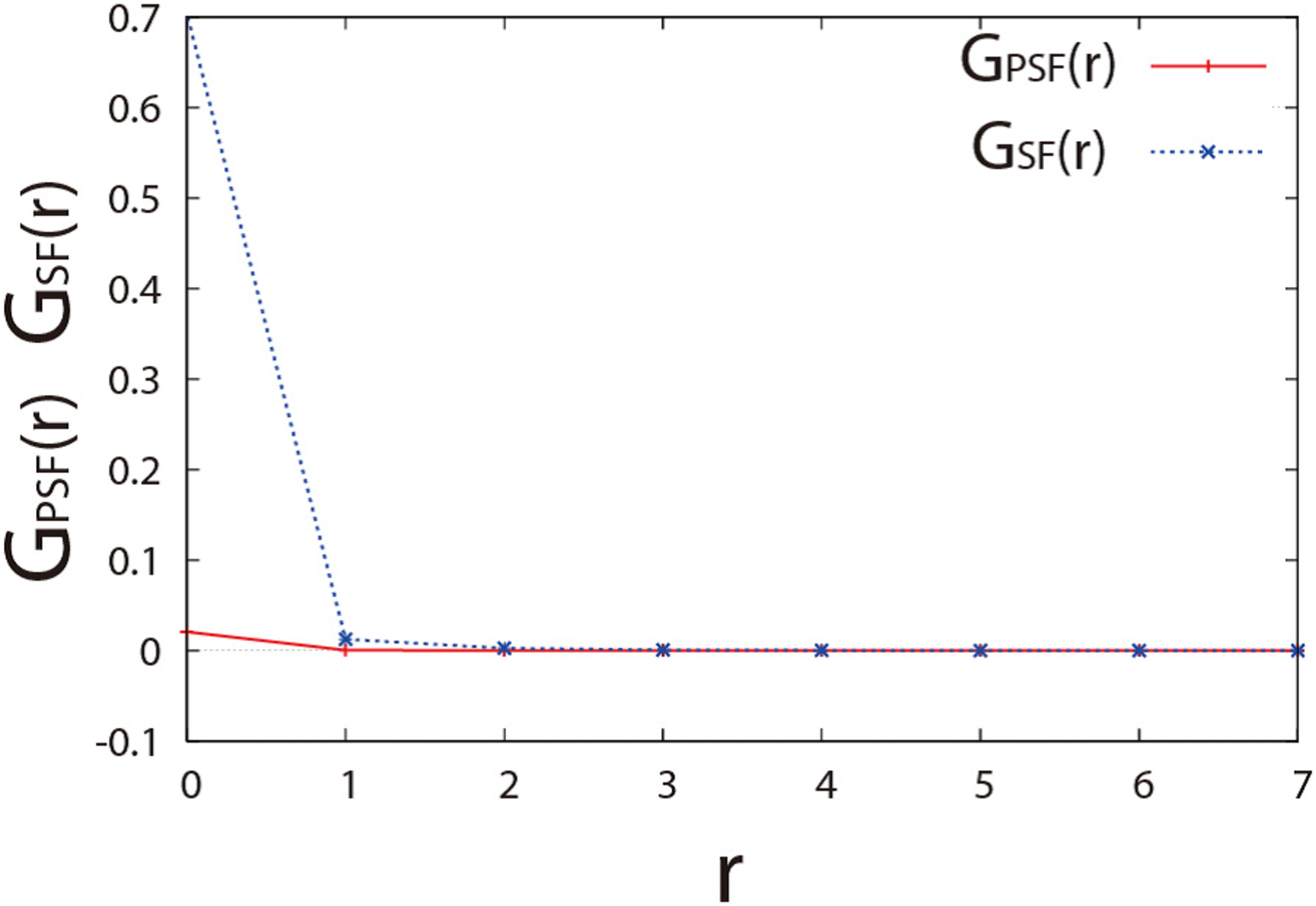}
\hspace{0.5cm}
\includegraphics[width=5cm]{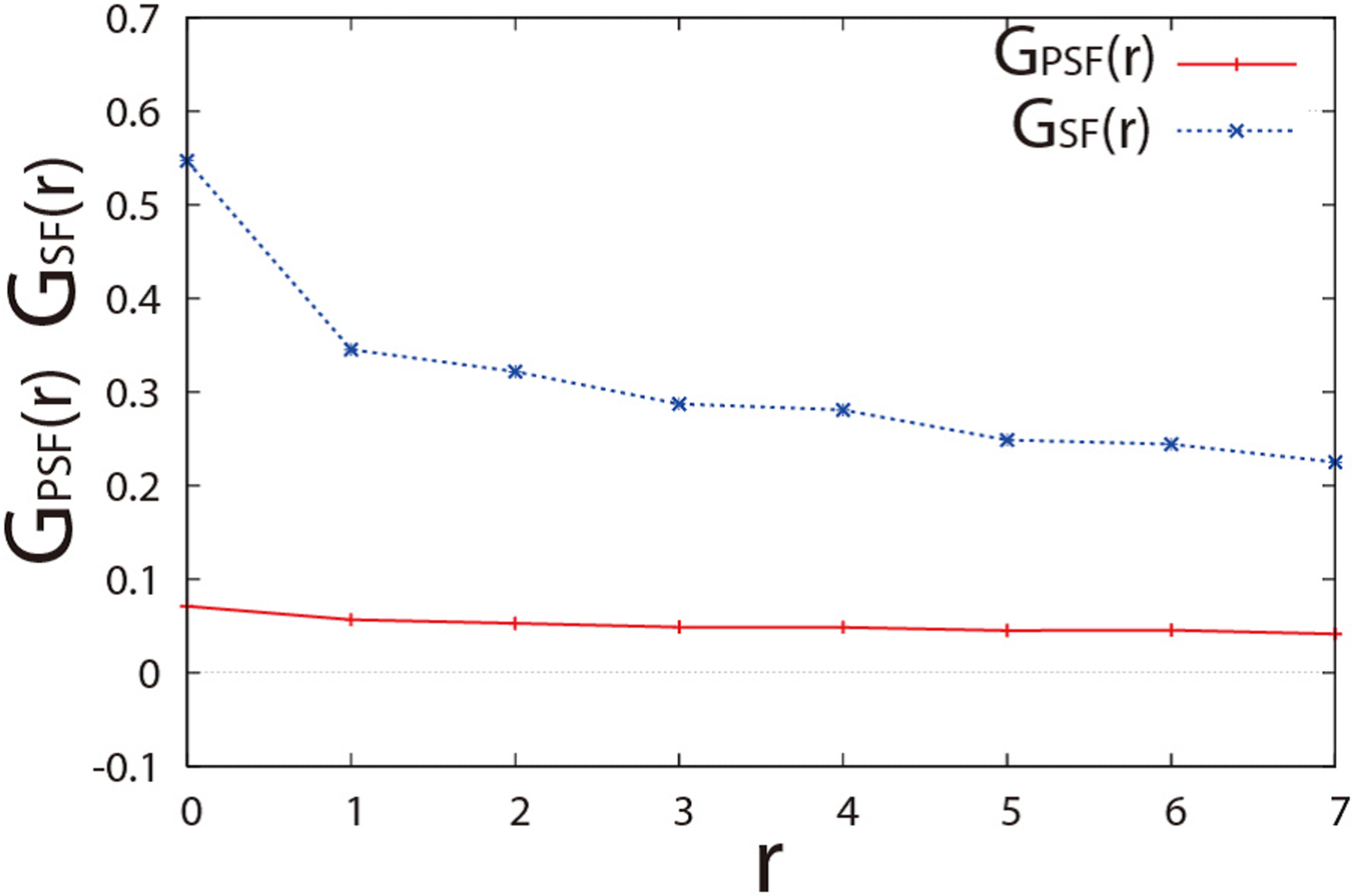} \\
\includegraphics[width=5cm]{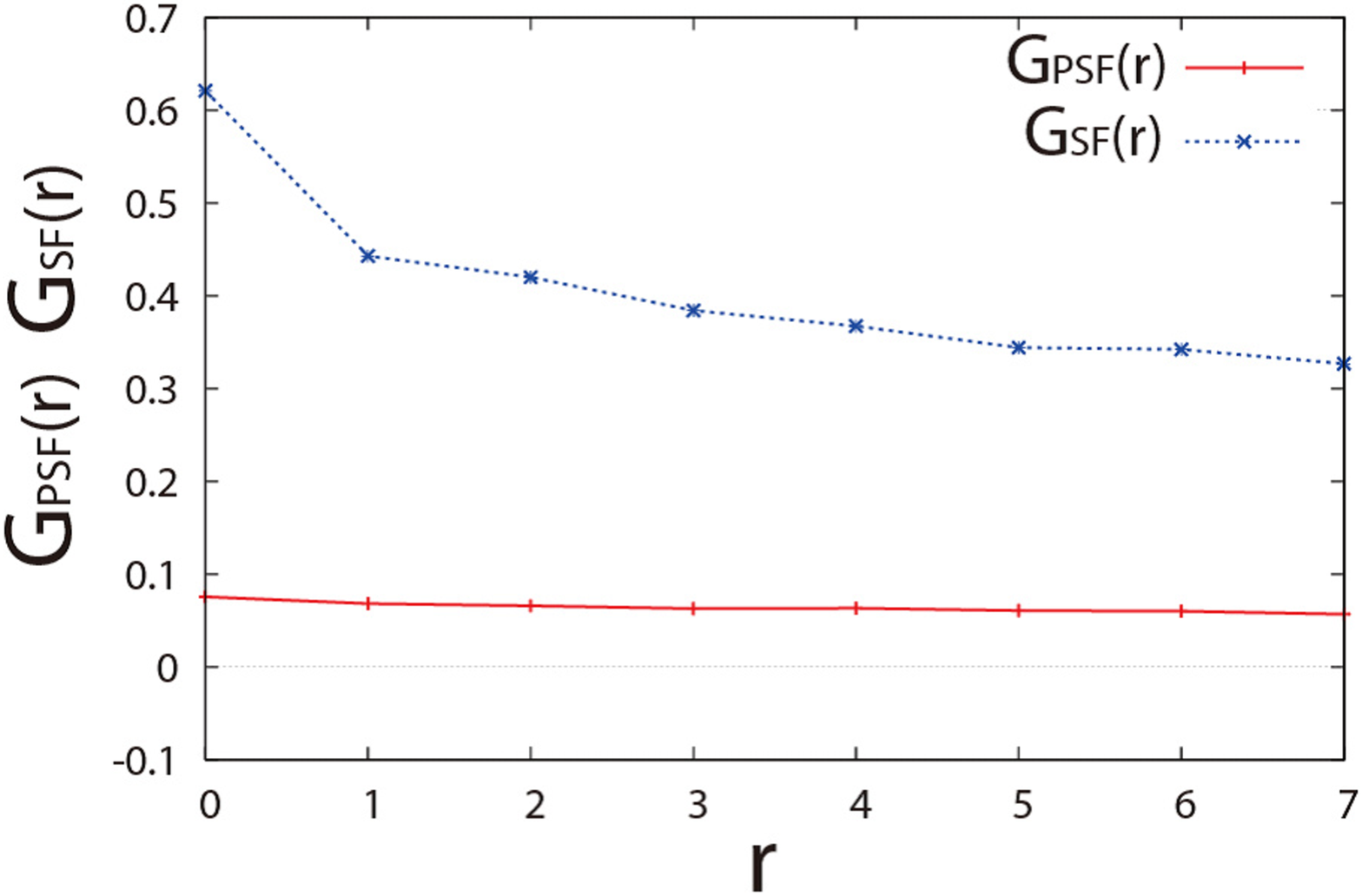}
\hspace{0.5cm}
\includegraphics[width=5cm]{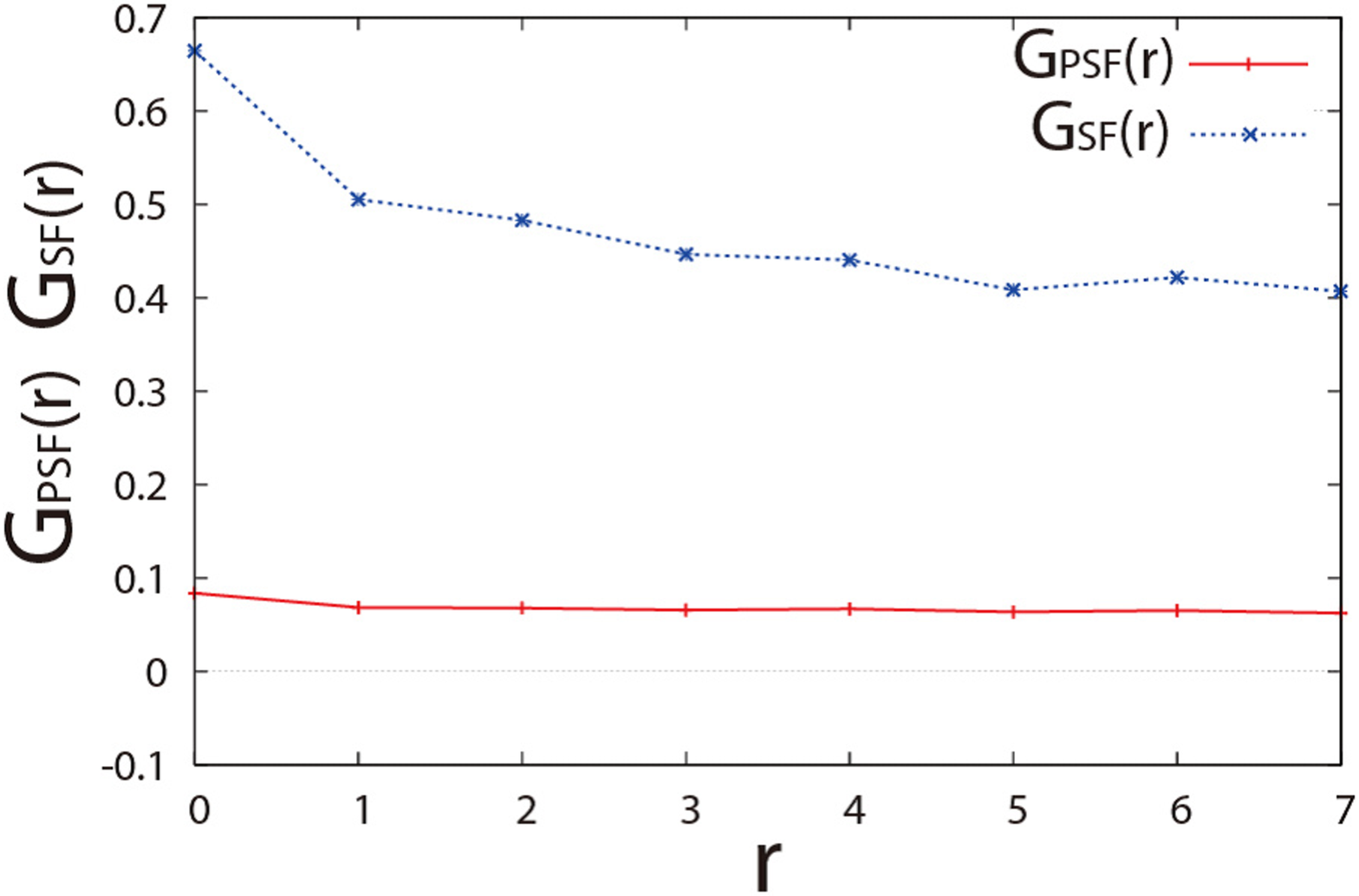}
%\vspace{-0.5cm}
\caption{%(Color online)
Correlation functions of atoms $G_{\rm SF}(r)$ and $G_{\rm PSF}(r)$ for
$c_3=10, \ 11.5, \ 13.5$ and $15$.
}
\label{fig:atomcor3}
\end{center}
\end{figure}
%%%%%%%%%%%%%%%%%%%%%%%%%%%%%%%%%%%%%%%%%%%%%%%%%%%%%%%%%%%
%%%%%%%%%%%%%%%%%%%%%%%%%%%%%%%%%%%%%%%%%%%%%%%%%%%%%%%%%%%%%%%
%FIG.25
\begin{figure}[h]
\begin{center}
\includegraphics[width=6cm]{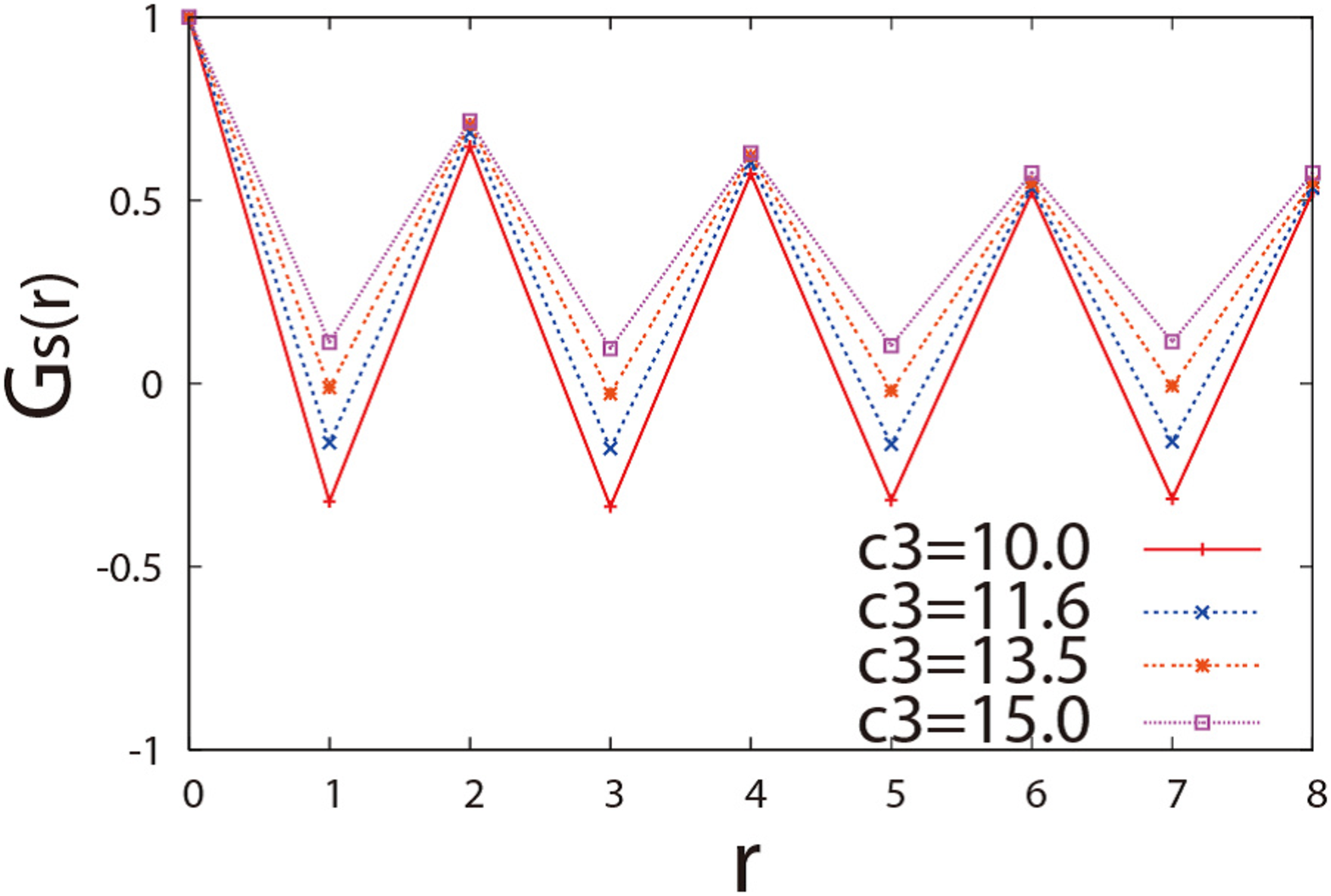}
\hspace{1cm}
\includegraphics[width=6cm]{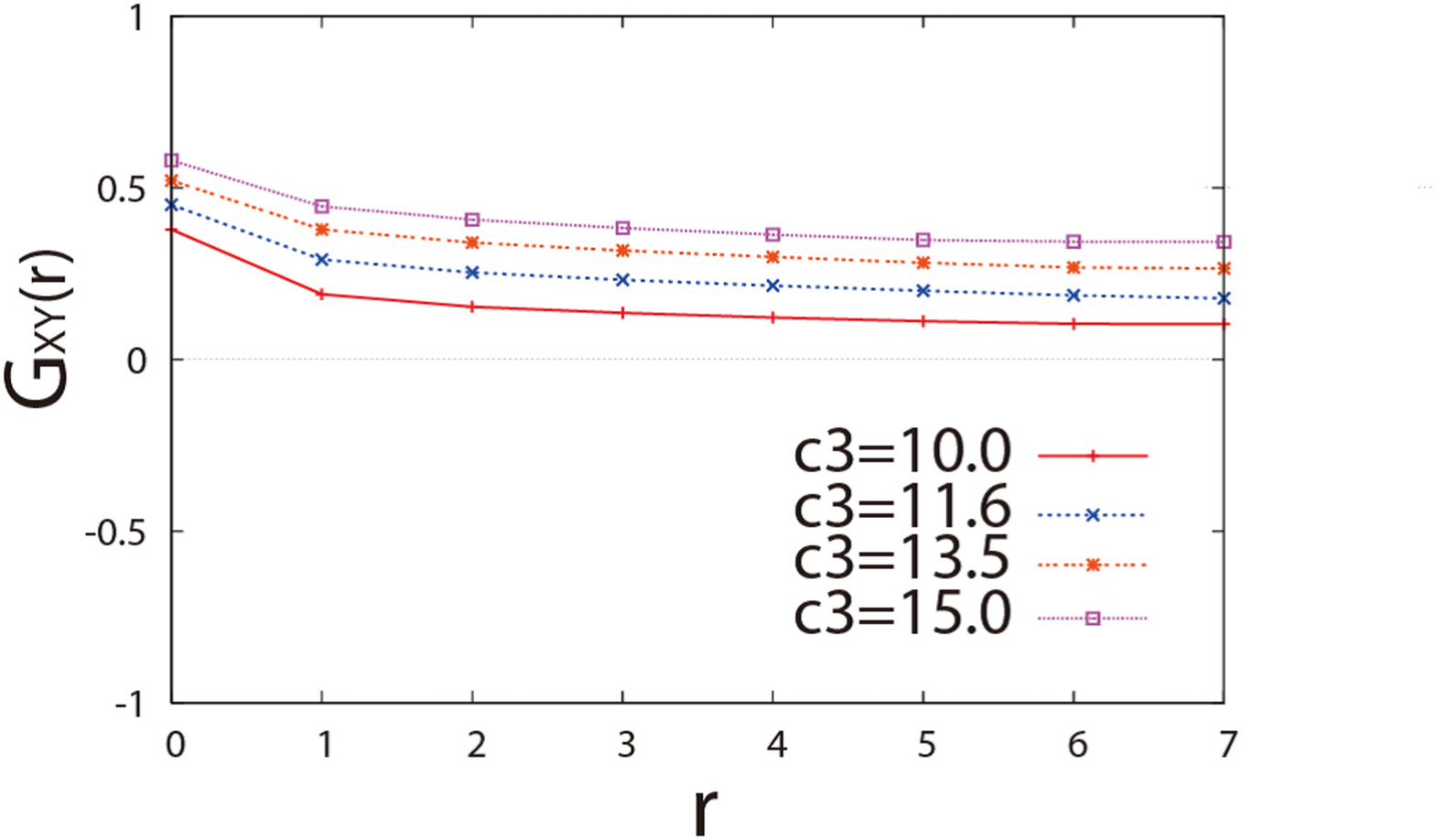}
%\vspace{-0.5cm}
\caption{%(Color online)
Correlation functions of pseudo-spins in each phase of model $g=3$.
There exists AF correlation in $z$-component, and also FM correlation
in $xy$-component.
}
\label{fig:atom_spin3}
\end{center}
\end{figure}
%%%%%%%%%%%%%%%%%%%%%%%%%%%%%%%%%%%%%%%%%%%%%%%%%%%%%%%%%%%
%%%%%%%%%%%%%%%%%%%%%%%%%%%%%%%%%%%%%%%%%%%%%%%%%%%%%%%%%%%%%%%
%FIG.26
\begin{figure}[h]
\begin{center}
\includegraphics[width=10cm]{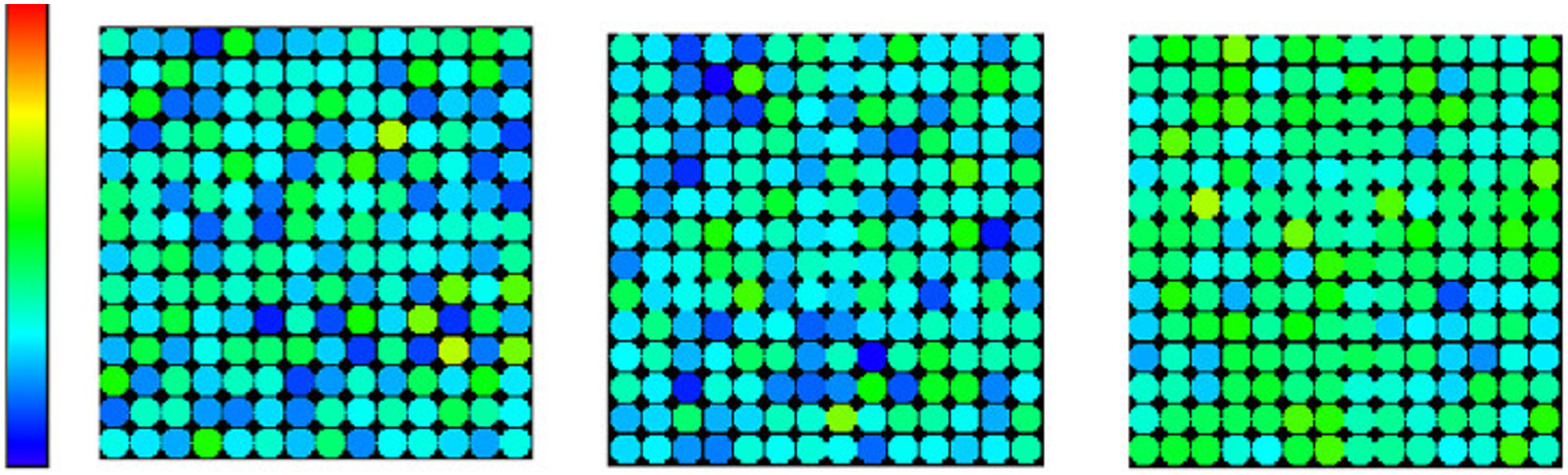}
%\hspace{0.5cm}
%\includegraphics[width=4cm]{fig26b.eps}
%\hspace{0.5cm}
%\includegraphics[width=4cm]{fig26c.eps}
%\vspace{-0.5cm}
\caption{%(Color online)
Snapshots of density of $a$-atom (left), $b$-atom (center) and hole (right) for 
$c_3=15$ and $g=3$.
The SS is realized in this parameter region.
Atoms are distributed rather homogeneously but there exist small regions
of the checkerboard pattern of the AF.
}
\label{fig:snap_g3}
\end{center}
\end{figure}
%%%%%%%%%%%%%%%%%%%%%%%%%%%%%%%%%%%%%%%%%%%%%%%%%%%%%%%%%%%

Then we studied the system with atomic density $\rho_a=\rho_b=34\%$,
i.e., in high-density region.
In Fig.\ref{fig:SHg=1}, we show the specific heat $C$ as a function of $c_3$ for $g=3$.
The result shows that there exist three phase transitions at $c_3\simeq 10.8$,
$12.5$ and $14.5$, respectively.
%We have verified that all of them are second-order phase transitions.
We also show the SF and the PSF correlation functions in Fig.\ref{fig:atomcor3}
(see also Fig.\ref{fig:PFSg=0}).
Contrary to the case $g=0$, all SF correlations have finite LRO
after the first phase transition at $c_3\simeq 10.8$.
Small but finite LRO of the PSF exists in all phases except the AF phase.

One may think that this difference in the SF behavior between in the $g=0$ and
$g=3$ cases, in particular (non)existence of the PSF,
is related to different behavior of spins in the two cases.
Then we studied the spin correlation and obtained the results shown
in Fig.\ref{fig:atom_spin3}, where
\begin{eqnarray}
%&&G_{\rm z}(r)
%={1\over L^3}\sum_{r_0}\langle S^z_{r_0}S^z_{r_0+r}\rangle, \nonumber \\
%&&
G_{\rm xy}(r)={1\over L^3}\sum_{r_0}\sum_{i=x,y}
\langle S^i_{r_0}S^i_{r_0+r}\rangle.
\label{GzGxy}
\end{eqnarray}
As expected, in the model $g=3$ there exists AF correlation in $z$-component, 
besides the FM correlation in $xy$-component.
We also studied the snapshots of each phase, which are quite helpful for
understanding properties of phases.
From these results, we found that the PS takes place even in the AF state
for $c_3<10.8$.
However in the present case, the phase is separated into empty phase with very
low particle density and the AF phase of checkerboard symmetry.
In the intermediate two phases, the AF solid and SF coexist but 
{\em they are immiscible}.
On the other hand in the SF for $c_3>14.5$, the global domain structure disappears,
and a homogeneous state is realized.
It is very interesting to see if the SS state is realized due to
the NN coupling $H_{\rm NN}$ in Eq.(\ref {HNN}).
Some previous study on the single-band Hubbard like model with a long-range
interaction suggested the existence of the SS in the square lattice.
The above results of the SFs and the spin correlation functions in the present case
show that {\em the state for $c_3>14.5$ is a SS and it appears as a result of 
the interaction $H_{\rm NN}$}.
Snapshots of the SS state in Fig.\ref{fig:snap_g3}
show that atoms are distributed rather homogeneously in the most spatial region 
but there also exist small domains of checkerboard symmetry.
The SS state is composed of quantum superposition of these inhomogeneous but 
structureless configurations.
It is easily seen that the (groundstate) wave function describing the 
observed SS in the MF level, $\Psi_{\rm SS}$, is given as follows,
\begin{eqnarray}
|\Psi_{\rm SS}\rangle&=&\prod_{r\in \mbox{\small even site}}[Aa^\dagger_rb^\dagger_{r+i}+B_ea^\dagger_r+B_2b^\dagger_r+C] \nonumber \\
&&\times\prod_{r\in \mbox{\small odd site}}
[Aa^\dagger_rb^\dagger_{r+i}+B_oa^\dagger_r+B_1b^\dagger_r+C]|0\rangle,
\label{WFSS}
\end{eqnarray}
where $A, B_e, B_o$ and $C$ are constants, and $B_e\neq B_o$ means the
existence of the AF order of the $z$-component besides the $xy$-FM.
The wave function (\ref{WFSS}) eloquently shows that the SS state
is the coherent superposition of the PSF, SF's and holes.
Each snapshot shows one component of the superposed state $\Psi_{\rm SS}$.

We expect that the obtained results in this section are quite instructive for the
mechanism of the high-$T_c$ SC of the cuprates.
It is expected that the single-particle SF in the bosonic t-J model
corresponds to the coherent hopping of electrons in the fermionic
t-J model, whereas the PSF corresponds to condensation of spin-singlet electron
pair at NN sites.
In other words, the single-particle SF {\em without} the PSF corresponds to
the (anomalous) metallic phase of the high-$T_c$ materials, whereas 
the single-particle SF {\em with} the PSF to the SC state.
Effective attractive force appears for a pair of holes 
sitting on NN sites in the AF background, as the NN hole pair break eleven 
AF bonds, whereas two holes separating more than one lattice spacing break 
twelve AF bonds\cite{IM}.
It is expected that this attractive force is an origin of the high-$T_c$ SC.
However the results of the present study show that the PSF does not exist
in the simple bosonic t-J model, though the phase transition from the insulating
AF state to the single-particle SF takes place as the hopping amplitude
is increased.
By adding the attractive force for particles sitting NN sites $H_{\rm NN}$ (\ref{HNN}), 
the PSF is actually realized. 
Furthermore the resultant PSF accompanies the AF order.
For the high-$T_c$ cuprates, coexistence of the AF and SC was actually
observed in the recent clean materials\cite{exp}.
This means that the attractive force coming from the magnetic terms in
$H_{\rm tJ}$
is not strong enough to generate the PSF in the present bosonic t-J model.
This conclusion may sound rather negative for the idea of AF force for 
the high-$T_c$ SC.
However, we found that the strong enough NN  attractive force generates 
both the AF order and the PSF.
This indicates a strong correlation between the above two phenomena.
Anyway, more precise study on difference between effect of coherent fermion
hopping and BEC is desired.

%%%%%%%%%%%%%%%%%%%%%%%%%%%%%%%%%%%%%%%%%%%%%%%%%%%%%%%%%%%%%%%%%%%%%%
\section{Conclusion}
\label{conclusion}
\setcounter{equation}{0}

\begin{figure}[h]
\begin{center}
%\vspace{0.7cm}
\includegraphics[width=7cm]{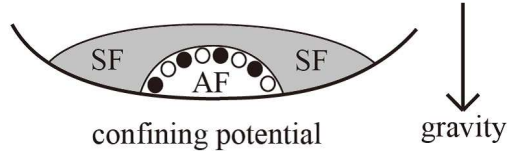}
%\vspace{-0.5cm}
\caption{%(Color online)
Intuitive picture of hetero-structure of AF crystal and SF,
which is expected to be observed by experiment.
}
\label{fig:hetero}
\end{center}
\end{figure}

In this paper we have studied the three-dimensional bosonic t-J model  at finite $T$.
This model corresponds to the large on-site repulsion limit of the
two-band Bose-Hubbard model.
We are particularly interested in the case of the fractional filling of atoms
and investigated the phase structure.
We studied the model both in the GCE and CE and found that the model has 
a very rich phase structure.
Besides the AF crystal, SF and SP state of the hetero-structure, there exist 
the inhomogeneous state with SF cloudlet in the AF solid.
We found that hetero-structure of the AF solid and SF has a sharp
phase boundary.
From this observation, we give an intuitive picture of the hetero-structure
that is expected to be observed by experiment in Fig.\ref{fig:hetero}.
%%%%%%%%%%%%%%%%%%%%%%%%%%%%%%%%%%%%%%%%%%%%%%%%%%%%%%%%%%%%%%%
%FIG.27
%%%%%%%%%%%%%%%%%%%%%%%%%%%%%%%%%%%%%%%%%%%%%%%%%%%%%%%%%%%

We also studied the effect of the NN attractive force between $a$ and
$b$ atoms.
We found that in certain parameter region there appears the SS and 
also PSF.
We think that this result gives an important insight into mechanism 
of high-$T_c$ SC of the cuprates.

In experiment, density and/or mass of two atoms can be varied.
Then study of the cases like $t_a\neq t_b$ and/or $\rho_a\neq \rho_b$
is important.
This problem is under study and results will be reported in a future publication.

\bigskip

\acknowledgments 
This work was partially supported by Grant-in-Aid
for Scientific Research from Japan Society for the 
Promotion of Science under Grant No.20540264
and No23540301.

%%%%%%%%%%%%%%%%%%%%%%%%%%%%%%%%%%%%%%%%%%%%%%%%%%%%%%%%%%%%%%%%%%


\begin{thebibliography}{99}
\bibitem{btJ2d}For the bosonic t-J model on a square lattice at $T=0$,
see Ref.\cite{btJ2}.
Some comments on three-dimensional system are also given there.
\end{thebibliography}
\end{document}